\documentclass[useAMS,usenatbib]{mnras}

\usepackage{booktabs,graphicx,amsmath,mhchem,dcolumn,xcolor,braket}
\newcommand{\Abinitio}{\emph{Ab initio}}
\newcommand{\abinitio}{\emph{ab initio}}
\newcommand{\cm}{cm$^{-1}$}
\newcommand{\um}{$\mu$m}
\newcommand{\X}{X $^4\Sigma^-$}
\newcommand{\Ap}{A$^{\prime}$ $^4\Phi$}
\newcommand{\A}{A $^4\Pi$}
\newcommand{\B}{B $^4\Pi$}
\newcommand{\C}{C $^4\Sigma^-$}
\newcommand{\D}{D $^4\Delta$}
\newcommand{\mc}{\multicolumn}
\newcolumntype{d}{D{.}{.}{-1}}
\newcommand{\Da}{a $^2\Sigma^-$}
\newcommand{\Db}{b $^2\Gamma$}
\newcommand{\Dc}{c $^2\Delta$}
\newcommand{\Dd}{d $^2\Sigma^+$}
\newcommand{\De}{e $^2\Phi$}
\newcommand{\Df}{f $^2\Pi$}
\newcommand{\Dg}{g $^2\Pi$}
\newcommand{\Dh}{h $^2\Pi$}
\newcommand{\Duo}{{\sc Duo}}

\newcommand{\pk}{k}
\newcommand{\m}{m}
\newcommand{\pt}{t}

\newcommand{\pll}{l_2}
\newcommand{\HSO}{\hat{H}_\text{SO}}
\newcommand{\Msol}{\ensuremath{M_{\odot}}}
\newcommand{\pp}{^{\prime\prime}}

\title[ExoMol line lists XVIII: VO]{ExoMol line lists XVIII. The high temperature spectrum of VO.}
\author[McKemmish et al.]{Laura K. McKemmish, Sergei N. Yurchenko and Jonathan Tennyson.
\\
Department of Physics and Astronomy, University College London, Gower Street, WC1E 6BT London
}
\begin{document}

\date{\today}

\pagerange{\pageref{firstpage}--\pageref{lastpage}} \pubyear{2002}

\maketitle

\label{firstpage}

\begin{abstract}
  An accurate line list, VOMYT, of spectroscopic transitions is presented for hot
  VO. The 13 lowest electronic states are
  considered. Curves and couplings are based on initial {\it ab
    initio} electronic structure calculations and then tuned using available experimental
  data.  Dipole moment curves, used to obtain transition intensities,
  are computed using high levels of theory (e.g. MRCI/aug-cc-pVQZ using state-specific or minimal-state CAS for dipole moments).
  This line list contains over 277 million transitions between almost 640,000 energy levels.
  It covers the wavelengths longer than 0.29 $\mu$m and includes all
  transitions from energy levels within the lowest nine electronic states which have energies less than 20,000 \cm{} to upper
  states within the lowest 13 electronic states which have energies below 50,000 \cm{}. The line lists give significantly increased
  absorption at infrared wavelengths compared to currently available
  VO line lists. The full line lists is made available in electronic
  form via the CDS database
  and at www.exomol.com.
\end{abstract}


\section{Introduction}


Vanadium oxide (VO) plays an important role in astrophysical
chemistry, particularly of cool stars, and is expected to also be
present in brown dwarfs and hot Jupiter exoplanets.  However, no
comprehensive, high quality line list has been published for this
molecule, limiting the potential information that can be obtained. The
ExoMol project \citep{jt528,jt569} aims to produce high
temperature line lists of spectroscopic transitions for key molecular
species likely to be significant in the analysis of the atmospheres of
extrasolar planets and cool stars. The molecular data is crucial for
accurate astrophysics models of the opacity, as discussed by
\citet{sb07} and \citet{09Bexxxx.exo},  and the spectroscopy of the
object. However, from a chemistry perspective, vanadium is a transition metal in terms of its electronic structure and spectroscopic properties. This makes the electronic structure calculations much more difficult and gives higher uncertainties \citep{jt632}.

VO absorption bands are generally present in cool late M
class stars with effective temperature on order 2500-3000 K, mass less
than 0.1 \Msol\ and are also expected to be observed in hot
Jupiter exoplanets \citep{08FoLoMa.VO}.  VO is generally present simultaneously with TiO and has
similar spectroscopic and thermodynamic properties, though its
abundance is about an order of magnitude less than TiO.  VO tends to
be more important in classifying slightly cooler (i.e. late) M dwarfs  \citep{93KiKeRi.VO,04McKiMc.dwarfs}.
VO is one of the dominant species in the spectra of young hot brown dwarfs \citep{04McKiMc.dwarfs,06KiBaBu.dwarfs,08PeMeLu.dwarfs}.

\begin{figure}
\includegraphics[width=0.5\textwidth]{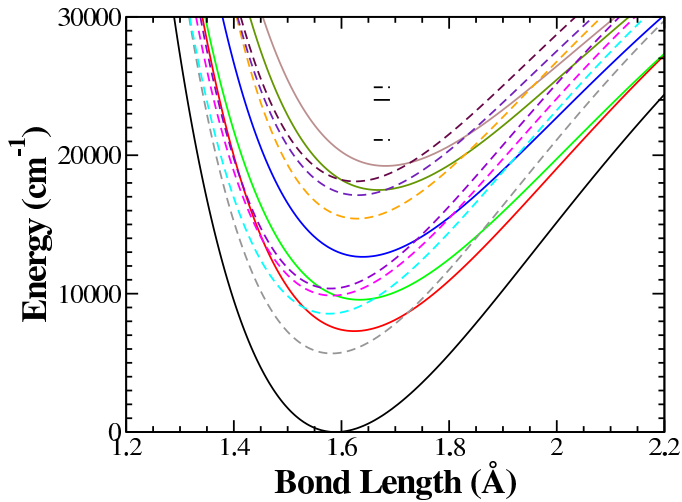}
\caption{\label{fig:PES} Potential Energy Curves. Curves in ascending order are: solid; \X, \Ap, \A, \B, \C, \D; dashed; \Da, \Db, \Dc, \Dd, \De, \Df, \Dg. The short horizontal lines above the full curves are indicative of the next highest  known quartet (solid) and doublet (dashed) states based on experimental data; there are many extra doublets believed to exist above 20,000 \cm{} not shown on this graph.}
\end{figure}

\begin{table*}
\def\arraystretch{1.2}

\caption{\label{tab:Morse} Specifications for extended Morse oscillator parameters (see Eq \ref{eq:EMO}) for the fitted potential energy
curves in {\Duo}. The dissociation energy, $D_e$, is 52288.4 \cm{} and the reduced mass of $^{51}$V$^{16}$O is 12.17296118 Da. }
\begin{tabular}{lrrrrrrr}
\toprule
State & \mc{5}{c}{{Morse  parameters}} & \mc{1}{c}{Properties} \\
\cmidrule(r){2-6} \cmidrule(r){7-7}
 & \mc{1}{c}{$T_e$/\cm}&\mc{1}{c}{$r_e$/\AA} & \mc{1}{c}{$b_0$} & \mc{1}{c}{$b_1$} & \mc{1}{c}{$b_2$} & \mc{1}{c}{$\omega_e$/\cm} \\
\midrule
\X & 0.000 & 1.589443 & 1.87974 & 0 & 0 & 1011.8\\
\Ap &7293.270 & 1.622990 &  1.89529 & 0 & 0 &946.4\\
\A & 9561.867 &1.633621  &  1.8200$^*$ & 0 & 0 & 890.3\\
\B &12655.372 & 1.640253 & 1.94733 & 0 & 0  & 912.4\\
\C & 17487.690 & 1.670620 & 1.94674 & -0.359 & 0.540 & 864.9\\
\D & 19229.786 & 1.683170 & 1.96500 & 0.060 & 0.900 & 840.9\\
\Da & 5630.00$^*$ & 1.58200$^*$ & 2.0100$^*$ & 0 & 0 & 1041.7\\
\Db &8551.49$^*$ & 1.57700$^*$ & 2.0900$^*$ & 0 & 0 & 1009.0\\
\Dc & 9860.107 & 1.582247 & 2.07538 & 0 & 0 & 1006.1\\
\Dd &10343.630 & 1.578560 & 2.16131 & 0 & 0  & 1041.7\\
\De & 15440.547 & 1.628968 & 2.0900$^*$ & 0 & 0 &  985.2\\
\Df & 17115.919 & 1.629330 & 2.12146 & 0.226 & 0.471  & 937.2\\
\Dg & 18108.500 & 1.635810 & 2.17648 & 1.157 & -1.850  &947.5\\
\bottomrule
\end{tabular}

$*$: Fixed based on theory or low-resolution experiment.
\end{table*}

The A-X transition of VO, which occurs at approximately
1.05 $\mu$m in the infrared, was first observed in the red giant
Mira-type variable stars Mira Ceti and R Leonis by \citet{47KuWiCa.VO},
and subsequently studied more extensively by
\citet{52KeShxx.VO,66SpYoxx.VO,67WiSpKu.VO,69SpWixx.VO,98AlPexx.VO} and \citet{00CaLuPi.VO}.
VO has also been observed in M red-dwarf stars \citep{98AlPexx.VO,09Bexxxx.exo,14RaReAl.VO}.
Molecular lines of VO have been detected in sunspot umbral spectra \citep{08SrBaRa.VO}.
\citet{08DeViLe.VO} found non-definitive evidence for VO in the
atmosphere of the hot Jupiter HD209458b. Tentative detection of VO and TiO in the hot atmosphere of the hot exoplanet WASP-121b were recently reported by \citet{16EvSiWa.exo}.

There has been considerable
recent debate
\citep{09SpSiBu.VO,10FoShSo.VO,10BoLiDu.VO,10MaSexx.VO,12HuSiVi.VO,13SpBuxx.VO,13GiAiBa.VO,13PaShLi.VO,14AgPaVe.VO,15HoKoSn.VO,15ScBrKo.VO,15HaMaMa.VO,15WaSixx.VO}
about a possible temperature inversion in hot Jupiters, potentially
caused by the presence of TiO and VO.
 \citet{15HoKoSn.VO} highlight the need for more accurate
line lists to resolve this issue; though they specifically mention TiO
in this paper, VO usually coexists, although it is generally thought to have a lower abundance.
Line lists for TiO \citep{98Plxxxx.TiO,98Scxxxx.TiO,VALD3},
despite their shortcomings, are still significantly more developed than those for VO.

\citet{Kurucz} and \citet{99Plxxxx.VO} have both circulated VO line lists. Both these line
 lists contained only transitions in the main A-X, B-X and C-X bands
 (in particular, no infrared X-X transitions were included). These line lists have been used extensively in stellar and planetary models.
\citet{89BuHuLu.VO} give an early study based on a simple model
atmospheres incorporating TiO and VO opacities. In particular, VO is
an important component of model atmospheres for M dwarfs
\citep{95AlHaxx.VO,12RaReSc.VO,14RaReAl.VO}. The more complex T Tauri
atmosphere models have also incorporated VO absorption bands
\citep{14HeHixx.VO}.

A good VO line list is especially important in light of the new
generation of proposed and planned satellites with the ability to take
high quality spectra of hot Jupiters. These are required both for missions
purpose-designed for studying exoplanet spectroscopy \citep{jt523,jt578,11TiChGr.exo}
or more general purpose satellites such as James Webb Space
Telescope (JWST) which will also have the capability to study atmospheres of
hot Jupiters in 0.6-- 28 \um\ region
\citep{14BeBeKn.Sa,15WaSixx.VO,15BaAiIr.Sa}.  

VO is generally critical in modelling oxygen-rich astronomical objects
with temperatures between 1500-3000 K: at lower temperatures, it
condenses to more complex oxides while at higher temperatures it partially
dissociates. VO may continue to be a
non-negligible source of opacity and absorption up to 5000 K;
therefore, we aim for a 90\% complete line list up to this
temperature. The resulting line list should automatically be valid for
any lower temperatures. 

Due to its astronomical importance in the spectroscopic analysis of M-dwarfs,
the spectroscopy of diatomic VO has been well studied
experimentally
\citep{68Kaxxxx.VO,68LaKaxx.VO,68RiBaxx.VO,68RiBaxy.VO,69HaNixx.VO,81ChHaLy.VO,81HoMeMi.VO,82ChHaMe.VO,82ChTaMe.VO,87MeHuCh.VO,89Mexxxx.VO,92HuMeCl.VO,94ChHaHu.VO,95AdBaBe.VO,02RaBeDa.VO,05RaBexx.VO,09HoHaMa.VO}.
A good summary of previous experimental results is given by
\citet{07MiMaxx.VO} and \citet{09HoHaMa.VO}. 
Generally, only the ground and first vibrational
energy levels are well characterised for observed electronic states. Fortunately, transitions
between quartet and
doublet electronic states have been observed; this enables
the relative positioning of the quartet and doublet manifold to be
fixed with reasonably accuracy (limited by the fact that the absolute
value of some spin-orbit terms is unknown experimentally).

The dipole moment of the ground state has been measured by
\citet{91SuFrLo.VO}.  There is no information on the transition dipole
moments of VO. However, lifetime measurements for levels in the A, B and C states
were performed by \citet{97KaLiLu.VO}.

The spectroscopy of VO has also been well studied theoretically
\citep{66CaMoxx.VO,77WoFaSh.VO,86BaLaxx.VO,87DoWeSt.VO,95BaMaxx.VO,96BaStTs.VO,00BrRoxx.VO,01BrBoxx.VO,03DaDeYa.VO,03PyWuxx.VO,04MaMiOw.VO,06DuWaSh.VO,07YaGuSo.VO,08DuWaSh.VO,08BaLuxx.VO,10KuMaxx.VO,11PrGuWe.VO}.
However, the challenging nature of theoretical treatments of excited states in
transition metal diatomics means that achieving quantitative accuracy is very
difficult, particularly for excitation energies \citep{jt632}.  Generally,
multi-reference configuration interaction (MRCI) approaches are essential.

The most detailed \abinitio\ electronic structure study was performed by
\citet{07MiMaxx.VO}, who looked at the lowest nine
electronic states, and also reviewed previous theoretical
studies. The quality of these calculations is high, but quantitative results are only given for equilibrium values.

Another important study was performed by \cite{15HuHoHi.VO}, who calculated
the energetics of a much larger number of electronic states, but did not
consider dipole moment or spin-orbit couplings (except for the X equilibrium
dipole moment). They investigate the effect of including $3p$ correlation on the
internally-contracted (ic) MRCI results; however, the accuracy of the potential energy surface parameters
does not show significant (if any) improvement. Their  icMRCI calculations
incorrectly predict the ordering of the \C{} and \D{} states; we find similar
difficulties in reproducing the correct ordering of these states.

The goal of this paper is to produce a comprehensive line list for the main isotopologue of VO accounting for all of the lowest 13 electronic states of VO. 

The structure of this article is as follows. In Section \ref{sec:SMforVO}, the
spectroscopic model for VO is developed. In Section \ref{sec:LL}, the line list
for VO, named VOMYT, is constructed based on the spectroscopic model developed here and earlier \citep{jt623}. 
In Section \ref{sec:comparisons}, we compare cross-sections using the new
VOMYT linelist against laboratory, observational, and previous line list
spectra.  


\section{Constructing the Spectroscopic Model for VO}
\label{sec:SMforVO}

\subsection{General Considerations}

Astrophysically, vanadium is predominantly ($>$99.7\%) in one isotopic
form, its only stable isotope, $^{51}$V.

There are three main electronic systems in VO; the A-X, B-X and
C-X bands with origin (0,0) bands starting around 1.05 \um{}
(9500 \cm{}), 0.79 \um{} (12,600 \cm{}) and 0.57 \um{} (17,400 \cm{})
respectively.

However, vanadium oxide, like most transition metal diatomics, has a large
number of low-lying electronic states, as shown in Fig. \ref{fig:PES}. The
relative energy positioning and identity of the 13 lowest electronic
states (pictured) are reasonably well established by current
experimental and theoretical evidence. We consider these 13 states, 
six quartet states and seven doublet states, in our spectroscopic model of VO;
the final parameters of these states in terms of extended Morse oscillators are
given in Table \ref{tab:Morse} for reference, with detailed equations later in
the manuscript.

\cite{15HuHoHi.VO} considered a larger number of electronic states than we use in our model; this
informs our considerations of the limitations of our 13 state electronic state
spectroscopy model for VO. Specifically, we have included all nearby quartet
states, but there are very many doublet states just about 20,000 \cm{} that will
be involved in perturbations, spin-forbidden transitions and hot bands. 

The results of \cite{15HuHoHi.VO} show that there is a significant energetic
gap between these six quartet states and the next lowest quartet state (more
than 12~000 \cm). 
This gives confidence that
there will not be additional strongly allowed electronic transitions from the
ground state in the visible region.

However,
there are numerous nearby higher doublet states which will certainly affect the
spectroscopy of VO by perturbing the quartet states, spin-forbidden transitions
and hot bands.   There are not sufficient experimental data to fit these extra
states and the accuracy of \abinitio\ electronic structure calculations for these high-lying levels
is not sufficient for spectroscopic purposes. It is clear that our 13
state model will not produce a full picture of the visible absorption within the
doublet manifold, especially not with spectroscopic accuracy. However, the transitions
involving the doublets are likely to be relatively weak, since they are either
spin-forbidden or originate from excited states lying at least 5000 \cm{} above the
ground state. Therefore, their absorption
will contribute predominantly to the underlying continuum of VO absorption
rather than to the strong signature bands; this absorption (which will particularly
affect the opacity of the atmosphere) is significantly easier to model and will
 be less affected by perturbations in the energy levels. 

We use some of the \citet{07MiMaxx.VO} results for potential energy curves
 as a source of data for our line list. In
particular, we use their results to fit the potential energy curves of
the \Da{} and \Db{} states and also use their calculation of the
diagonal spin-orbit splitting of the \Dc{} and \De{} state to fix the
spin-orbit splitting of the \Dc{}, \De{} and \Df{} states (the
relative size of the spin-orbit coupling is known from experiment, but
not their absolute values). However, \citet{07MiMaxx.VO} did not consider  dipole
moment (diagonal or off-diagonal), spin-orbit or
electronic angular momentum coupling curves. New calculations for
these properties are presented here. The spin-orbit
couplings in particular can drastically affect the energy levels of
the molecule.

\subsection{Methodology}

\subsubsection{Electronic Structure Calculations}

We use high level \abinitio\ electronic structure calculations for VO by \cite{jt623}, which include consideration of both static and dynamic electron correlation.  In brief, the electronic structure calculations for this molecule were generally performed using  {\sc Molpro}
\citep{MOLPRO_brief} with the  internally-contracted multi-reference configuration interaction (icMRCI) \citep{88KnWexx.ai,88WeKnxx.ai,92KnWexx.ai} level of theory
with  the large aug-cc-pVQZ basis set
\citep{89Duxxxx.ai,05BaPexx.ai} which incorporates diffuse functions to accurately describe the electronically excited states. The orbitals used in the icMRCI calculation were obtained using state-specific (SS) or minimal-state (MS) complete active space self-consistent field (CASSCF) calculations. 

The \abinitio\ results are shown as data points in Figs.~
\ref{fig:DiagSO},\ref{fig:ODSO},\ref{fig:ODLx},\ref{fig:DDM} and \ref{fig:ODDM}.
Note that the \abinitio\ results often do not extend beyond about 1.8 \AA. This
is due to convergence difficulties associated with an ionic/covalent avoided crossing;
even if calculations converged, the dipole moment obtained was often
unpredictable and not smooth in this region. Changes in basis set and/or method
 did not significantly improve smoothness and convergence. Thus, we choose
to use \abinitio\ points only where the calculations were trusted and the curves are
smooth 
for accurate calculation of absorption intensities below 5000 K. Smoothness is an essential feature of dipole curves if physically correct results are to be obtained \citep{jt573}.

 The \abinitio\ electronic structure results obtained using the above methodologies are not continuous or on a sufficiently fine grid to be used directly by a nuclear motion code to compute rovibronic energies, wavefunctions and transition intensities. Therefore, we need to interpolate and extrapolate the \abinitio\ results to form our final spectroscopic model for VO. Due to the small number of points,
simple cubic spline gave unphysical curves, particularly for the dipole moments. We thus chose to fit to physically-motivated functional forms. The choice of functional form for each property of interest is discussed as relevant.

\subsubsection{Nuclear Motion Calculations}
\Duo\ is a new program \citep{jt609} written by the ExoMol group that solves the
nuclear motion problem for diatomics with multiple highly-coupled electronic
states. 
The program \Duo\ was used to produce rovibronic energy levels for VO. Input to
this program were potential energy curves, spin-orbit, spin-spin and
spin-rotation coupling curves and dipole moment curves.  \Duo\ allows
optimisation of parameters or morphing of  input curves to minimise the difference between the
calculated and observed energy levels or transition frequencies that are provided.


 \begin{table}
 \def\arraystretch{1.2}
\caption{\label{tab:doublets} Overview of empirical energy levels of the doublet states
used to refine the spectroscopic model of VO, and quantification of the quality of this fit. MH indicates a Model Hamiltonian fit was used to evaluate the energy levels. 
RMS and Max are root-mean-squared-error and maximum deviations respectively against
our VO spectroscopic model in \cm{}. `No' gives the number of energy levels considered.}
\begin{tabular}{llrrlrrl}
\toprule
State & $v\pp$ & $\Omega\pp$ & $J\pp$ range & No & {RMS} & {Max} \\
\toprule
\multicolumn{7}{l}{\underline{MH from \cite{02RaBeDa.VO,05RaBexx.VO}}} \\

\Dc{} &  0  & 1.5 & 4.5 - 39.5  & 72 &   0.272 & 0.423 \\
 &  & 2.5 & 4.5 - 46.5  & 86 &   0.473 & 1.757 & \\
 &  1  & 1.5 & 4.5 - 39.5  & 72 &   1.078 & 3.295 &  \\
\De{} &  0  & 2.5 & 6.5 - 39.5  & 68 &   0.109 & 0.117 \\
 &  & 3.5 & 4.5 - 41.5  & 76 &   0.107 & 0.108 \\
\Df{} &  0  & 0.5 & 5.5 - 39.5  & 70 &   0.371 & 0.617 \\
 &  & 1.5 & 4.5 - 46.5  & 86 &   0.100 & 0.227 \\
 &  1  & 0.5 & 4.5 - 39.5  & 72 &   0.493 & 1.072 \\
 &  & 1.5 & 4.5 - 33.5  & 60 &   0.948 & 1.007 \\
 &  2  & 0.5 & 4.5 - 39.5  & 72 &   0.467 & 0.966 \\
 &  3  & 0.5 & 6.5 - 35.5  & 59 &   0.572 & 0.976 \\
\Dg{} &  0  & 0.5 & 0.5 - 19.5  & 40 &   0.192 & 0.376 \\
 &  & 1.5 & 1.5 - 19.5  & 38 &   0.140 & 0.391 \\
 &  1  & 0.5 & 0.5 - 19.5  & 39 &   0.271 & 0.541 \\
 &  4  & 0.5 & 0.5 - 19.5  & 40 &   0.821 & 2.112 \\
 \multicolumn{7}{l}{\underline{MH from \cite{87MeHuCh.VO} using perturbations in B-X}} \\
\Dd{} &  2  & 0.5 & 0.5 - 39.5  & 82 &   1.079 & 8.534 \\
 &  3  & 0.5 & 0.5 - 39.5  & 79 &   1.577 & 4.106 \\
 \bottomrule
\end{tabular}
\end{table}

\begin{table}
\caption{\label{tab:XABCenergies} Overview of the empirical energy levels of the \X{}, \A{}, \B{} and \C{} state used in fitting, and quantification of the quality of this fitting. MH stands for Model Hamiltonian and CD stands for Combination Differences. Also included are the RMS and Max deviations against VO spectroscopic model in \cm{}. The `No' column gives the number of energy levels considered.  }
\def\arraystretch{1.2}
\begin{tabular}{llrrlrr}
\toprule
State & $v"$ & $\Omega"$ & $J"$ range & No & {RMS} & {Max} \\
\midrule
 \multicolumn{7}{l}{\underline{MH from \cite{95AdBaBe.VO}}} \\

\X{} & 0 & 0.5 & 0.5 - 50.5  & 101 &   0.014 & 0.040 \\
 &  & 1.5 & 1.5 - 50.5  & 100 &   0.015 & 0.041 \\
 & 1 & 0.5 & 0.5 - 50.5  & 102 &   0.017 & 0.034 \\
 &  & 1.5 & 1.5 - 50.5  & 100 &   0.028 & 0.051 \\
 \multicolumn{7}{l}{\underline{CD from \cite{82ChTaMe.VO} }} \\
\A{} &  0  & -0.5 & 13.5 - 63.5  & 96 &   0.494 & 1.989 \\
 &  & 0.5 & 20.5 - 68.5  & 67 &   0.415 & 0.618 \\
 &  & 1.5 & 20.5 - 49.5  & 35 &   0.228 & 0.302 \\
&  & 2.5 & 7.5 - 66.5  & 84 &   0.259 & 0.355 \\
 \multicolumn{7}{l}{\underline{CD from \cite{94ChHaHu.VO,95AdBaBe.VO}}} \\
\B{} &  0  & -0.5 & 5.5 - 47.5  & 80 &   0.972 & 1.402 \\
 &  & 0.5 & 4.5 - 45.5  & 68 &   1.628 & 2.380 \\
 &  & 1.5 & 7.5 - 40.5  & 51 &   0.505 & 1.242 \\
&  & 2.5 & 7.5 - 47.5  & 76 &   0.108 & 0.393 \\
 &  1  & -0.5 & 5.5 - 24.5  & 23 &   2.421 & 3.732 \\
 &  & 0.5 & 10.5 - 35.5  & 27 &   1.220 & 1.830 \\
 & & 1.5 & 9.5 - 31.5  & 20 &   0.221 & 0.300 \\
 &  & 2.5 & 7.5 - 33.5  & 43 &   0.213 & 0.404 \\
 \multicolumn{7}{l}{\underline{CD from \cite{82ChHaMe.VO}}} \\
\C{} & 0 & 0.5 & 0.5 - 41.5  & 53 &   0.062 & 0.3455 \\
 &  & 1.5 & 1.5 - 38.5  & 60 &   0.260 & 1.9935 \\
 \multicolumn{7}{l}{\underline{MH from \cite{09HoHaMa.VO}}} \\

\C{}   & 1 & 0.5 & 0.5 - 9.5  & 19 &   1.277 & 1.432 \\
 &  & 1.5 & 1.5 - 9.5  & 18 &   1.425 & 1.570 \\
 & 2 & 0.5 & 0.5 - 9.5  & 20 &   3.286 & 3.43 \\
 &  & 1.5 & 1.5 - 9.5  & 18 &   2.937 & 2.964 \\
 & 3 & 0.5 & 0.5 - 9.5  & 10 &   1.655 & 1.857 \\
 &  & 1.5 & 1.5 - 9.5  & 9 &   1.872 & 2.040 \\
 & 4 & 0.5 & 0.5 - 9.5  & 20 &   1.705 & 1.906 \\
 &  & 1.5 & 1.5 - 9.5  & 18 &   1.873 & 2.093 \\
 & 5 & 0.5 & 0.5 - 9.5  & 20 &   2.140 & 2.796 \\
 &  & 1.5 & 1.5 - 9.5  & 18 &   2.007 & 2.480 \\
 & 6 & 0.5 & 0.5 - 9.5  & 18 &   0.949 & 1.466 \\
 &  & 1.5 & 1.5 - 9.5  & 18 &   0.998 & 1.203 \\

\bottomrule
\end{tabular}
\end{table}

\begin{table}
\caption{\label{tab:DApenergies} Overview of the empirical energy levels of the \D{} and \Ap{} states used in fitting, and the RMS and Max deviations against VO spectroscopic model (in \cm{}). The `No' column gives the number of energy levels considered.}
\def\arraystretch{1.2}
\begin{tabular}{llrrlrr}
\toprule
State & $v"$ & $\Omega"$ & $J"$ range & No & {RMS} & {Max} \\
\toprule
\multicolumn{7}{l}{\underline{CD from A fit and D-A in \cite{87MeHuCh.VO}}} \\
\D{} &  0  & 0.5 & 6.5 - 23.5  & 10 &   2.325 & 2.446 \\
 &  & 1.5 & 8.5 - 25.5  & 34 &   0.186 & 0.468 \\
 &  & 2.5 & 14.5 - 14.5  & 2 &   0.460 & 0.469 \\
&  & 3.5 & 6.5 - 24.5  & 32 &   0.441 & 0.484 \\
\multicolumn{7}{l}{\underline{CD from D fit and D-A$^{\prime}$ in \cite{87MeHuCh.VO} }} \\
\Ap{} &  0  & 1.5 & 5.5 - 6.5  & 2 &   17.448 & 17.506 \\
 &  &  2.5  & 8.5 - 28.5  & 8 &   8.435 & 10.269 \\
 &  & 3.5   & 8.5 - 47.5  & 21 &   1.498 & 2.588 \\
&  & 4.5   & 13.5 - 42.5  & 7 &   4.072 & 5.710 \\
 & 1 & 2.5 & 15.5 - 18.5  & 2 &   2.023 & 4.246 \\
 & & 3.5 & 13.5 - 46.5  & 30 &   1.108 & 2.272 \\
 &  &  4.5 & 19.5 - 31.5  & 7 &   3.641 & 5.106 \\
 & 2& 3.5 & 15.5 - 39.5  & 21 &   0.662 & 1.324 \\
 &  &4.5 & 19.5 - 39.5  & 20 &   1.583 & 2.746 \\
 \bottomrule
 \end{tabular}
 \end{table}

\begin{table*}
\def\arraystretch{1.2}
\caption{
\label{tab:AXBXCX} Overview of the experimental frequencies for the
\A{}-\X{}, \B{}-\X{} and \C{}-\X{}
transitions from \protect\cite{82ChHaMe.VO,82ChTaMe.VO,94ChHaHu.VO,95AdBaBe.VO}  used in fitting, and
the RMS and Max deviations against VO spectroscopic model in \cm.
The `No' column gives the number of frequencies considered.
}
\begin{tabular}{llrrlrrrrr}
\toprule
Transition & $v"$ & $\Omega"$ & $J"$ range & $v'$ & $\Omega'$ & $J'$ range & No & {RMS} & {Max} \\
\midrule
\A{}-\X{} &  0  & -0.5 & 17.5 - 49.5  & 0  &  0.5 & 16.5 - 50.5 & 42 &   0.376 & 0.633 \\
 &  & 0.5 & 30.5 - 63.5  & &  1.5 & 31.5 - 63.5 & 25 &   0.510 & 0.662 \\
 &  & 1.5 & 17.5 - 49.5  & &  1.5 & 16.5 - 49.5 & 20 &   0.237 & 0.315 \\
 &  & 2.5 & 28.5 - 62.5  & &  0.5 & 29.5 - 63.5 & 27 &   0.183 & 0.414 \\
 &  & 2.5 & 8.5 - 65.5  & &  1.5 & 8.5 - 66.5 & 46 &   0.076 & 0.260 \\
 &    & -0.5 & 19.5 - 49.5  & 1  &  0.5 & 18.5 - 50.5 & 35 &   0.393 & 0.593 \\
 &  & -0.5 & 30.5 - 63.5  & &  1.5 & 31.5 - 63.5 & 15 &   0.396 & 0.686 \\
 &  & 1.5 & 17.5 - 49.5  & &  1.5 & 16.5 - 49.5 & 20 &   0.228 & 0.318 \\
 &  & 2.5 & 8.5 - 33.5  & &  1.5 & 8.5 - 32.5 & 24 &   0.086 & 0.232 \\
\B{}-\X{} &  0  & -0.5 & 6.5 - 46.5  & 0  &  0.5 & 7.5 - 46.5 & 38 &   0.932 & 1.290 \\
 &  & 0.5 & 4.5 - 46.5  & &  1.5 & 4.5 - 46.5 & 96 &   0.970 & 1.412 \\
 &  & 1.5 & 7.5 - 37.5  & &  0.5 & 6.5 - 37.5 & 36 &   0.517 & 0.959 \\
 &  & 1.5 & 6.5 - 40.5  & &  1.5 & 5.5 - 39.5 & 47 &   0.463 & 1.227 \\
 &  & 2.5 & 8.5 - 36.5  & &  0.5 & 7.5 - 35.5 & 15 &   0.077 & 0.102 \\
 &  & 2.5 & 6.5 - 47.5  & &  1.5 & 5.5 - 47.5 & 105 &   0.107 & 0.226 \\
 &  1  & -0.5 & 5.5 - 23.5  & 0  &  1.5 & 5.5 - 23.5 & 21 &   2.429 & 3.442 \\
 &  & 1.5 & 8.5 - 28.5  & &  1.5 & 7.5 - 29.5 & 22 &   0.212 & 0.293 \\
 &  & 2.5 & 6.5 - 33.5  & &  1.5 & 5.5 - 33.5 & 60 &   0.220 & 0.429 \\
\C{}-\X{} & 0 & 0.5 & 4.5 - 40.5 &0 & 0.5 & 4.5 - 40.5 & 44 &   0.041 & 0.138 \\
 &  & 1.5 & 2.5 - 38.5  &  & 1.5 & 2.5 - 38.5 & 79 &   0.042 & 0.103 \\
\bottomrule
\end{tabular}
\end{table*}

\begin{table*}
\def\arraystretch{1.2}
\caption{\label{tab:ExpDADAp} Overview of the empirical Duo-derived frequencies for the \D{}-\A{} and \D{}-\Ap{} transitions from \protect\cite{87MeHuCh.VO} used in fitting, and the RMS and Max deviations against VO spectroscopic model in \cm{}. The `No' column gives the number of frequencies considered.}
\begin{tabular}{llrrlrrrrr}
\toprule
Transition & $v"$ & $\Omega"$ & $J"$ range & $v'$ & $\Omega'$ & $J'$ range & No & {RMS} & {Max} \\
\midrule
\D{}-\A{} &  0  & 0.5 & 6.5 - 23.5  & 0  &  -0.5 & 6.5 - 24.5 & 10 &   0.204 & 0.421 \\
&    & 1.5 & 8.5 - 24.5  &   &  0.5 & 7.5 - 25.5 & 47 &   0.309 & 0.682 \\
 &    & 2.5 & 14.5 - 14.5  &   &  1.5 & 13.5 - 14.5 & 3 &   0.604 & 0.643 \\
 &    & 3.5 & 6.5 - 24.5  &   &  2.5 & 5.5 - 23.5 & 41 &   0.442 & 0.500 \\
\D{}-\Ap{} &  0  & 0.5 & 2.5 - 7.5  & 0  &  1.5 & 3.5 - 6.5 & 6 &   11.201 & 11.266 \\
&   & 1.5 & 7.5 - 28.5  &   &  2.5 & 8.5 - 28.5 & 13 &   3.069 & 4.039 \\
&    & 2.5 & 7.5 - 48.5  &  &  3.5 & 8.5 - 47.5 & 49 &   0.922 & 1.877 \\
&    & 3.5 & 12.5 - 43.5  &   &  4.5 & 13.5 - 42.5 & 14 &   1.413 & 2.194 \\
&    & 1.5 & 14.5 - 18.5  & 1  &  2.5 & 15.5 - 18.5 & 4 &   0.588 & 0.636 \\
&    & 2.5 & 12.5 - 47.5  &   &  3.5 & 13.5 - 46.5 & 75 &   0.756 & 1.190 \\
&    & 3.5 & 18.5 - 31.5  &   &  4.5 & 19.5 - 31.5 & 14 &   7.292 & 8.219 \\
&  1  & 1.5 & 11.5 - 19.5  & 0  &  2.5 & 12.5 - 19.5 & 3 &   3.632 & 4.092 \\
&    & 2.5 & 13.5 - 39.5  & 2  &  3.5 & 14.5 - 39.5 & 53 &   0.785 & 1.018 \\
&    & 3.5 & 18.5 - 39.5  &   &  4.5 & 19.5 - 39.5 & 47 &   4.818 & 6.060 \\

 \bottomrule
\end{tabular}
\end{table*}

\subsubsection{Collation and Selection of Experimental Data}
The use of experimental data is  imperative to ExoMol's methodology,
particularly for molecules like VO where current \abinitio\ electronic structure methods do not
deliver sufficient accuracy as discussed by \citet{jt623} and \citet{jt632} in detail. An important
component of the line list generation is thus the collation and selection of a
set of empirical energies and frequencies. The spectroscopic model has been
refined to minimize the root mean squared deviation of the \Duo\ predicted
(calculated) energy levels and/or frequencies against the empirical (observed)
values.

Tables \ref{tab:doublets},
\ref{tab:XABCenergies}, \ref{tab:DApenergies}, \ref{tab:AXBXCX} and
\ref{tab:ExpDADAp} detail the  major sources of experimental data used to
construct the \Duo\ spectroscopic model of VO. 

Experimental transitions involving the doublet \Dc{}, \De{}, \Df{} and \Dg{}
states have been measured and used to construct model Hamiltonians fits for each
spin-vibronic band. Some spin-forbidden doublet-to-quartet transitions have
recently been observed (\Dg{}-\X{}) by \citet{09HoHaMa.VO} which are extremely useful
in fixing the relative positions of the doublet and quartet states. However,
these observed transitions do not quite provide sufficient information to determine
the energy of all spin rovibronic bands relative to the zero of the \X{} state.
Specifically, there is a spectroscopic network connecting the \{\Dc$_{5/2}$,
\De$_{7/2}$, \Df$_{3/2}$ and \Dg$_{3/2}$\} states from the experiments reported by
\citet{87MeHuCh.VO}, \citet{02RaBeDa.VO} and \citet{05RaBexx.VO} via the observed \Dg$_{3/2}$-\X
transition. This network  can be connected to the ground state of VO.
However, the spectroscopic network comprising \{\Dc$_{3/2}$, \De$_{5/2}$, \Df$_{1/2}$\}
is not connected to \Dg$_{1/2}$ and therefore also not to the \X{} state and the
absolute energy scale of VO. Nevertheless, the relative spin-orbit coupling
matrix elements of the \Dc{}, \De{} and \Df{} are known from experiment, as
detailed in Table \ref{tab:diagSOfit}. We can thus use a single magic number \citep{jt412}
to connect the two spectroscopic networks. Based on \abinitio\ electronic structure predictions by
\cite{07MiMaxx.VO} and our own calculations, we choose this magic number by
setting the spin-orbit coupling constant of the \Dc{} state to 180 \cm{}.
Using this magic number and the spin-forbidden transitions, we can then set
the absolute $T_e$ for each doublet state. The model Hamiltonians given in the
original experimental papers can then be used to produce empirical energies
using PGopher \citep{PGopher}. The $J$ range of the energy levels we used was
informed by the experimental data available in the original paper.  The use of
experimental energies from model Hamiltonian fits was judged sufficient for
doublet states for two reasons: (1) the intensity of their transitions do not
contribute significantly to the final absorption spectra and (2) the spin and
rotational structure of doublet spectra are easier to fit (and hence more
reliable) than for quartet spectra. Furthermore, manual checks of the reported
frequencies against our calculated PGopher empirical energy levels showed good
agreement. Table \ref{tab:doublets} details all the spin-vibronic bands of
doublet states which we include in to refine the spectroscopic model. Note that
we also include data on some \Dd{} bands; these were not directly measured but
inferred from perturbations to the B-X transitions.

For quartet states, we found some significant discrepancies between the
calculated PGopher empirical energy levels and the frequencies reported in the original experimental
manuscripts, particularly for the A-X transition. The origin of the differences
could be typographic errors or different definitions of model Hamiltonian
constants. Instead of retracing this error, we decided to use the
frequencies directly as our source of experimental data. Assuming the model
Hamiltonian representation of the \X{} state (reasonable given the significant
amount of data involving this state, and the consistency with latter combination
differences), we produced empirical energy levels for the \X{} state using
PGopher. Based on these lower state energies, we could derive the upper state
energies (e.g. the \A{}, \B{} and \C{} state energies) from the experimental
frequencies based on the assignments given in the
original paper. The `combination differences' method was used in which an upper
state energy was only trusted if two or more transitions gave the same energy
for a particular upper state (within a threshold; here selected to be the
relatively loose 0.2 \cm{}). This process yielded a set of \A{}, \B{}
and \C{} upper state rovibronic energy levels and their associated quantum
numbers. We also had a set of A-X, B-X and C-X frequencies for which the
assigned quantum numbers had been verified through combination differences.
Finally, \citet{09HoHaMa.VO} measured some vibrational excited and
overtone bands of the C-X transition. We use the model Hamiltonian fits and
PGopher to produce empirical energy levels of these vibrationally excited \C{}
states. Table \ref{tab:XABCenergies} details all the \X{}, \A{}, \B{} and \C{}
spin-vibronic band energies which we include in the model refinement, while
Table \ref{tab:AXBXCX} details the transitions used.

The \D{} and \Ap{} energy levels are obtained indirectly using the A-X, D-A,
D-A$^{\prime}$ transitions. Given the difficulties in the model Hamiltonian parameters for
the \A{} state, we decided to not use the energy levels from \D{} and \Ap{} model Hamiltonian
parameters. Instead, we used the \Duo\ energy levels for the \A{} state in
combination with D-A transition frequencies to obtain combination differences
for the \D{} state. After fitting to the \D{} state in Duo, we used the \Duo\
\D{} state energies and the D-A$^{\prime}$ transition frequencies to obtain combination
differences for the \Ap{} state. We refine our model to both these \D{} and
\Ap{} combination difference energies (see Table \ref{tab:DApenergies}) and the
D-A and D-A$^{\prime}$ frequencies (see Table \ref{tab:ExpDADAp}) that produced these
combination differences.

For such a complex molecule, the refinement of the theoretical model
to experimental data needs to be performed iteratively.

In the first step, we used the independent electronic state
approximation to find potential energy curves and diagonal spin-orbit
couplings for each of the electronic states assuming that there was no
interaction between electronic states. 
The independent state approximation was then relaxed through the
inclusion of off-diagonal spin-orbit coupling and electronic angular
momentum coupling terms.

The final parameters are detailed in the next section and provided in \Duo\ input format as part of the
supplementary information for this article. 

\begin{table}
\def\arraystretch{1.2}
\caption{\label{tab:abinitioexp} Other data used to refine the spectroscopic model}
\begin{tabular}{lllll}
\toprule
State &  Method/ Source & Property & Value \\
\midrule
\mc{4}{l}{\underline{Experimental data}} \\
\A & Tentative assignment$^1$& $\omega_e$ & 884 \cm{} \\
\Da & Low-res. Photoelectron$^2$  & $T_e$ & 5630 \cm{} \\
\mc{4}{l}{\underline{\Abinitio\ data$^3$}} \\
\Da & C-icMRCI+DKH2+Q/BP & $r_e$ & 1.582 \AA  \\
& & $\omega_e$ & 1020 \cm{}   \\
\Db & C-icMRCI+DKH2+Q/BP & $T_e$ & 8849 \cm{}\\
& & $r_e$ & 1.577 \AA \\
& icMRCI+Q/B & $\omega_e$ & 1025 \cm{} \\
\De & icMRCI+Q/B & $\omega_e$ & 934 \cm{} \\
 \Dc & icMRCI/BP & SO & 180 \cm{} \\
  \bottomrule
\end{tabular}

$^1$\cite{89Mexxxx.VO}, $^2$\cite{98WuWaxx.VO}, $^3$ \cite{07MiMaxx.VO}
\end{table}

\subsection{Results}

\subsubsection{Potential Energy Curves}

The potential energy curves of the 13 electronic states is the most important component of the VO spectroscopic model. It controls the main rovibronic energy structure (i.e. the gaps between electronic states, the vibrational spacing and the rotational constants). 

The high dissociation energy of VO means that the Morse oscillator is a good
representation of the potential energy curves (PEC) for
the low-lying electronic states of VO in the region of interest. This
observation combined with the lack of reliable \abinitio\ electronic structure results at long bond
lengths/ higher energies meant that we did not use \abinitio\ PECs as input into
\Duo.  Instead we directly used the extended Morse oscillator (EMO) potential \citep{EMO}
 \begin{equation}
\small
\label{eq:EMO}
V(r) = T_e + (D_e - T_e) \left(1- \exp\left[\left(\sum_{i=0}^2 b_i\left(\frac{r^4-r_e^4}{r^4+r_e^4} \right)^i \right) (r-r_e)  \right] \right)^2.
\end{equation}
We started by setting the $b_1$ and $b_2$ parameters to zero, reducing the EMO to a simple Morse oscillator. We fixed the dissociation energy of the \X{} state to
52290 \cm{} based on experiment \citep{83BaGiGu.VO}; $T_e$+$D_e$ for all
other states was also fixed to this value, since all electronic  states considered
dissociate to the same atomic limit.  Then the empirical excitation energies, harmonic frequencies and equilibrium bond lengths were used to find initial parameters. Note that in \Duo\ default inverse length units, \AA$^{-1}$, the Morse oscillator parameter $a=K\omega\sqrt{\mu/D_e}$ where if $D_e$ and $\omega_e$ are in \cm{} and the reduced mass $\mu$ is in Dalton then $K=0.121778815$.
Where the term energies, harmonic frequencies and/or equilibrium bond lengths were unknown, we used \abinitio\ electronic structure calculations data from \citet{07MiMaxx.VO} as summarised in Table \ref{tab:abinitioexp}.

The $T_e$, $a$ and $r_e$
parameters were then modified to reproduce the energy levels and frequencies in Tables \ref{tab:doublets}, \ref{tab:XABCenergies}, \ref{tab:AXBXCX}, \ref{tab:DApenergies}, \ref{tab:ExpDADAp}. For the \C{}, \D{}, \Df{} and \Dg{} states, there was sufficient experimental data available to also optimise $b_1$ and $b_2$. The final parameters for the PEC of the spectroscopic model of VO are given in Table \ref{tab:Morse}.

\begin{table*}
\def\arraystretch{1.2}

\caption{\label{tab:diagSOfit} Diagonal spin-orbit coupling matrix elements in \cm.
The `Fit?' column provides information about the degree to which the \abinitio\ results could be modelled by the functional form in Eq. \ref{eq:SOfitform}: GF means good fit, OF means satisfactory fit with minor deviations, WC means the results showed the run concavity (i.e. not going towards an asymptote), TPS means there is a turning point in the \abinitio\ data but the magnitude of the changes is very small - data is fit by constraining $k=0$.  Most \abinitio\ results are from 1.41-1.80 \AA; Fig. \ref{fig:DiagSO} provide the data points explicitly.  The  equilibrium spin-orbit matrix elements are evaluated at 1.58 \AA{} for the b and c states, at 1.63 \AA{} for the A$^{\prime}$, e, f, g states, at 1.64 \AA{} for the A and B states and  at 1.69 \AA{}  for the D state. $f$ refers to the multiplicative factor described in Eq. \ref{eq:scaling}.}
\begin{tabular}{lrcrrrcdrrlrrr}
\hline\hline
 & \mc{1}{c}{\emph{Ab Initio}}&  \mc{5}{c}{\emph{Extrapolation Parameters}} &  \mc{3}{c}{\emph{Duo Fit}} & \mc{1}{c}{\emph{Exp}} \\
& {$\text{SO}^\text{abinitio}_{r_{eq}}$} &   Fit?  &   {$\text{SO}^\text{fit}_{\infty}$}  &  {$k$} & {$m$} & {$\text{SO}^\text{fit}_{r_{eq}}$}   &  \mc{1}{c}{{$f$}} & $|\text{SO}|_{\infty}^\text{Duo}$ & $|\text{SO}|_{r_{eq}}^\text{Duo}$ & \mc{1}{c}{$|S A \Lambda|$} \\
\hline
A$^{\prime}$&  256.26 & OF &  181.8 & 0.66 &  1$^*$ & 255.16 & 0.987i &   179.36&  251.71 &  256.2 \\
A &  68.99  & GF &  140.3 & -0.97 &  1.31 & 69.06 & 0.762i &  106.98 & 52.64 & 52.8 \\
B  & 90.63 & GF &  63.5  & 1.21 & 2.11 & 90.63 & 1.070i & 67.98   & 97.02  & 96.9 \\
D  &169.81 &  WC &  155.2  & 0.15 & 1$^*$ & 169.28 & 0.849i  & 131.70 & 143.68  & 143.5 \\
\hline
b  &  0.71 &  GF& 2.1 & -1.05 & 1$^*$ & 0.72 & 1  \\
c  &  188.02 &  TPS& 187.7  & 0$^*$ & N/A & 187.73 & 0.958 & 179.81 &179.81  & $p$$^\#$\\
e  &  176.54 &  WC& 140.5  & 0.41 & 1$^*$ & 175.62 & 1.055i &   148.28 & 185.34   & $p$$^\#$ + 6.1 \\
f  &  & Exp & 126.4 & 0$^*$ & N/A & 126.41 & -i &  126.41 & 126.41 & $p$$^\#$ $-$ 54.4\\
g &   61.96 & OF& 76.8  & -1.47 & 4$^*$ & 60.76 & -0.949i &  72.82 & 57.65  & 59.8\\
\hline\hline
\end{tabular}
\small
\begin{flushleft}
$^*$ Parameter constrained.
$^\#$ Experiment constrains the ratio between these three spin-orbit matrix elements, but not their exact value. Based on the theory
calculations of \citet{07MiMaxx.VO}, we constrained $p=180$ \cm{}.
\end{flushleft}
\end{table*}

\begin{figure*}
\centering
\includegraphics[width=0.45\textwidth]{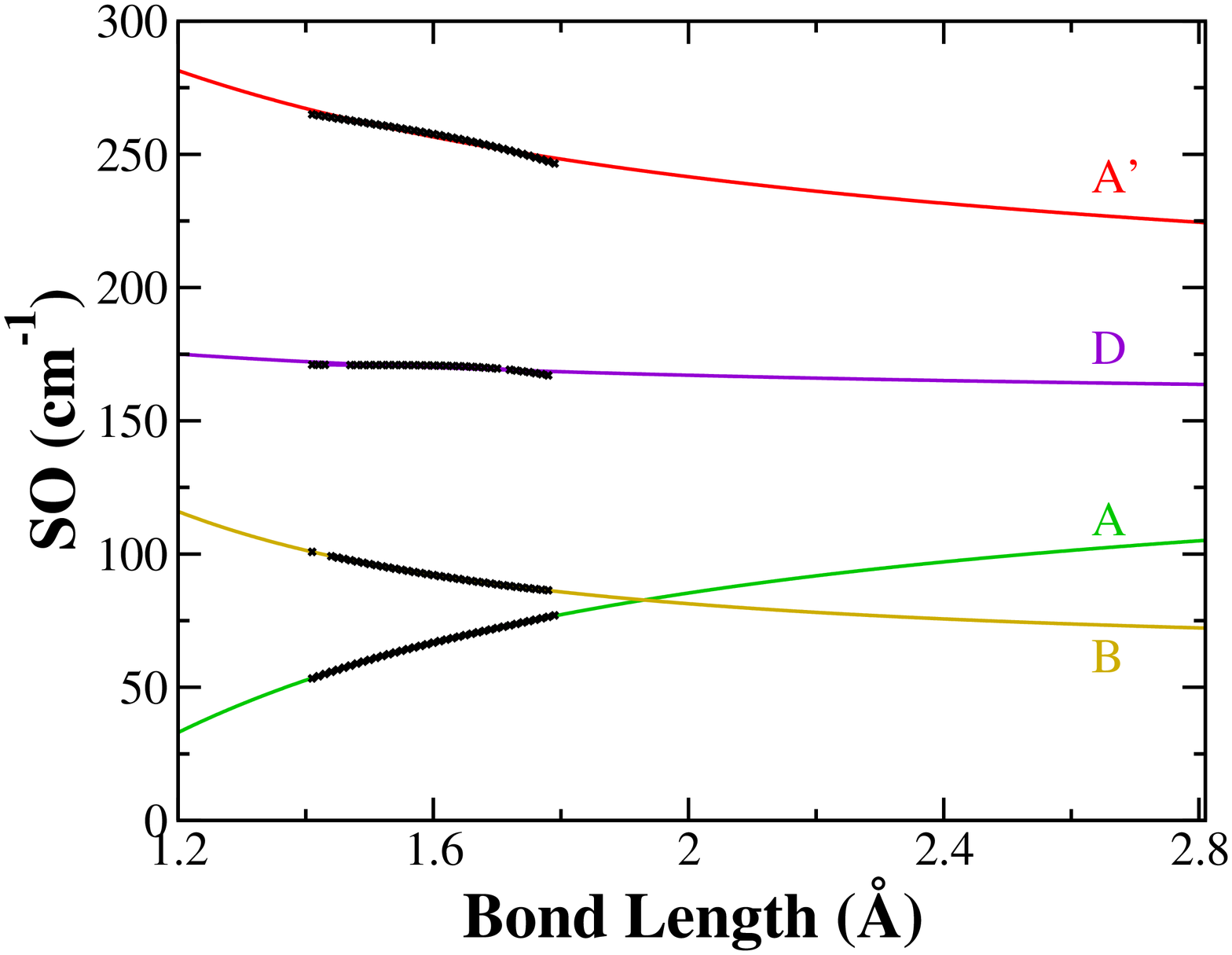}
~
\includegraphics[width=0.45\textwidth]{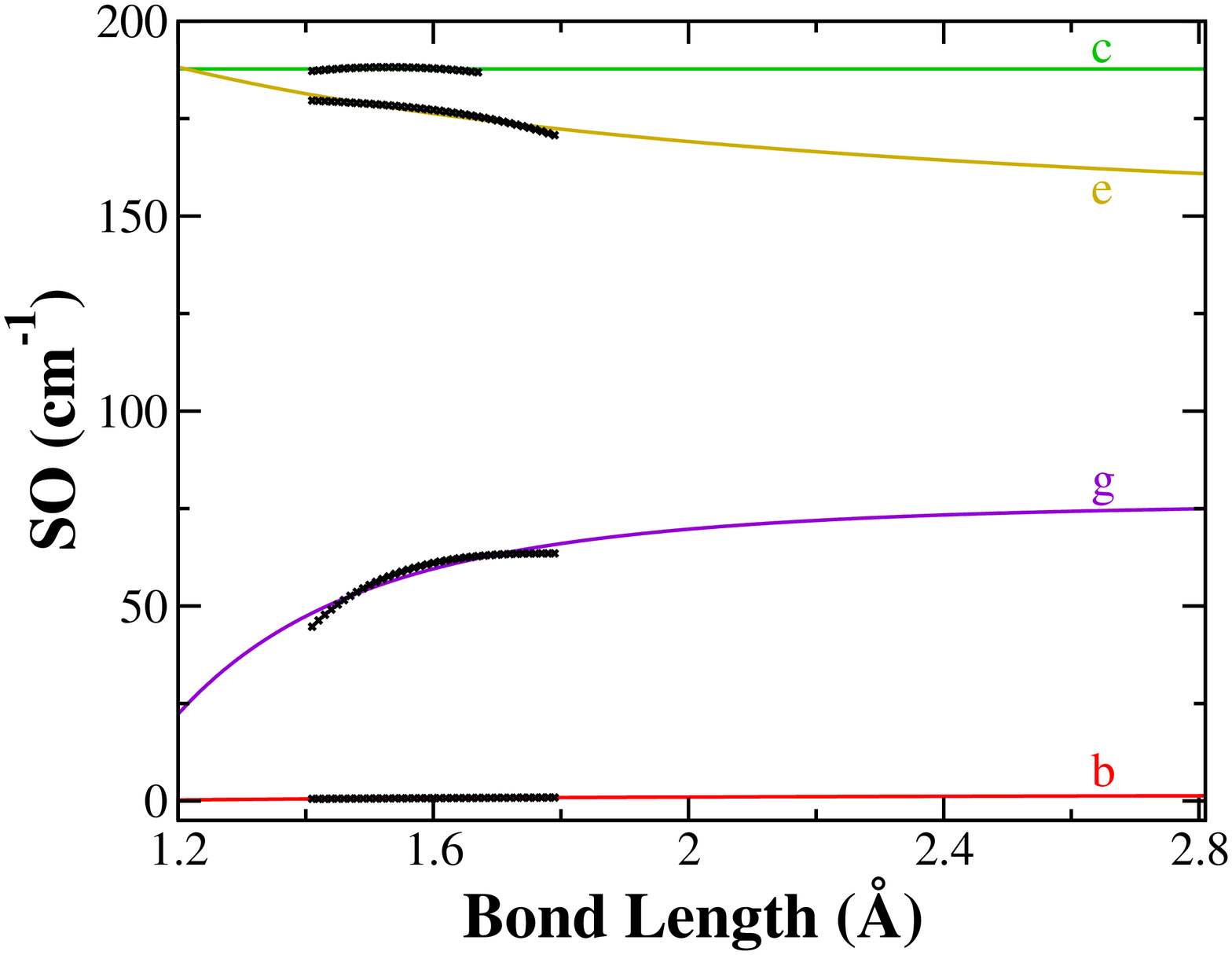}
\caption{\label{fig:DiagSO} Diagonal spin-orbit coupling matrix elements. The darker small crosses are \abinitio\ data from \citet{jt623}. The lighter continuous lines are fitted curves as described in the text.}
\end{figure*}

\begin{table*}
\caption{\label{tab:offdiagSO} Off-diagonal spin-orbit coupling matrix elements in \cm\ evaluated at $r=1.59$~\AA\ and $r\to \infty$. 
See the caption for Table \ref{tab:diagSOfit} for the meanings of the abbreviations in the comments column. Most \abinitio\ results are from 1.41-1.80 \AA; Fig. \ref{fig:ODSO} provide the data points explicitly. The matrix element specifies the $\Sigma$ quantum number for which the coupling element is evaluated, with a negative sign on either electronic state indicating that the electronic angular momentum, $L_z$, expectation value for that electronic state is negative for the signs of the Duo wavefunction. }
\def\arraystretch{1.2}
\begin{tabular}{llcrrrrdrrrrrrr}
\hline\hline
 & \mc{1}{c}{\emph{Ab Initio}}&  \mc{5}{c}{\emph{Extrapolation Parameters}} &   \\
& {$|\text{SO}|^\text{abinito}_{r=1.59}$} &   Fit?  &   {$|\text{SO}|^\text{fit}_{\infty}$}  &  {$k$} & {$m$} & {$|\text{SO}|^\text{fit}_{r=1.59}$}   & \mc{1}{c}{$f$} & {\emph{Matrix Element}}\\ 
\hline
X-A  &   65.98& GF& 91.6 & -0.45 &  1$^*$ & 65.82 & -1 &   $\braket{- {}^3\Pi_x,0.5|\HSO|^3\Sigma^-,1.5}$\\
X-B& 7.89 &OF  & 86.1 & -1.43 &  1$^*$ & 8.70 & -1 & $\braket{^3\Pi_x,0.5|\HSO|^3\Sigma^-,1.5}$\\
A$^{\prime}$-D  & 43.03& OF  & 41.5 & 1.31 &  7.70 & 43.06 & i &  $\braket{^3\Phi_x,1.5|\HSO|^3\Delta_y,0.5}$\\
A-B  & 2.70 &  TPS  & 2.2 & 0$^*$ &  N/A & 2.22 & i &  $\braket{- {}^3\Pi_x,1.5|\HSO|^3\Pi_y,1.5}$\\
A-C&  6.39 &  WC  & 54.6 & -1.41 &  1$^*$ & 6.29 & 1 &  $\braket{- {}^3\Pi_x,0.5|\HSO|^3\Sigma^-,1.5}$\\
A-D &  45.37 & WC  & 20.5 & 1.90 &  1$^*$ &44.89 & -i &  $\braket{- {}^3\Pi_x,0.5|\HSO|^3\Delta_y,1.5}$\\
B-C & 41.70$^{r=1.57}$ &  TPS  & 41.8 & 0.0 &  N/A & 41.83 & 0 &  $\braket{^3\Pi_x,0.5|\HSO|^3\Sigma^-,1.5}$ \\
B-D&  4.73& OF  & 39.4 & -1.38 &  1$^*$ & 5.32& i  & $\braket{^3\Pi_x,0.5|\HSO|^3\Delta_y,1.5}$\\
\hline
X-d &  265.95&  OF  & 277.6 & -0.29 &  4$^*$ & 265.19 & -i &  $\braket{^1\Sigma^+,-0.5|\HSO|^3\Sigma^-,0.5}$\\
X-f &  57.79&  OF  & 60.6 & -0.29 &  4$^*$ & 57.88  & 0&  $\braket{- {}^1\Pi_x,0.5|\HSO|^3\Sigma^-,1.5}$ \\
X-g &  78.27&  WC  & 89.4 & -0.20 &  1$^*$ & 78.40 & 0 &  $\braket{- {}^1\Pi_x,0.5|\HSO|^3\Sigma^-,1.5}$\\
A$^{\prime}$-b &  82.28&  TPS  & 82.0 & 0$^*$ &  {N/A} & 81.96  & i &  $\braket{^3\Phi_x,1.5|\HSO|^1\Gamma_y,0.5}$\\
A$^{\prime}$-c &  32.39 &  GF  & 140.6 & -1.20 & 1$^*$ & 32.35 & i  &  $\braket{^3\Phi_x,1.5|\HSO|- {}^1\Delta_y,0.5}$\\
A$^{\prime}$-e &  122.27&  WC  & 80.1 & 0.83 &  1$^*$ & 121.75  & -i &  $\braket{^3\Phi_x,0.5|\HSO|^3\Phi_y,0.5}$\\
A-a &  48.76& TPS  & 48.5 & 0$^*$ &  N/A  & 48.48& -1 &  $\braket{^1\Sigma^-,-0.5|\HSO|- {}^3\Pi_y,1.5}$\\
A-c  &8.89 & TPS  & 8.6 & 0$^*$ &  N/A & 8.63& i &  $\braket{- {}^1\Delta_x,0.5|\HSO|- {}^3\Pi_y,1.5}$\\
A-d&  62.02&  WC  & 10.6 & 7.58 &  1$^*$ & 60.92 & 1 &  $\braket{^1\Sigma^+,-0.5|\HSO|- {}^3\Pi_y,1.5}$\\
A-f &  22.33& TPS  & 25.0 & 0$^*$ &  N/A & 25.01& 0 &  $\braket{- {}^3\Pi_x,0.5|\HSO|- {}^1\Pi_y,0.5}$\\
A-g & 216.32& WC  & 171.5 & 0.40 &  1$^*$ & 215.13& 1 &  $\braket{- {}^3\Pi_x,0.5|\HSO|- {}^1\Pi_y,0.5}$\\
B-a  &16.37 & GF  & 48.8 & -2.72 &  3.05 & 16.45 & 1  & $\braket{^3\Pi_x,1.5|\HSO|^1\Sigma^-,0.5}$\\
B-c  &8.89 & TPS  & 8.6 & 0$^*$ &  N/A & 8.63 & -i &  $\braket{- {}^1\Delta_x,0.5|\HSO|^3\Pi_y,1.5}$\\
B-d & 6.96 & GF  & 13.8 & -1.80 &  2.79 & 6.98 & -2.17i &  $\braket{^1\Sigma^+,0.5|\HSO|^3\Pi_y,1.5}$\\
B-f  &  12.14$^{r=1.65}$& WC  & -20.8 & -2.49 &  1$^*$ & 11.82& 0 &  $\braket{^3\Pi_x,0.5|\HSO|- {}^1\Pi_y,0.5}$\\
B-g &  12.86$^{r=1.47}$&  WC  & 25.9 & -0.74 &  1$^*$  & 13.85& 0 &  $\braket{^3\Pi_x,0.5|\HSO|- {}^1\Pi_y,0.5}$\\
C-d  & 18.85 & GF  & 77.2 & -1.19 &  1$^*$ & 19.22& -i & $\braket{^3\Sigma^-,0.5|\HSO|^1\Sigma^+,0.5}$\\
C-f  & 6.40$^{r=1.61}$  & OF  & 17.4 & -4.11 &  4$^*$ & 6.20& 0 &  $\braket{- {}^1\Pi_x,0.5|\HSO|^3\Sigma^-,1.5}$ \\
C-g & {Not calculated}  &   & 0$^*$ & 0$^*$ &  N/A & & 0  & \\
D-c &    14.53$^{r=1.58}$ & OF  &102.3 & -1.34&  1$^*$  &  15.78& i & $\braket{^3\Delta_x,0.5|\HSO|- {}^1\Delta_y,0.5}$\\\
D-e &    32.21 &  WC& 35.2  & -0.13 &  1$^*$ & 32.26  & -i & $\braket{^3\Delta_x,1.5|\HSO|^1\Phi_y,0.5}$\\
D-f &    36.49 & TPS & 36.0 & 0$^*$ &  N/A & 36.04 & 0  & $\braket{^3\Delta_x,1.5|\HSO|- {}^1\Pi_y,0.5}$\\
D-g &   34.53 & GF &23.1& 3.62 &  4.32 & 34.41 & 0 & $\braket{^3\Delta_x,1.5|\HSO|- {}^1\Pi_y,0.5}$\\
\hline
a-d&  192.33& OF & 200.7 & -0.28 & 4$^*$ & 191.80 & i &  $\braket{^1\Sigma^-,0.5|\HSO|^1\Sigma^+,0.5}$ \\
a-f&  78.62& GF & 94.9 & -0.27& 1$^*$ & 78.64 & 0&  $\braket{^1\Sigma^-,-0.5|\HSO|- {}^1\Pi_y,0.5}$ \\
a-g & 68.86& OF & 94.9 & -0.27 & 1$^*$ & 68.55 & 0&  $\braket{^1\Sigma^-,-0.5|\HSO|- {}^1\Pi_y,0.5}$ \\
b-e& 69.01 & TPS & 68.8 & 0$^*$ & N/A &68.47 & -i &  $\braket{- {}^1\Gamma_x,-0.5|\HSO|^1\Phi_y,0.5}$\\
c-e& 12.35 & OF & 76.4 & -1.32 & 1$^*$ & 13.21& -i &  $\braket{- {}^1\Delta_x,-0.5|\HSO|^1\Phi_y,0.5}$\\
c-f& 19.21 & OF & 60.3 & -1.09 & 1$^*$  & 19.14& 0 &  $\braket{- {}^1\Delta_x,-0.5|\HSO|- {}^1\Pi_y,0.5}$\\
c-g & 34.66$^{r=1.61}$ & OF & 134.4 & -1.20 & 1$^*$ & 33.36 & 0 &  $\braket{- {}^1\Delta_x,-0.5|\HSO|- {}^1\Pi_y,0.5}$\\
d-f&  53.52& WC & 21.7 & 2.30 & 1$^*$  & 52.99& 0  &  $\braket{^1\Sigma^+,-0.5|\HSO|- {}^1\Pi_y,0.5}$\\
d-g & 2.97 & TPS & 2.7 & 0$^*$ & N/A & 2.67 & 0 &  $\braket{^1\Sigma^+,-0.5|\HSO|- {}^1\Pi_y,0.5}$\\
f-g &150.95$^{r=1.56}$& GF & 111.86 & 0.54 & 1$^*$  & 150.19 & i  &  $\braket{- {}^1\Pi_x,-0.5|\HSO|- {}^1\Pi_y,0.5}$\\
\hline\hline
\end{tabular}
\small
\begin{flushleft}
$^*$ Parameter constrained.

\end{flushleft}
\end{table*}

\begin{figure*}

\centering
\includegraphics[width=0.45\textwidth]{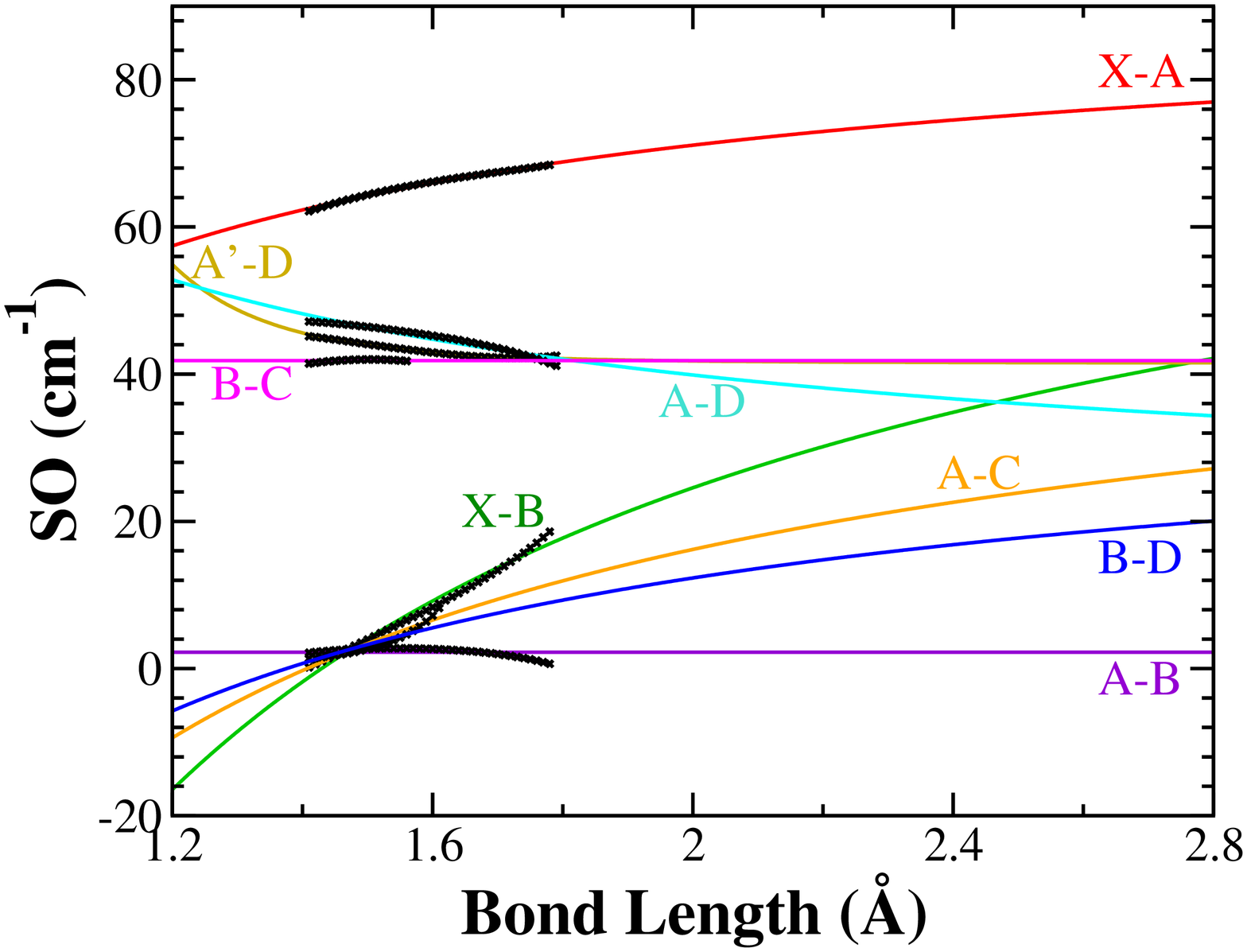}
~
\includegraphics[width=0.45\textwidth]{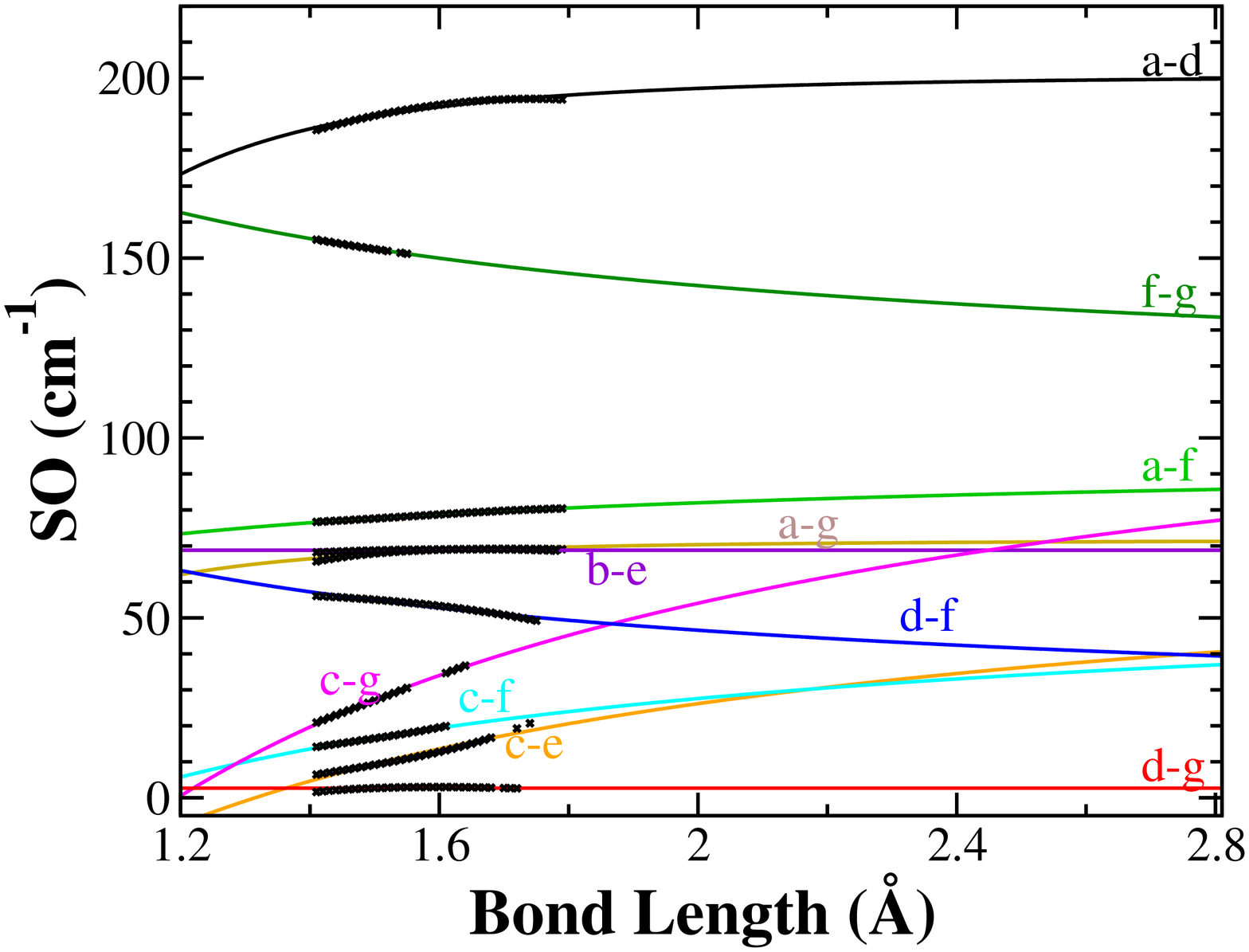}
~
\includegraphics[width=0.45\textwidth]{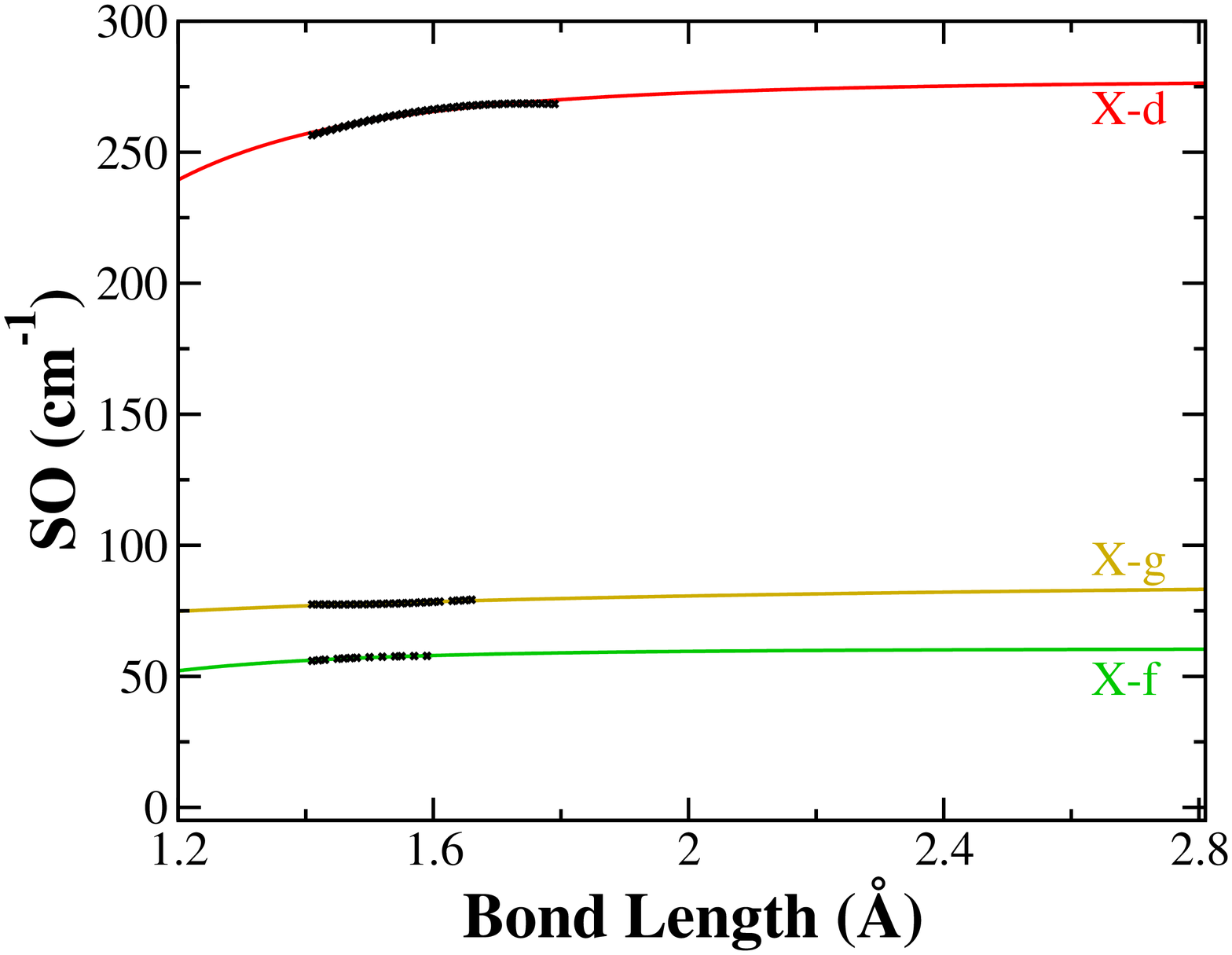}
~
\includegraphics[width=0.45\textwidth]{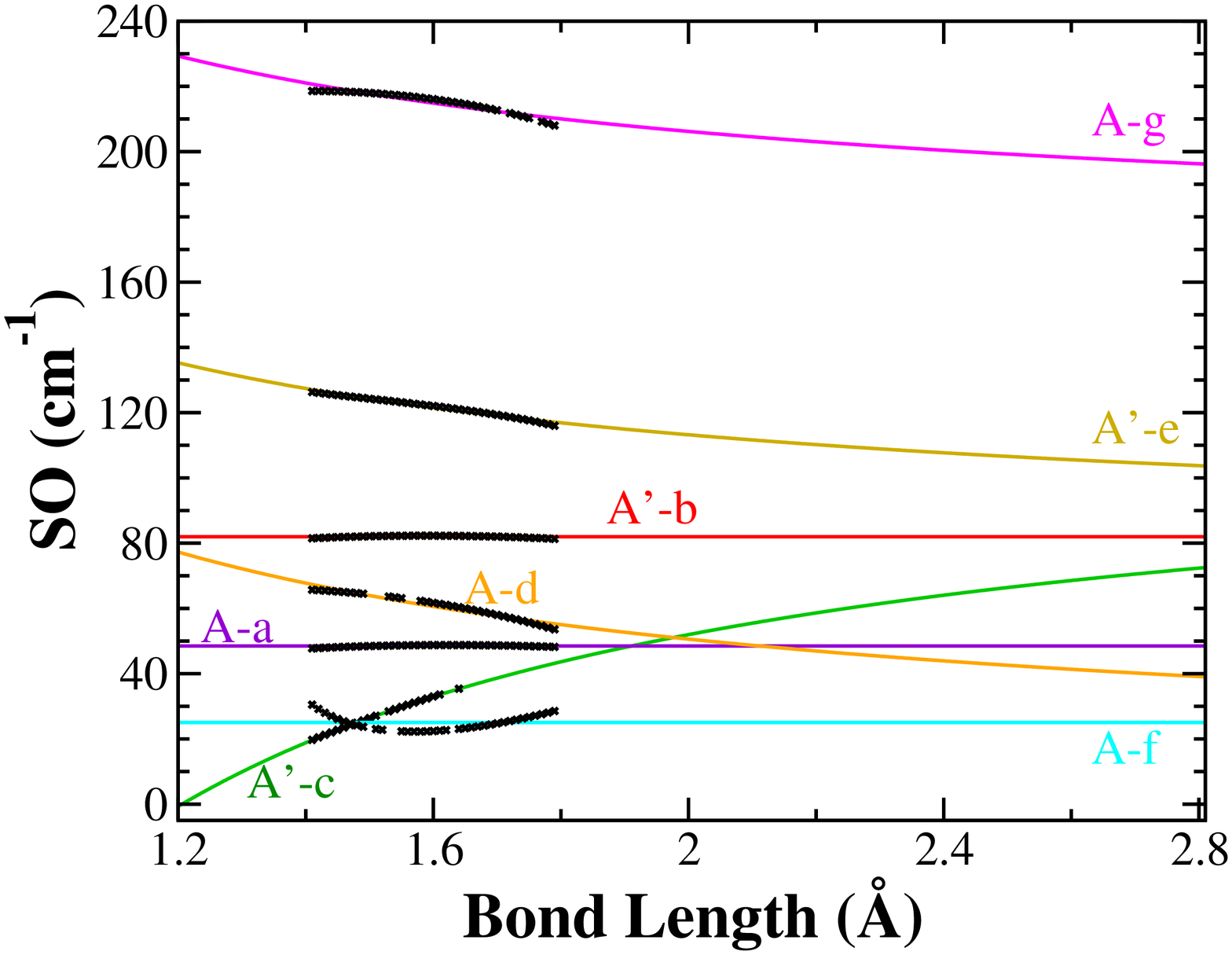}
~
\includegraphics[width=0.45\textwidth]{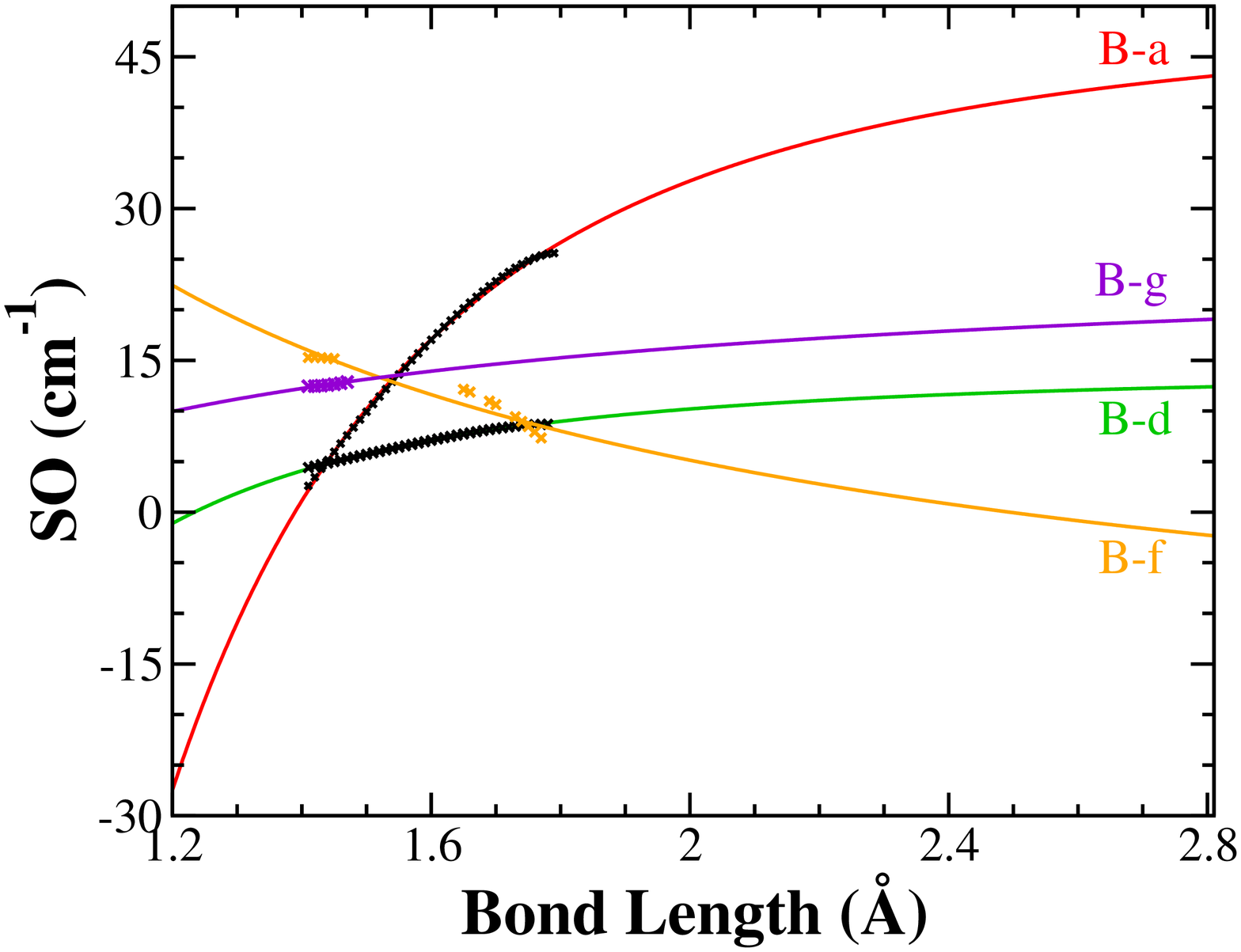}
~
\includegraphics[width=0.45\textwidth]{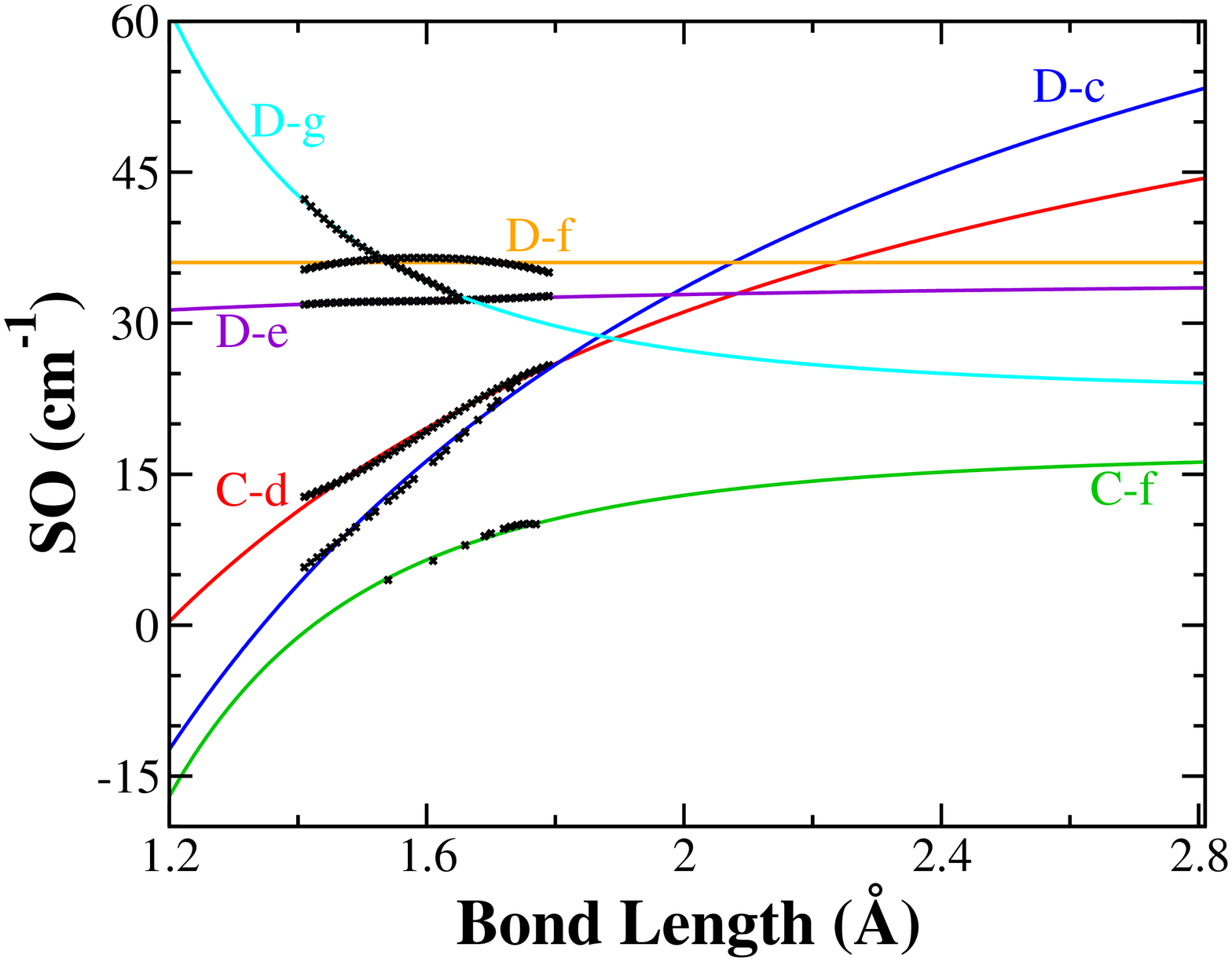}
\caption{\label{fig:ODSO} \Abinitio\ off-diagonal spin-orbit coupling matrix elements calculated by \citet{jt623} in darker small crosses, and the fitted curves in lighter continuous curves.}
\end{figure*}

\subsubsection{Spin-Orbit Coupling Curves}
The spin-orbit coupling constants have two main effects in the spectroscopy of VO. First, diagonal coupling terms split the energies of the different spin components of the quartet or doublet electronic state with non-zero electronic angular momentum $\Lambda$. This significantly increases the complexity of the spectra of molecules with high spin states such as VO compared to more common main-group chemical molecules like \ce{H2O}. Second, off-diagonal spin-orbit coupling causes mixing between different electronic states of both the same and different spins. This mixing gives intensity to spin- and symmetry-forbidden transitions, significantly increasing the complexity of the final spectra, though most of these new lines are quite weak compared to the allowed bands. In fact, the inclusion of off-diagonal spin-orbit coupling in our spectroscopic model is the major contributor (with Lx coupling playing a smaller but important role) to the very large number of transitions in the ExoMol VOMYT line list compared to the earlier Plez and Kurucz line lists. 

We fit the absolute value of the \abinitio\ data to the form:
\begin{equation}
\label{eq:SOfitform}
\text{SO}^\text{fit} = \text{SO}^\text{fit}_{\infty}\left(1 + \frac{k}{R^m}\right)
\end{equation}
where $R$ is the bond distance, $k,m$ are parameters, and $m\ge1$ is used to ensure sufficiently fast convergence to the atomic value $\text{SO}_\infty$.
For many coupling elements, the data is insufficient to provide a strong constraint on the variable parameters and consequently there can be a lot of variance in the fitted value of $\text{SO}^\text{fit}_{\infty}$. Thus, different spin-orbit curves should be compared by the equilibrium value not the asymptotic value because the equilibrium value is generally interpolated not extrapolated.

To fit experimental data, and introduce the correct sign and phase information, we included a further multiplicative value, $f$, i.e.
\begin{equation}
\label{eq:scaling}
\text{SO}^\text{Duo} = \text{SO}^\text{Duo}_{\infty}\left(1 + \frac{k}{R^m}\right) = f \text{SO}^\text{fit}_{\infty}\left(1 + \frac{k}{R^m}\right).
\end{equation}
$f$ may be complex. 

The specifications for the diagonal spin-orbit coupling curves are tabulated in Table \ref{tab:diagSOfit}, while the \abinitio\ data points and fits are plotted for quartets and doublets in Fig. \ref{fig:DiagSO}.
The specifications for the off-diagonal spin-orbit coupling curves are tabulated in Table \ref{tab:offdiagSO}, while the \abinitio\ data points and fits are plotted in Fig. \ref{fig:ODSO}. One result worth commenting about is the \Df{} and \Dg{} spin-orbit coupling. Experimentally, there is little evidence of coupling between the \Df{} and \Dg{} states; the fact that our \abinitio\ electronic structure calculations give a large spin-orbit coupling constant is indicative of mixing between the \Df{} and \Dg{} in the \abinitio\ calculation (since the two states share the same spin and symmetry). We choose to include this spin-orbit coupling because it should improve dipole moment matrix elements involving the \Df{} and \Dg{} states.


\begin{table*}
\caption{\label{tab:Dgerrors} Comparison of quality of the fit for \D{} state energy levels with and without
the inclusion of the \D-\Dg{} spin-orbit coupling constants.}
\def\arraystretch{1.2}
\begin{tabular}{lrrlrrrr}
\toprule
& &  & & \mc{2}{c}{{With D-g SO}} & \mc{2}{c}{{No D-g SO}}  \\
\cmidrule(l){5 - 6}\cmidrule(l){7-8}
State &  $\Omega$ & $J$ range & No.  & {RMS}/\cm & {Max}/\cm & {RMS}/\cm & {Max}/\cm \\
\midrule
\D{}   & 0.5 & 6.5 - 23.5  & 10  &  15.95 & 18.00 & 1.11 & 2.02 \\
  & 1.5 & 8.5 - 25.5  & 35  &  7.43 & 10.24 & 0.65 & 1.33 \\
  & 2.5 & 14.5 - 14.5  & 2  &  0.42 & 0.42 & 0.35 & 0.36 \\
 & 3.5 & 6.5 - 24.5  & 32  &  0.56 & 1.00 & 0.56 & 1.00 \\
\bottomrule
\end{tabular}
\end{table*}
In most cases, \abinitio\ spin-orbit coupling constants were found to
be generally reliable; in particular, our diagonal spin-orbit
constants were usually very accurate.  However, calculations involving
states \A{}, \B{}, \Df{} or \Dg{} showed much larger errors. In
particular the \Df{}, \Dg{} state were often not in the right order
with respect to each other and the \Dh{} state. These errors partially arise
due to significant mixing between nearby states of the same spin and
symmetry: the \A{} and \B{} state share the same spin and symmetry and
are only approximately 3000 \cm{} apart, while the \Df{} and \Dg{}
states are even closer at approximately 1000 \cm{} apart and within
3000 \cm{} of a third state of the same spin and symmetry,
\Dh{}.
A specific example of the errors this caused in the final line list can be seen in the energy levels of the D state given in
Table \ref{tab:Dgerrors} with and without the inclusion of the D-g spin-orbit coupling constants (with other parameters optimised).
The inclusion of the \D-\Dg{} spin-orbit coupling constant significantly deteriorates the quality with which the \Duo\ model reproduces the
\D{} state combination differences.
Given these {\it ab initio} difficulties combined with the fact there are other very closely lying $^2\Pi$ states
that are not considered in our model, we choose not to include many diagonal spin-orbit coupling elements involving the \Df{} and \Dg{} states.
The only exceptions are the \Df{}-\Dg{} and \A-\Dg{} spin-orbit coupling elements.

\begin{table*}
\caption{\label{tab:offdiagLx} Off-diagonal Lx coupling constants evaluated at $r=1.59$~\AA\ and $r\to \infty$. . See the caption for Table \ref{tab:offdiagSO} for the meanings of the abbreviations in the comments column. Most \abinitio\ results are from 1.41-1.80 \AA; Fig \ref{fig:ODLx} provide the data points explicitly. In cases of TPS, we fit to the data based on the nearest logical $\text{Lx}_\infty$. }
    \def\arraystretch{1.2}
    \begin{tabular}{llclrrrdl}
\hline\hline
 &\mc{1}{c}{\emph{Ab Initio}} &  \mc{5}{c}{\emph{Extrapolation Parameters}} &  \\
&   {$|\text{Lx}|^\text{abinito}_{r=1.59}$} &   Fit?  &   {$|\text{Lx}|^\text{fit}_{\infty}$}  &  {$k$} & {$m$} & {$|\text{Lx}|^\text{fit}_{r=1.59}$}   & \mc{1}{c}{$f$} \\ 
\hline
X-A  &    1.320& GF& 1.5$^*$ & 1.34 &  4.34& 1.321 &  -1 \\
X-B& 0.016 &GF  &  0.5$^*$ & -0.76 &  1$^*$ & 0.020  & -1 \\
A$^{\prime}$-D &  0.512& OF  & 1$^*$ & -0.78 &  1$^*$ & 0.510 & +i \\
A-C&   1.063 &  TPS  & 1$^*$ & 0.08 &  1$^*$ & 1.049 & +1 \\
A-D &   0.549 & TPS  & 0.5$^*$ & -0.72 &  4$^*$ &0.557 & -i \\
B-C   & 0.128 &  OF  & 0.5$^*$ & -0.86 &  1.76 & 0.121 & -1 \\
B-D   & 0.131$^{r=1.61}$ & GF  & 0.5$^*$ & -1.44&  2.86 & 0.117 & +i \\
\hline
a-f   & 0.301 &  OF  & 1$^*$ & -1.54 &  1.70 & 0.300  & -1 \\
a-g & 0.072 & TPS & 0$^*$ & 0.10 & 1$^*$ & 0.064  & -1 \\
b-e   & 0.016 &  TPS  & 0$^*$ & 0.03 &  1$^*$ & 0.016  & -i \\
c-e   & 0.217$^{r=1.58}$ &  GF  & 1$^*$ & -1.46 &  1.36 & 0.223 & -i \\
c-f &   0.301 & GF & 1$^*$ & -1.54 & 1.70 & 0.300  & +i \\
c-g &  0.175$^{r=1.61}$ & OF & 1$^*$ & -1.32 & 1$^*$ & 0.172 & -i\\
d-f   & 0.049 &  TPS  & 0$^*$ & 0.07 &  1$^*$ & 0.041  & -i \\
d-g & 1.538 & GF & 2$^*$ & -1.12 & 1.93 & 1.542 & -i \\
\hline\hline
\end{tabular}
$^*$ Parameter constrained.
\end{table*}

\begin{figure*}
\includegraphics[width=0.45\textwidth]{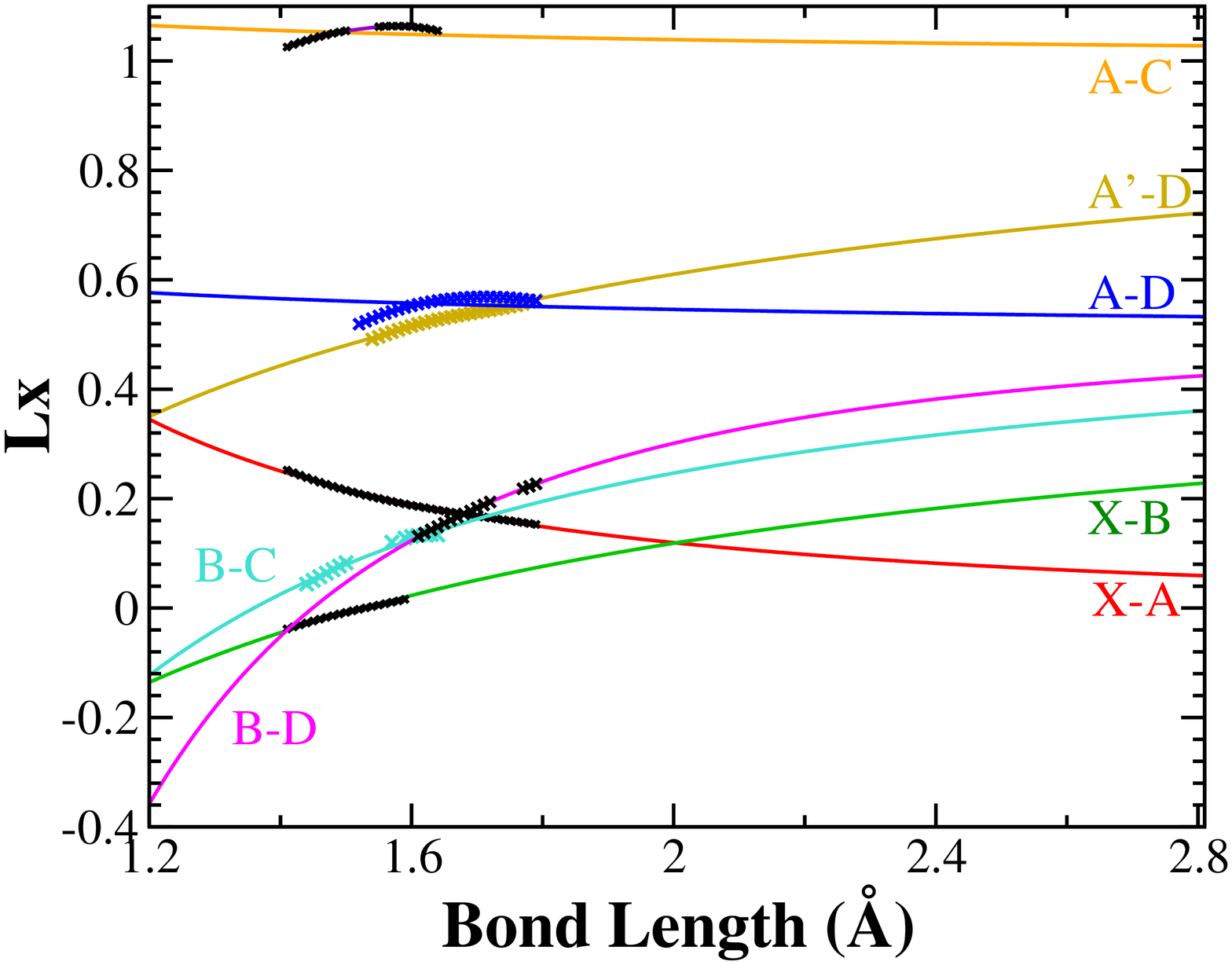}
~
\includegraphics[width=0.45\textwidth]{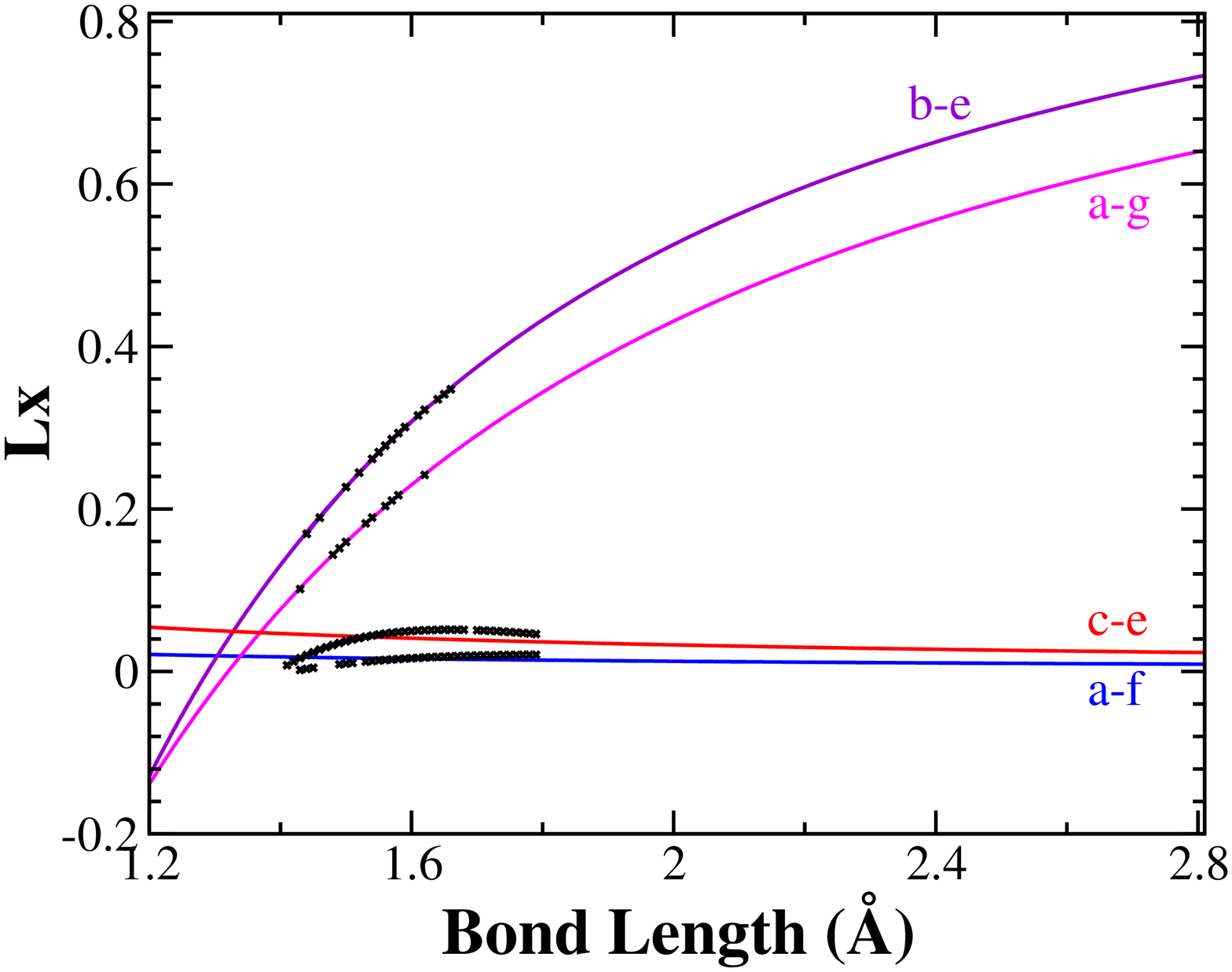}
\caption{\label{fig:ODLx} \Abinitio{} off-diagonal Lx coupling matrix elements as calculated by \citet{jt623} in darker small crosses, and the fitted curves in lighter continuous curves.}
\end{figure*}

\subsubsection{Lx coupling}

The Lx coupling is the electronic angular momentum coupling between electronic states. It acts in a similar way to the spin-orbit coupling to give rise to mixing of electronic states and the weak occurence of some previously forbidden spectral lines. However, Lx coupling has a much smaller effect because diagonal elements and coupling between electronic states of different spins is strictly zero. 

We fit the absolute of the \abinitio\ data to the form:
\begin{equation}
\label{eq:Lxfitform}
\text{Lx}^\text{fit} = \text{Lx}^\text{fit}_{\infty} + \frac{k}{R^m}
\end{equation}
where $m\ge1$ to ensure sufficiently fast convergence to the atomic limiting value $\text{Lx}_\infty$. When the data supports a particular logical rational value of $\text{Lx}_\infty$ (e.g. 0 or 1), this value is fixed.

The input to the \Duo\ program was given by
\begin{equation}
\text{Lx}^\text{Duo} = f \text{Lx}^\text{fit}.
\end{equation}
The parameter $f$ was used to provide the correct sign and phase information appropriate to Duo and may be complex; getting the sign correct from Molpro output to \Duo\ output is non-trivial \citep{jt589}. It could also be used in fitting; however, this was not done in the case of VO.

The \abinitio\ points and the fit are shown graphically in Fig. \ref{fig:ODLx}. The specifications of the \abinitio\ data, the fit and the \Duo\ input are given in Table \ref{tab:offdiagLx}.

%

It is difficult to judge the accuracy of the theoretical electronic
angular momentum terms. For VO, inclusion of all of these terms did
not adversely affect the fit; therefore we included non-zero terms
involving the lowest 13 electronic states. Note, however, that there
are a significant number of electronic states just above 20,000 \cm{}
that will couple to lower states,  in particular the \Df{}, \Dg{}, \C{} and \D{} states.

\begin{table}
    \def\arraystretch{1.2}

\caption{\label{tab:sssr} Spin-spin and spin-rotational constants, both experimental and fitted, in \cm{}.
Good agreement is not expected in many cases as the experimental constants also account for other effects (see text). }
\begin{tabular}{llrr}
\toprule
State & &  \mc{1}{c}{Fitted$^*$} & \mc{1}{c}{Exp} \\
\hline
\X{} & $\lambda_\text{SS}$  &0.0529 & 2.0300  \\
 & $\gamma_\text{SR}$ & 0.0191& 0.0225 \\
\Ap{} & $\lambda_\text{SS}$&  -0.6030 & \\
& $\gamma_\text{SR}$ & 0.4400\\
\A{} & $\lambda_\text{SS}$ & 1.0410 & 1.8670 \\
& $\gamma_\text{SR}$ & 0.0065 & 0.0038\\
\B{} & $\lambda_\text{SS}$ & 2.5869& 2.6579 \\
& $\gamma_\text{SR}$ & 0.0452 & 0.0336\\
\C{}& $\lambda_\text{SS}$ & 0.8037 &0.7469 \\
& $\gamma_\text{SR}$  & -0.0155 & -0.0184 \\
\D{} & $\lambda_\text{SS}$&0.2625 & \\
& $\gamma_\text{SR}$ & 0.0160 &  \\
\Dc{} & $\gamma_\text{SR}$ &0.2340 & \\
\De{} & $\gamma_\text{SR}$ & -0.0864 \\
\Df{} & $\gamma_\text{SR}$ & -0.1520& \\
\Dg{} & $\gamma_\text{SR}$& 1.2500 & \\
\bottomrule
\end{tabular}

$^*$ Note that \Duo\ actually takes as input $\frac{2}{3}\lambda_\text{SS}$; $\lambda_\text{SS}$  values
are tabulated to match experiment.
\end{table}

\subsubsection{Spin-Spin and Spin-Rotational Constants}
Traditionally, the model Hamiltonians used to fit observed data often included empirical spin-spin and spin-rotation elements. These matrix elements always involve only one electronic state. They quantify how the energy of a particular energy level depends on its spin state and the rotational levels. The effect of the spin-spin coupling is much smaller than the spin-orbit coupling, but can be important for high accuracy and when there is no orbital angular momentum (and therefore no diagonal spin-orbit coupling). The spin-rotation element is again generally a small effect needed only for very high accuracy reproduction of energy levels.   The spin-spin and spin-rotation elements are physical components of the total spin-rovibronic Hamltonian \citep{jt632}. 
Though it is theoretically possible to calculate these terms, this is not routinely done due to computational difficulty and also because these terms are usually also  empirical corrections for interactions that have been neglected in the model.

Since these are empirical terms that asympotically go to zero, we decided to use the Surkus-polynomial expansion \citep{84SuRaBo.method}
formula for their matrix elements:
\begin{equation}
V(r) = \left(1-\left(\frac{r^p-r^p_e}{r^p+r^p_e}\right)\right) \alpha  
\end{equation}
where $r_e= 1.6$~\AA, $p=2$ and $\alpha$ is an adjustable parameter, given by $\lambda_\text{SS}$ for spin-spin constants and $\gamma_\text{SR}$ for spin-rotation constants. Note that the form of this expression is such that the matrix elements go to zero asymptotically. 
The value of the included spin-spin and spin-rotation terms are specified in
Table \ref{tab:sssr}, and compared against previously empirical (equilibrium)
values where available. In many cases, the previous empirical values agree quite
well to our values. The major difference is the spin-spin constant of the \X{}
state which is only about 2.5\% of the experimental value. This reduction occurs
because the spin-orbit couplings (particularly between the X-d states) produce
the same type of perturbation to the \X{} state energy levels as the spin-spin
coupling constant. However, the effect on other energy levels will be different,
and some spin-forbidden transitions may be given intensity through this
mechanism. The importance of the X-d spin-orbit coupling explaining the
splitting of the \X{} state energy levels could not be predicted by experiment,
as the \X{} and \Dd{} states are more than 10,000 \cm{} apart; it is only the
very large magnitude (approximately 250 \cm{}) of the spin-orbit coupling
constant that makes this element important. This example illustrates that the
spin-spin coupling constants are sometimes used as empirical corrections rather
than a fundamental characteristic. Proper accounting for important spin-orbit
coupling can provide a more accurate picture of the molecule's spectroscopy.
unfortunately, for higher lying electronic states, this is currently impossible for
transition metal diatomics with the high density of electronic states and the
quality of modern \abinitio\ calculations \citep{jt632}.

\begin{figure*}
\includegraphics[width=0.45\textwidth]{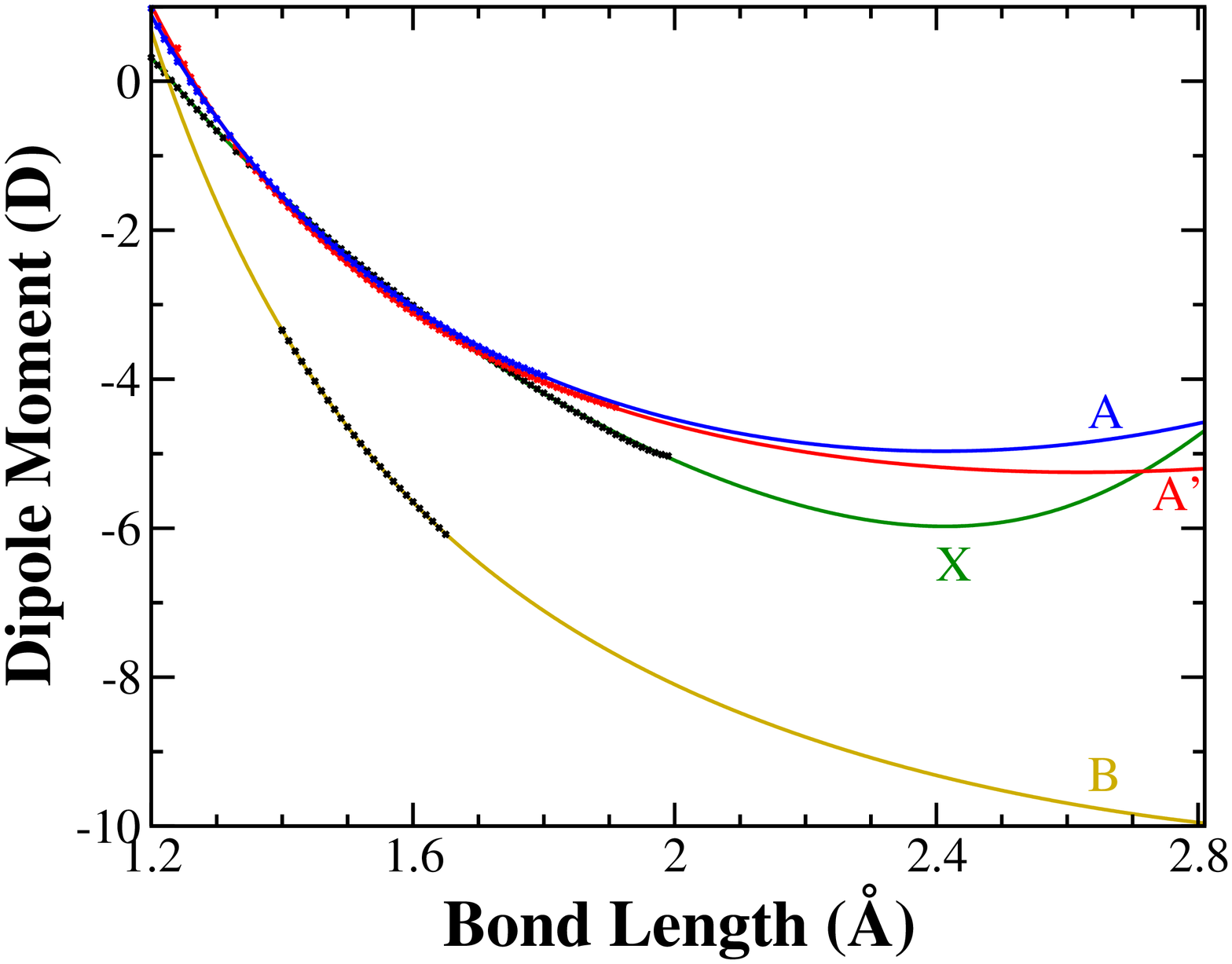}
~
\includegraphics[width=0.45\textwidth]{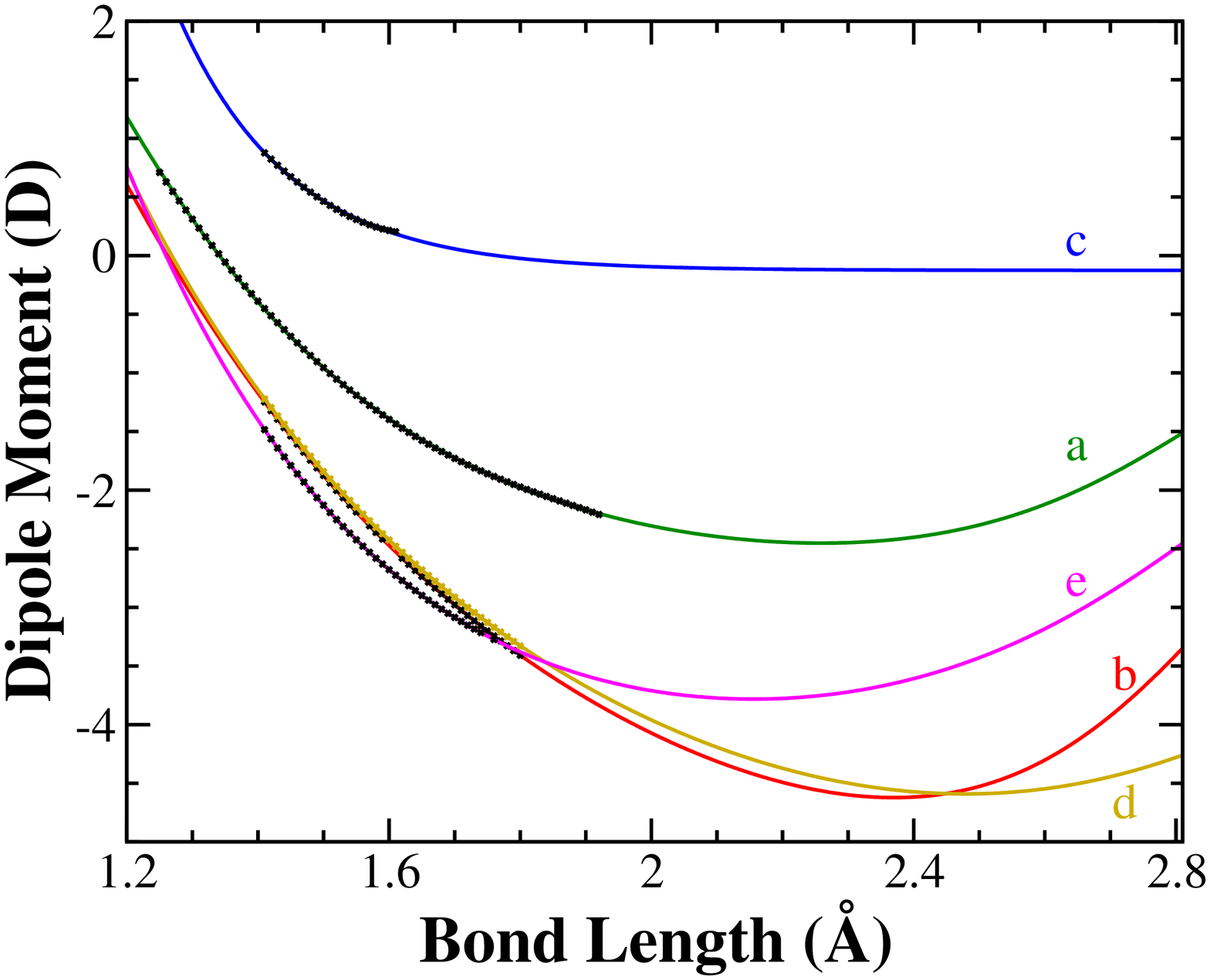}
\caption{\label{fig:DDM} \Abinitio{} diagonal diagonal dipole moment curves as calculated by \citet{jt623} in darker small crosses, and the fitted curves in lighter continuous curves.}
\end{figure*}

\begin{table*}
\caption{\label{tab:diagDMmethod} Diagonal dipole moment in \Duo, where dipole moment is given in Debye. The  equilibrium dipole moment for the a, b, c and d states is evaluated at 1.58 \AA{}, at 1.63 \AA{} for the A$^{\prime}$ and e states, at 1.64 \AA{} for the A and B states and at 1.59 \AA{} for the X state.}
    \def\arraystretch{1.2}

    \resizebox{\textwidth}{!}{%
\begin{tabular}{lcrrrrrrrrrrrrrrrrrrrrr}
\hline\hline
State & \mc{2}{c}{\emph{Ab Initio}} & \mc{5}{c}{{Extrapolation Parameters}} & \mc{4}{c}{{Fit Characteristics}}\\
& {Range (\AA)} & $\mu_{r_\text{eq}}^\text{abinitio}$ &   {$j$} & {$\m$} & {$\pt$} & {$l_1$} & {$\pll$}  & {$\mu_{r_\text{eq}}^\text{fit}$}  & {${\mu^\prime}_{r_\text{eq}}^\text{fit}$}  & {$|\mu|^\text{fit}_\text{max}$} & {$r$ for $|\mu|^\text{fit}_\text{max}$} \\
\hline
\X{}	&	1.15 - 1.99	&	-2.943	&	3$^*$	&	-13.440	&	3.830	&	0.000	&	-8.812	&	-2.958	&	-6.656	&	-5.975	&	2.413	\\
\Ap{}	&	1.24-1.91	&	-3.281	&	3.493	&	-2.801	&	2.503	&	13.519	&	-28.013	&	-3.279	&	-5.401	&	-4.623	&	2.370\\	
\A{}	&	1.20-1.80	&	-3.263	&	2.617	&	-4.356	&	2.320	&	11.368	&	-23.962	&	-3.253	&	-5.302	&	-4.967	&	2.408	\\
\B{}	&	1.40-1.65	&	-5.996	&	3$^*$	&	-0.846	&	6.123	&	18.798	&	-42.106	&	-5.993	&	-8.209	&	-10.171	&	3.237	\\
\Da{}	&	1.25 - 1.92	&	-1.319	&	0.859	&	-5.042	&	0.630	&	6.478	&	-13.689	&	-1.311	&	-4.008	&	-2.453	&	2.259	\\
\Db{}	&	1.41-1.80	&	-2.359	&	1.876	&	-9.668	&	2.468	&	3.058	&	-11.127	&	-2.361	&	-5.673	&	-4.623	&	2.37	\\
\Dc{}	&	1.41-1.61	&	0.244	&	\mc{5}{l}{Fit to 				$		32298.656	\exp[-7.443	Q]$}	&	0.252	&	-1.878	&	\\			
\Dd{}	&	1.41-1.80	&	-2.318	&	1.368	&	-2.849	&	2.819	&	4.671	&	-14.822	&	-2.315	&	-5.520	&	-4.590	&	2.479\\	
\De{}	&	1.41-1.76	&	-2.814	&	1.039	&	-4.333	&	0.719	&	12.126	&	-21.054	&	-2.857	&	-4.189	&	-3.782	&	2.155	\\
\hline\hline
\end{tabular}}
 $^*$: constrained to ensure reasonable fit.
\end{table*}

\begin{table}
\caption{\label{tab:offdiagDMmethod} Off-diagonal dipole moments in \Duo; values are given in Debye.}
    \def\arraystretch{1.2}

\begin{tabular}{llrrrrrrrrr}
\hline\hline
 & \mc{1}{c}{\emph{Ab Initio}} & \mc{3}{c}{\emph{Extrapolation Parameters}} \\
 &  \mc{1}{c}{{$|\mu|^\text{ab initio}_{r=1.59}$}} &  \mc{1}{c}{{$k$}} & \mc{1}{c}{{$m$}} & \mc{1}{c}{{$|\mu|^\text{fit}_{r=1.59}$}} \\
\hline
X-A	&	0.19	&	1.306	&	2.122	&	0.192\\	
X-B	&	0.613	&	6.451	&	3.054	&	0.616\\	
X-C	&	1.149	&	7.627	&	2.079	&	1.145	\\
A$^{\prime}$-D	&	0.106	&	1.195	&	3$^*$	&	0.112	\\
A-B	&	0.067	& \mc{3}{c}{Inbuilt Duo interpolation} \\
A-C	&	0.186	&	1.294	&	2.172	&	0.186	\\
A-D	&	0.11	&	0.452	&	1$^*$	&	0.11	\\
B-C	&	0.06	&	0.480	&	2.428	&	0.061	\\
B-D	&	0.018$^{r=1.61}$	&	1.657	&	7.596	&	0.019	\\
a-f	&	0.025	&	1.454	&	6.75	&	0.025	\\
a-g	&	0.011	&	0.033	&	0.397	&	0.011	\\
b-e	&	0.061	&	0.346	&	1.733	&	0.061	\\
c-e	&	0.275$^{r=1.58}$	&	2.384	&	2.68	&	0.27	\\
c-f	&	0.276	&	2.539	&	2.772	&	0.276	\\
c-g	&	0.071$^{r=1.61}$	&	1.408	&	4.299	&	0.076	\\
d-f	&	0.067	&	0.407	&	1.848	&	0.068	\\
d-g	&	0.052	&	1.045	&	4.371	&	0.054	\\
\hline\hline
\end{tabular}

\begin{flushleft}
$^*$ Parameter constrained.
\end{flushleft}
\end{table}

\begin{figure*}
\centering
\includegraphics[width=0.7\textwidth]{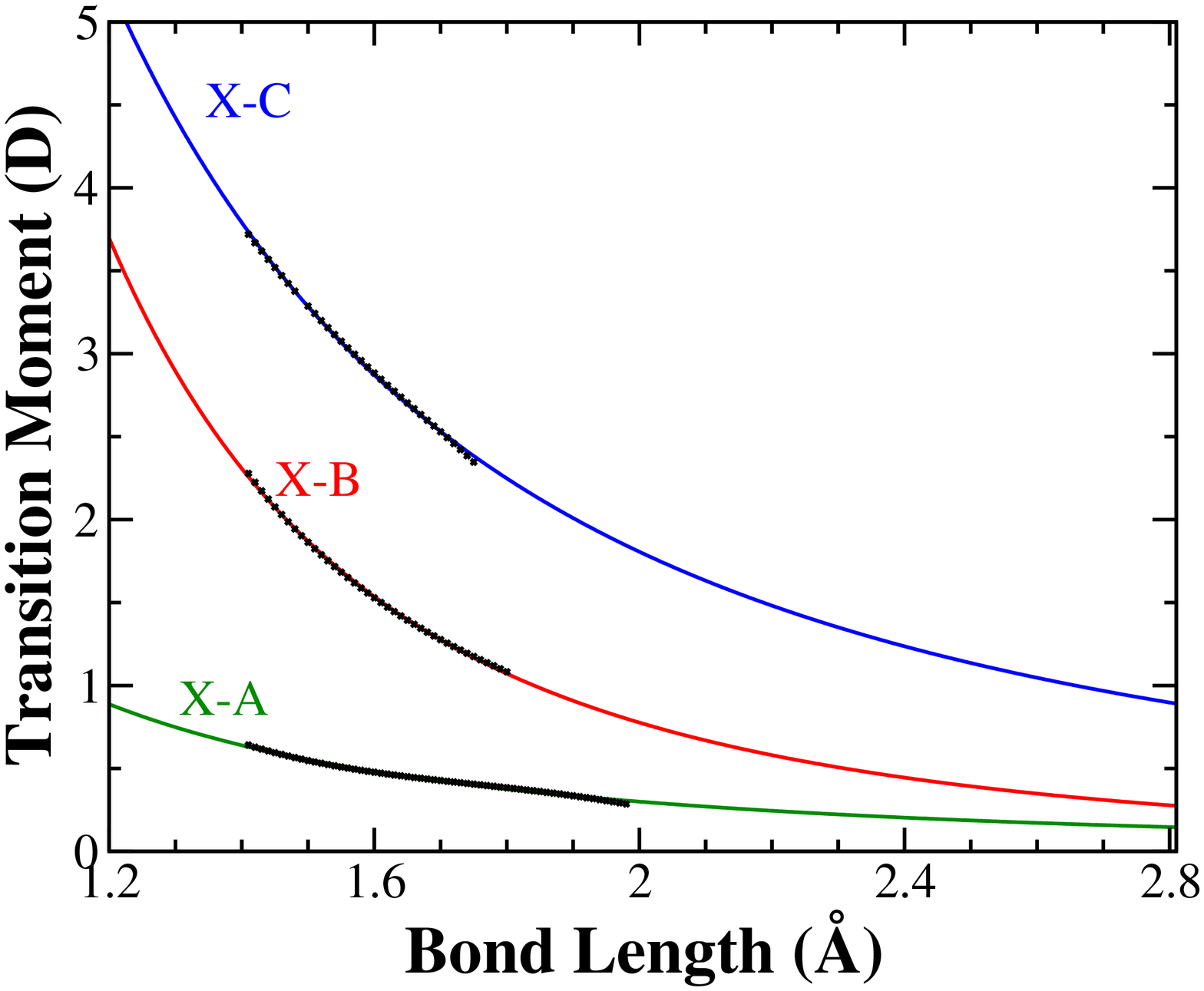}
~
\includegraphics[width=0.45\textwidth]{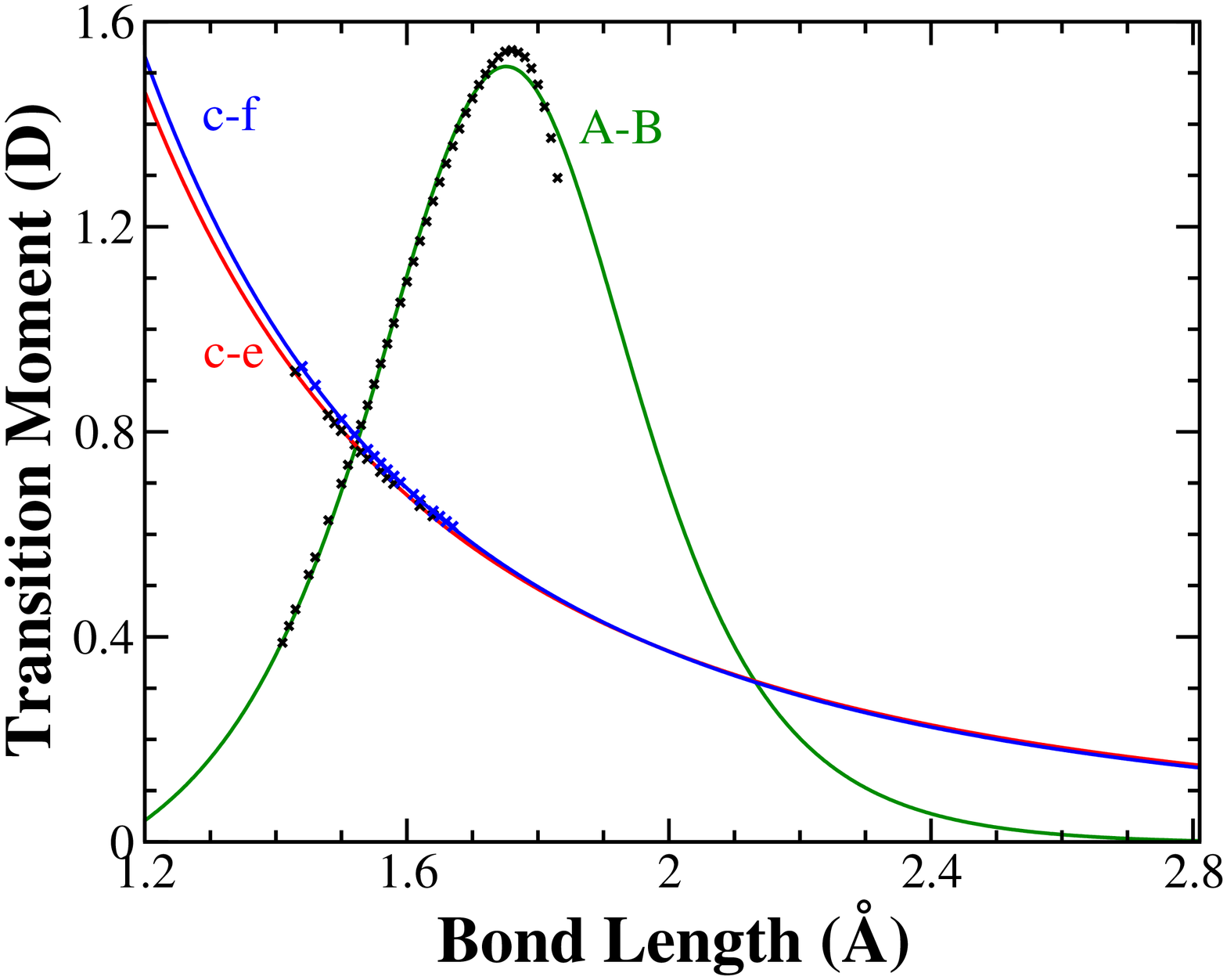}
~
\includegraphics[width=0.45\textwidth]{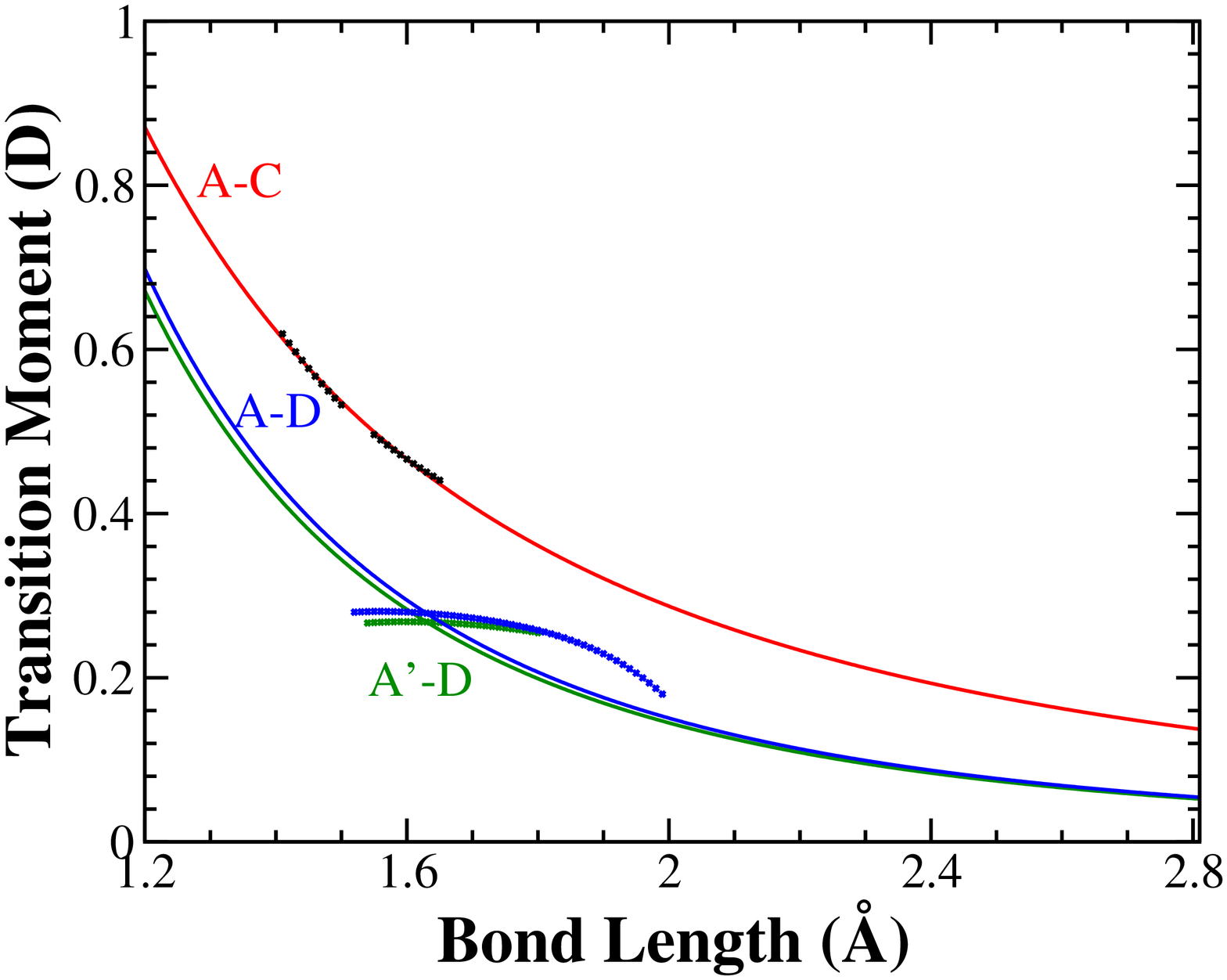}
~
\includegraphics[width=0.45\textwidth]{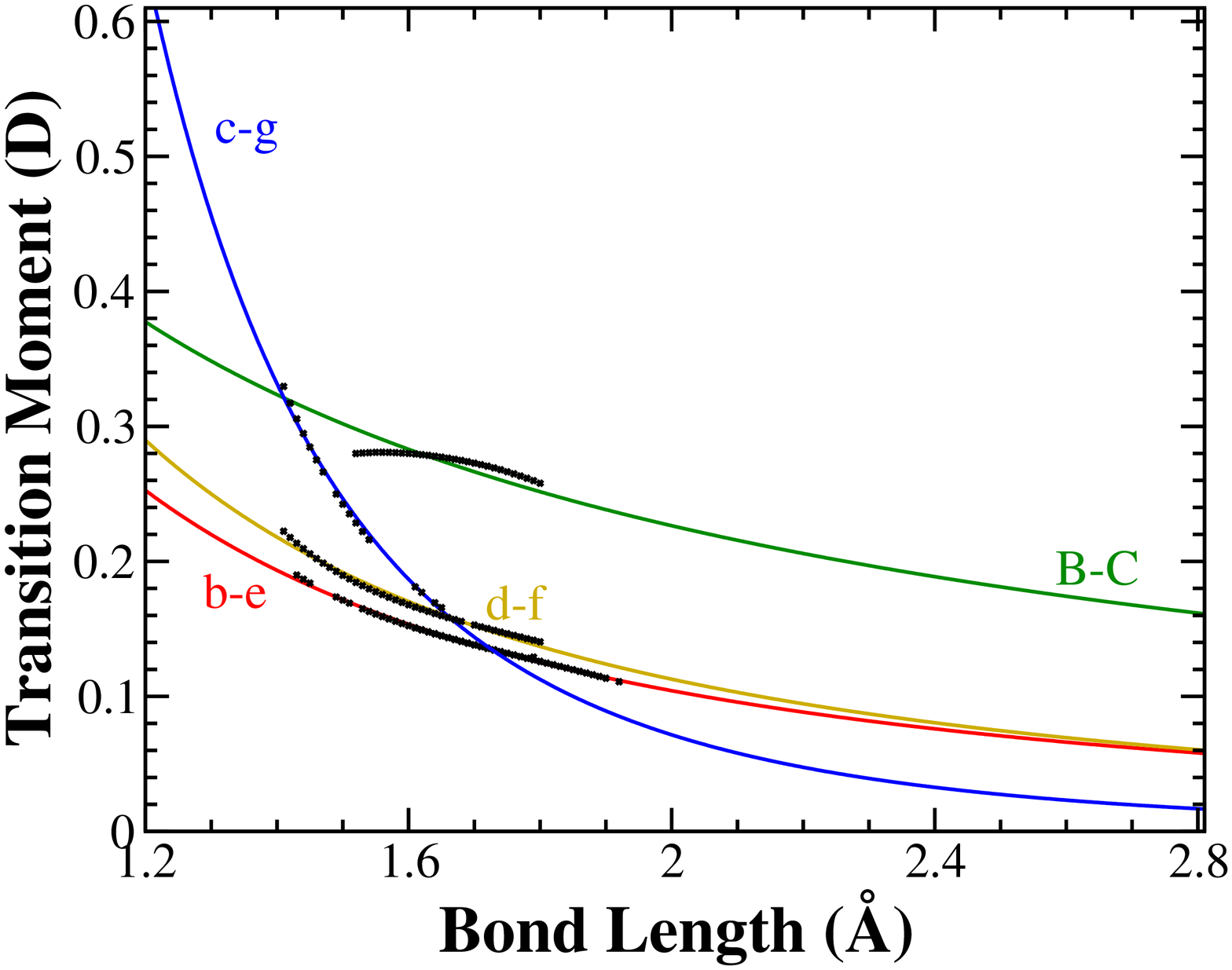}
~
\includegraphics[width=0.45\textwidth]{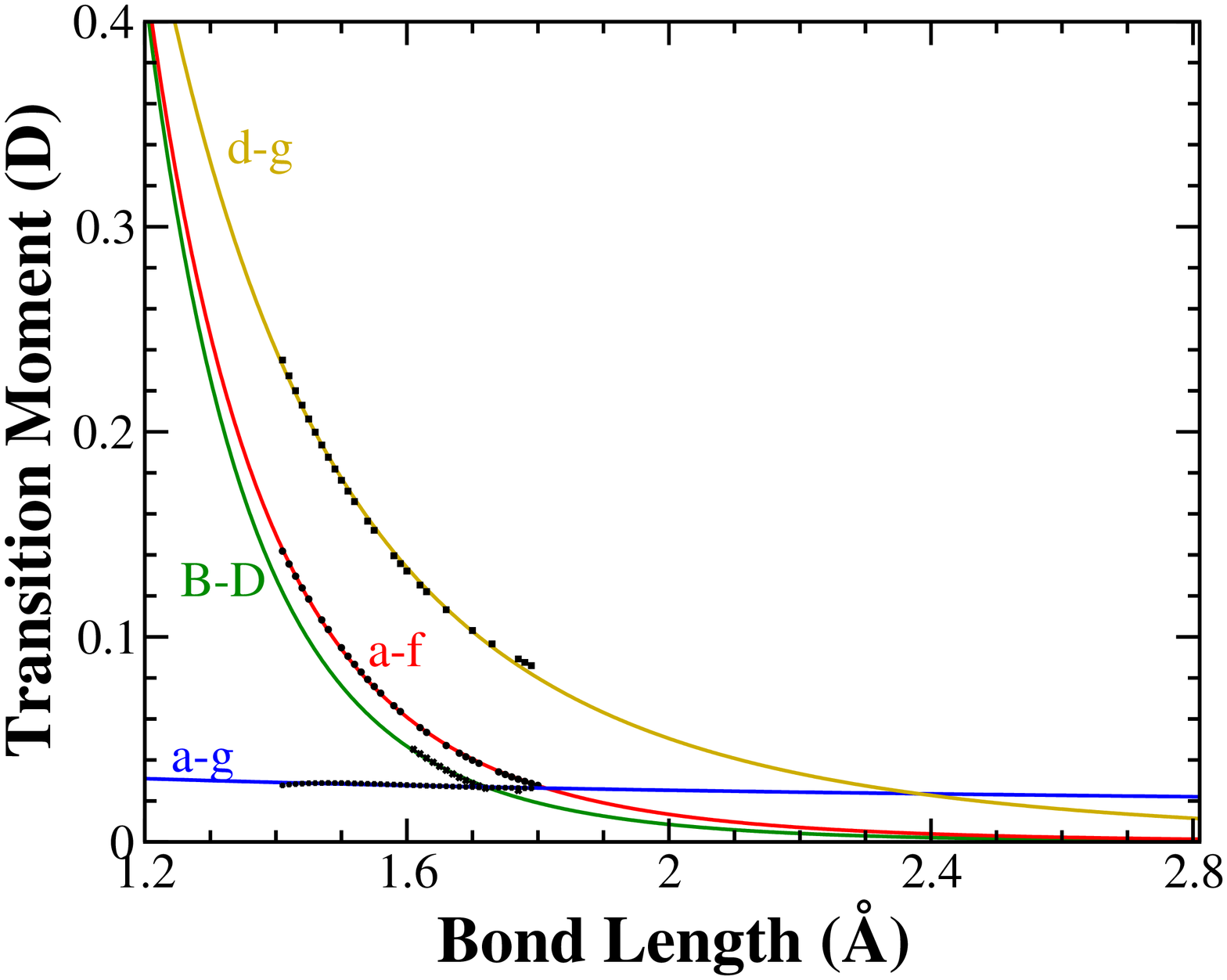}
\caption{\label{fig:ODDM}\Abinitio{} off-diagonal diagonal dipole moment curves as calculated by \citet{jt623} in darker small crosses, and the fitted curves in lighter continuous curves.}
\end{figure*}

\subsubsection{Diagonal Dipole Moment Curves. }
All diagonal dipole moment curves use icMRCI/aug-cc-pVQZ wavefunctions, evaluated using finite-field methodology \citep{jt623}.  As previously discussed, there were significant convergence difficulties at
long bond lengths. We thus choose to use \abinitio\ points only where the
calculations were trusted and the dipole moment curves smooth; this range is
given in Table \ref{tab:diagDMmethod}. Fortunately, the most physically
important bond lengths were generally stable. However, we needed smooth,
physically reasonable dipole moment curves to the boundary of the \Duo\ grid to
avoid adding spurious peaks to the spectrum (which can occur when the dipole moment
curves or its derivatives have discontinuities). Physically, at long bond
lengths, the dipole moment goes to zero when the system dissociates to neutral species; the unknown is how quickly this process
occurs and the form of the intermediate dipole moment curve.

We choose to fit to a newly developed functional form of the dipole moment curve near a long bond length neutral/ionic crossing.  This form will be fully derived, justified and possibly refined in a future theoretical paper, but is sufficiently accurate for current purposes.
In basic terms, the model involves two diabatic states, the ionic and covalent diabats, that interact to form an avoided crossing.
The two diabatic states mix to form an adiabatic ground state which is ionic at short bond lengths, covalent at long bond lengths and has mixed character near the avoided crossing (note there is also an adiabatic excited state with the opposite characteristics that is not considered here). The dipole moment curve of this adiabatic ground state depends on the contribution of the ionic state and the dipole moment of the ionic state; this yields the functional form:
\begin{equation}
\mu_{GS}(R) = \frac{\left(\sqrt{4 j^2+\lambda ^2}+\lambda \right)^2 \mu _{\text{Ionic}}(R)}{\left(\sqrt{4 j^2+\lambda ^2}+\lambda \right)^2+4 j^2}
\end{equation}
The $\mu_\text{ionic}$ is a functional form of the dipole moment of the ionic diabatic state, empirically expanded. 
\begin{equation}
\mu_\text{ionic}(R) = \pt R + l_1 R^{-1} + \pll R^{-3}.
\end{equation}
Note that the second and third terms are found empirically to be necessary for a good fit,
indicating the ionic state is not quite a true diabatic state in the traditional definition.
Nevertheless, the use of pseudo-diabatic representation is useful.
The $j$ parameter
controls the interaction between the ionic and covalent diabatic potentials and therefore how sharp the transition is.
 The $\lambda$ parameter controls the distance between the two diabatic states and their crossing points and is of the form
$\lambda=\m(R-d)(R-\pk)$,
where $\pk$ and $d$ are the crossing points of the ionic and covalent diabatic
curves (assuming that the ionic curve is a quadratic function of energy and the
covalent curve is a linear or constant function) and $m$ together with $j$
controls the depth of the ionic well.
We constrain $\pk=$2.75 \AA{} ; this is
the predicted location of the avoided crossing of the
\X{} state \citep{07MiMaxx.VO}. Physically, $d$ is approximately the short bond
length where the energy of the harmonic potential of the ionic state equals the
energy of the covalent state (which is to a first approximation the dissociation
energy); this was constrained to $d=1$~\AA\ without significantly affecting the
quality of the fit. %

The \abinitio\ points and final fitted curves are shown graphically in Fig. \ref{fig:DDM}, with quantitative data in Table \ref{tab:diagDMmethod}. 

The \Df{}, \C{}, \Dg{} and \D{} diagonal dipole moments are not needed because
the $T_e$ of these states are above 16,000 \cm{} and thus their thermal
population is negligible at the temperatures considered for this line list.

\subsubsection{Off-diagonal Dipole Moments}
Off-diagonal dipole moment curves were generally well fitted by the form:
\begin{equation}
\mu_{OD} = \frac{k}{R^m}
\end{equation}
However, for some cases, generally involving higher electronic states, this fit was not good. For the A-B dipole moment where very atypical behaviour with bond length was observed (probably due to state mixing), we put the \abinitio\ points into Duo and then used the inbuilt interpolation routine. For other results, we determined that the quality of the \abinitio\ calculations were not sufficiently reliable to trust the non-typical behaviour; however, the magnitude of the transition moment is probably reasonably reliable (and in any case, there is no other source of data). In these cases, we fixed $m$ and fit $k$ to the available data. 

All results are tabulated in Table \ref{tab:offdiagDMmethod} and shown graphically in Fig. \ref{fig:ODDM}.

The f-g transition moment is not included in the calculation because the $T_e$ of the \Df{} state is above 16,000 \cm{} and thus its thermal population is negligible even at 5000 K.

\subsubsection{Summary of Spectroscopic Model}

Our \Duo\ model for VO includes the following coupling terms:
\begin{itemize}
\item 13 potential energy curves
\item 4 diagonal quartet spin-orbit coupling curves
\item 5 diagonal doublet spin-orbit coupling curves
\item 5 off-diagonal quartet-quartet spin-orbit coupling curves
\item 4 off-diagonal doublet-doublet spin-orbit coupling curves
\item 15 off-diagonal quartet-doublet spin-orbit coupling curves
\item 7 off-diagonal quartet-quartet electronic angular momentum coupling curves
\item 8 off-diagonal doublet-doublet electronic angular momentum coupling curves
\item 6 diagonal quartet spin-spin coupling curves
\item 6 diagonal quartet spin-rotation coupling curves
\item 3 diagonal doublet spin-rotation coupling curves
\item 2 diagonal lambda doubling curves
\item 9 diagonal dipole moment curves
\item 9 quartet off-diagonal dipole moment curves
\item 8 doublet off-diagonal dipole moment curves
\end{itemize}

This is the most sophisticated \Duo\ spectroscopic model of a diatomic system produced so far.
The model, as specified by input to \Duo, is given in the supplementary data.

\subsection{Assessing Quality of Spectroscopic Model Against Experimental Energies}
For the X state, we were able to accurately reproduce the
empirical energy levels.
This is evidenced by the relatively low obs-calc residues (less than 0.05 \cm{} in all cases) for
the X state shown in Table \ref{tab:XABCenergies}.

For higher lying electronic states, however, the choice and magnitude of spin-orbit and electronic angular momentum terms were not supported by as much experimental evidence. This is partially because there are many couplings to higher electronic states not considered in the \Duo\ model for VO.

Quartet states have four spin-orbit components that couple differently to
other electronic states to produce very distinctive splitting between
the components.
Nevertheless, sub \cm{} accuracy in the
fits for the quartet states was generally achieved, as shown in
Tables \ref{tab:XABCenergies},\ref{tab:AXBXCX},\ref{tab:DApenergies} and \ref{tab:ExpDADAp}. In examining these tables, it is important to note that the optimisation did not weight all empirical data equally. The highest weights were put on the \A{}, \B{} and \C{} transitions confirmed by combination difference. The associated combination difference energies were 
attributed a higher weight than the PGopher energy levels. This accounts for the larger errors in the higher vibrational levels of the \C{} state compared to energies associated with the \C{} state ground vibrational levels. Furthermore, the \D{} and \Ap{} energy levels were optimised significantly less than the \X{}, \A{}, \B{} and \C{} energy levels as the former do not contribute to the primary absorption bands of VO; thus errors in their positions will not significantly effect the utility of the line list.  This accounts for the larger errors in the \Ap{} state compared to the other quartets, particularly for those spin-rovibronic bands for which very few lines are assigned.

For the \Dc{}, \De{}, \Df{} and \Dg{} doublets, as demonstrated by Table \ref{tab:doublets}, a very
good fit could usually be obtained because there was only one
spin-orbit interval to reproduce; the biggest difficulties occurred
with high vibrational states of \Df{} and \Dg{}. Though more
sophisticated potential energy curves might reduce these
errors, we believe that they probably originate from inadequate
treatment of spin-orbit coupling, particularly with respect to electronic states not considered in our spectroscopic models. We thus chose to use a smaller
number of parameters.

The \Dd{} state parameters were optimised predominantly to the frequencies and energies of the \B{} state, as this was the experimental origin of 
the PGopher model Hamiltonian parameters. This accounts for the larger errors in this state.

\begin{table*}
\caption{\label{t:Energy-file}  Extract from the *.states file for $^{51}$V$^{16}$O.}
\footnotesize \tabcolsep=5pt
\renewcommand{\arraystretch}{1.2}
\begin{tabular}{clccccclcccccccc}
\toprule
  1     &        \mc{1}{c}{2}        &  3  &  4  &  5  &  6  &  7  &  8  &  9  & 10  & 11 & 12 \\
    \midrule
     $n$ & \multicolumn{1}{c}{$\tilde{E}$} &  $g$    & $J$  & $\tau$ &  \multicolumn{1}{c}{$+/-$} &  \multicolumn{1}{c}{$e/f$} & State & $v$    & $\Lambda$& $\Sigma$ & $\Omega$& \\
     \midrule
1453	&	19124.63877	&	32	&	1.5	&	1.03E-07	&	+	&	f	&	C4Sigma-	&	2	&	0	&	0.5	&	0.5	\\
1454	&	19199.40966	&	32	&	1.5	&	1.77E-04	&	+	&	f	&	d2Sigma+	&	9	&	0	&	0.5	&	0.5	\\
1455	&	19219.56157	&	32	&	1.5	&	6.42E-03	&	+	&	f	&	Ap4Phi	&	14	&	3	&	-1.5	&	1.5	\\
1456	&	19363.88832	&	32	&	1.5	&	1.96E-05	&	+	&	f	&	A4Pi	&	12	&	1	&	-1.5	&	-0.5	\\
1457	&	19397.77872	&	32	&	1.5	&	1.88E-05	&	+	&	f	&	A4Pi	&	12	&	1	&	-0.5	&	0.5	\\
1458	&	19430.9299	&	32	&	1.5	&	1.81E-05	&	+	&	f	&	A4Pi	&	12	&	1	&	0.5	&	1.5	\\
1459	&	19444.43327	&	32	&	1.5	&	7.31E-07	&	+	&	f	&	B4Pi	&	8	&	1	&	-1.5	&	-0.5	\\
1460	&	19499.13805	&	32	&	1.5	&	7.21E-07	&	+	&	f	&	B4Pi	&	8	&	1	&	-0.5	&	0.5	\\
1461	&	19561.62306	&	32	&	1.5	&	7.11E-07	&	+	&	f	&	B4Pi	&	8	&	1	&	0.5	&	1.5	\\
1462	&	19627.44267	&	32	&	1.5	&	5.70E-04	&	+	&	f	&	a2Sigma-	&	15	&	0	&	0.5	&	0.5	\\
1463	&	19695.12998	&	32	&	1.5	&	1.59E-05	&	+	&	f	&	f2Pi	&	3	&	1	&	-0.5	&	0.5	\\
1464	&	19782.22031	&	32	&	1.5	&	3.21E-03	&	+	&	f	&	X4Sigma-	&	22	&	0	&	0.5	&	0.5	\\
1465	&	19788.59904	&	32	&	1.5	&	3.21E-03	&	+	&	f	&	X4Sigma-	&	22	&	0	&	1.5	&	1.5	\\
1466	&	19844.13679	&	32	&	1.5	&	2.13E-05	&	+	&	f	&	D4Delta	&	1	&	2	&	-1.5	&	0.5	\\
1467	&	19899.60278	&	32	&	1.5	&	1.28E-04	&	+	&	f	&	g2Pi	&	2	&	1	&	-0.5	&	0.5	\\
1468	&	19939.21599	&	32	&	1.5	&	2.13E-05	&	+	&	f	&	D4Delta	&	1	&	2	&	-0.5	&	1.5	\\
1469	&	19944.05045	&	32	&	1.5	&	1.87E-05	&	+	&	f	&	f2Pi	&	3	&	1	&	0.5	&	1.5	\\
1470	&	19952.35583	&	32	&	1.5	&	1.02E-07	&	+	&	f	&	C4Sigma-	&	3	&	0	&	1.5	&	1.5	\\
1471	&	19955.48362	&	32	&	1.5	&	1.02E-07	&	+	&	f	&	C4Sigma-	&	3	&	0	&	0.5	&	0.5	\\
1472	&	19964.05228	&	32	&	1.5	&	3.68E-03	&	+	&	f	&	c2Delta	&	11	&	2	&	-0.5	&	1.5	\\
1473	&	20017.11981	&	32	&	1.5	&	NaN	&	+	&	f	&	g2Pi	&	2	&	1	&	0.5	&	1.5	\\
1474	&	20017.3805	&	32	&	1.5	&	NaN	&	+	&	f	&	Ap4Phi	&	15	&	3	&	-1.5	&	1.5	\\
1475	&	20096.78221	&	32	&	1.5	&	NaN	&	+	&	f	&	d2Sigma+	&	10	&	0	&	0.5	&	0.5	\\
1476	&	20140.40162	&	32	&	1.5	&	NaN	&	+	&	f	&	A4Pi	&	13	&	1	&	-1.5	&	-0.5	\\
1477	&	20171.64156	&	32	&	1.5	&	NaN	&	+	&	f	&	A4Pi	&	13	&	1	&	-0.5	&	0.5	\\
1478	&	20197.23409	&	32	&	1.5	&	NaN	&	+	&	f	&	A4Pi	&	13	&	1	&	0.5	&	1.5	\\
1479	&	20262.64513	&	32	&	1.5	&	NaN	&	+	&	f	&	B4Pi	&	9	&	1	&	-1.5	&	-0.5	\\
1480	&	20317.01479	&	32	&	1.5	&	NaN	&	+	&	f	&	B4Pi	&	9	&	1	&	-0.5	&	0.5	\\
1481	&	20379.26176	&	32	&	1.5	&	NaN	&	+	&	f	&	B4Pi	&	9	&	1	&	0.5	&	1.5	\\
1482	&	20468.80376	&	32	&	1.5	&	NaN	&	+	&	f	&	a2Sigma-	&	16	&	0	&	0.5	&	0.5	\\
\bottomrule
\end{tabular}

\begin{tabular}{cll}
\\
             Column       &    Notation                 &      \\
\midrule
   1 &   $n$              &       Energy level reference number (row)    \\
   2 & $\tilde{E}$        &       Term value (in \cm) \\
   3 &  $g_{\rm tot}$     &       Total degeneracy   \\
   4 &  $J$               &       Rotational quantum number    \\
   5 & $\tau$ & Radiative lifetime \\
   6 & $+/-$ & Total parity  \\
   7 & $e/f$ & Rotationless parity \\
   8 & State & Electronic state \\
   9 & $v$ & State vibrational quantum number \\
  10 &  $\Lambda$ &   Projection of the electronic angular momentum \\
 11 & $\Sigma$ &  Projection of the electronic spin \\
12 & $\Omega$ &   $\Omega=\Lambda+\Sigma$ (projection of the total angular momentum) \\
\bottomrule
\end{tabular}
\end{table*}

  \begin{table}
\caption{\label{t:Transit-file} Extract from the transition file for $^{51}$V$^{16}$O.}
\begin{center}
\renewcommand{\arraystretch}{1.2}
\begin{tabular}{rrc}
\toprule
         $F$  &  $I$  & A$_{\rm IF}$ / s$^{-1}$ \\
\midrule
881 & 23 & 8.8193E-09 \\ 
641 & 23 & 6.1704E-03 \\ 
544 & 23 & 8.5161E-04 \\ 
761 & 23 & 5.6700E-03 \\ 
882 & 23 & 1.5569E-09 \\ 
642 & 23 & 1.4461E+02 \\ 
545 & 23 & 1.9240E-11 \\ 
762 & 23 & 5.1683E-08 \\ 
883 & 23 & 6.8306E-12 \\ 
643 & 23 & 1.5453E-08 \\ 
546 & 23 & 3.3650E-04 \\ 
763 & 23 & 1.1708E-08 \\ 
884 & 23 & 4.1096E-12 \\ 
644 & 23 & 1.1716E-07 \\ 
547 & 23 & 2.3980E-02 \\ 
764 & 23 & 3.5896E-09 \\ 
885 & 23 & 1.0346E-03 \\ 
\bottomrule
\end{tabular}

\noindent
 $F$: Upper state counting number

$I$:      Lower state counting number

$A_{IF}$:  Einstein $A$ coefficient in s$^{-1}$

\end{center}

\end{table}

\section{The Line List}
\label{sec:LL}
\subsection{Methodology}
Direct solution of the nuclear-motion Schr\"odinger
equation for diatomics systems with many coupled electronic states,
and a variety of spins and symmetries has recently been made possible
by the development of the program \Duo\ \citep{jt609}. \Duo\ has previously been
used to model
three electronic states for AlO by \citet{jt589}, who
computed a line list using this model \citep{jt598}. \citet{jt599}
consider six coupled electronic states in their study on ScH and \citet{jt618}
constructed a line list for CaO using 5 electronic states. This is the first
example where such a large number (13) of states have been considered.
The above references give full details
of the underlying theory used here to solve the rovibronic problem.
All VO calculations were made without
modification to the main \Duo\ code base. As time considerations for
diatomic molecules are minimal, for our final line list we used 301
grid points between 1.2 \AA{}  and 4.0 \AA{}  in these calculations.

\begin{figure}
\includegraphics[width=0.5\textwidth]{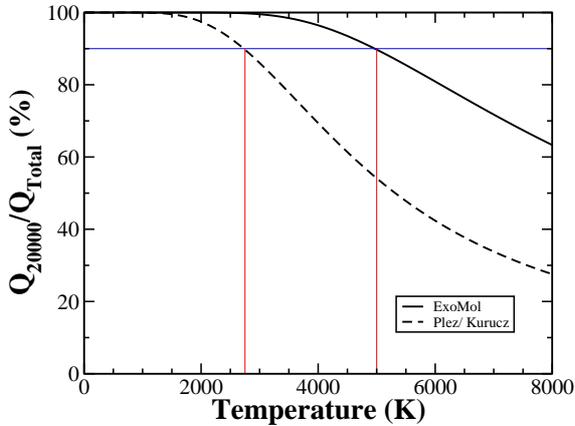}
\caption{\label{Qratio}Ratio of effective partition function used in the VO line list, $Q_{20000}$, to the converged total value, $Q_{\rm total}$.
This ratio gives a measure of completeness of the VO line list as a function of temperature.}
\end{figure}

The lower energy threshold was
set to 20,000 \cm{}. This means our line list is 90\% complete at 5000
K, as shown in Fig. \ref{Qratio}.

However, dipole moments were not included for transitions that
originated in the \C{}, \Df{} and \Dg{} states (i.e. states with $T_e$
above 16,000 \cm{}) because these would be negligible given the very
small population in these electronic states even at 5000 K.

The upper energy threshold was set at 50,000 \cm{}, just below the
dissociation energy at 52290 \cm{} \citep{83BaGiGu.VO}. The frequency range considered
is up to 35,000 \cm{}; however, the line list will not fully describe
emission spectroscopy of high lying energy levels due to the lower
energy threshold.

At this stage, we have produced a Duo VO linelist from the VOMYT spectroscopic model (and it is these energies and frequencies that are used in the comparisons in Tables 4 to 8). To increase the linelist's accuracy, we will make one final modification. 781 Duo energies were substituted with the \A{}, \B{} and \C{} experimental energies derived from combination differences as specified above to form the final VOMYT linelist. 

\subsection{Partition Function}

\begin{table*}
\caption{\label{tab:partition} Partition function, $Q(T)$, for ${}^{51}$V$^{16}$O, as a function of temperature. The literature partition function are multiplied by the nuclear degeneracy factor (8) to match ExoMol's standard definitions.}
    \def\arraystretch{1.2}
\begin{tabular}{rrrrr}
\toprule
T/K & \mc{1}{c}{VOMYT} & \mc{1}{c}{VOMYT} & {\citet{84SaTaxx.partfunc}} & {\citet{16BaCoxx.partfunc}}  \\
& \mc{1}{c}{Final states} & \mc{1}{c}{Up to J=270.5} & \\
\midrule
0	&	16.0	&	&	&  \\
10	&	314.5	& &	&417.8	\\
20	&	691.0	&&	&	824.8\\
50	&	1888.3	&&	&2045.9	\\
100	&	3915.2	&&	&4081.6	\\
300	&	12159.9	&&	&	12332.3\\
500	&	21428.4	&&	&21608.9	\\
1000	&	53462.1	&	53462.1	&	72288.9	& 53665.6\\
1500	&	100352.2	&	100352.1		&	142806.4	&100104.8\\
2000	&	165542.6	&	165542.0	&	237928.8	&161561.6\\
3000	&	384291.3	&	384314.8	&	518396.0	& 332684.8\\
4000	&	808962.6	&	809705.9	&	951848.0	& \\
5000	&	1566676.4	&	1573175.4	&	1584840.0	&925184.0 \\
8000	&	7045027.9	&	7293154.8	&	5289232.0	& 2859736.0\\
\bottomrule
\end{tabular}
\end{table*}

The partition function for $^{51}$V$^{16}$O were calculated by summing
all the calculated energy levels given by the \Duo\ calculation up to both $J=$ 200.5 and 270.5. The partition function is increased by 0.34\% at 5000 K for the latter more thorough calculation; this indicates convergence of the partition function from our model with respect to increased number of energy levels.
These results are compared with the result of \citet{84SaTaxx.partfunc} and \citet{16BaCoxx.partfunc} in
Table \ref{tab:partition}. Both these previous results rely on data from \citet{HerzHub}.
There are significant differences between the two literature values. Barklem \& Collet agree much better with our new ExoMol data up to around 2000 K,
while Sauval \& Tatum are closer at the very high temperature range.
In both cases, the partition function from the VOMYT spectroscopic model is significantly different from the pre-existing partition functions and should be used in preference to either of the former.  Note that the VOMYT partition function relies primarily on the potential energy curves, one of the most well-known aspects of the VO spectroscopic model due to the availability of significant experimental data. This gives higher reliability to our partition function.

\subsection{Lifetimes}

\begin{table}
\caption{\label{tab:lifetimes} Comparison of computed lifetimes, in $\mu$s, with the measurements of \citet{97KaLiLu.VO}.}
\begin{tabular}{llddd}
\toprule
& $v$ & \mc{1}{c}{\citet{97KaLiLu.VO}} & \mc{1}{c}{VOMYT} \\
\midrule
\A$_{5/2}$ & 0 & 7.0(4) & 18.7 \\
\A$_{3/2}$ & 0 & 5.2(3) & 19.1 \\
\A$_{1/2}$ & 0 & 5.3(3) & 19.6 \\
\A$_{-1/2}$ & 0 & 5.7(5) & 19.2 \\
\B$_{5/2}$ & 0 & 0.348(20) &  0.823\\
\B$_{5/2}$ & 1 & 0.328(22) &  0.803\\
\B$_{3/2}$ & 0 & 0.397(15) & 0.838 \\
\B$_{3/2}$ & 1 & 0.380(18) & 0.816 \\
\B$_{1/2}$ & 0 & 0.346(15) & 0.851\\
\B$_{1/2}$ & 1 & 0.420(35) & 0.85\\
\B$_{-1/2}$ & 0 & 0.406(18) & 0.870 \\
\B$_{-1/2}$ & 1 & 0.50(6) & 1.16  \\
\C & 0 & 0.073(2) & 0.103 \\
\C & 1 & 0.063(4) & 0.103 \\
\bottomrule
\end{tabular}
\end{table}

One important external check for our transition moments is the
comparison of our lifetimes against experimental values. This is shown
in Table \ref{tab:lifetimes}.  The ordering of lifetimes between the different electronic bands is definitely preserved, as is the rough order of magnitude of the result. However, there are discrepancies, particularly for the A-X transition. At this level, it is somewhat unclear whether this indicates errors in the \abinitio\ or experimental results. There are certainly errors in the \abinitio\ calculations, which we estimate to be approximately 10\%\ of the total value \citep{jt623}. However, there is no theoretical support for an approximately 40\%\ increase in the off-diagonal dipole moments which would be needed to bring the \abinitio\ and \citet{97KaLiLu.VO} experimental results into agreement. Therefore, we have decided to use pure \abinitio\ values for all transition moments. 
Further lifetimes measurements would be much appreciated to help resolve these discrepancies.

\subsection{Results}

A line list, known as VOMYT,  was calculated for ${}^{51}$V$^{16}$O which
 contains 277 million transitions, For
compactness and ease of use, it is divided into separate energy level
and transition files.
 This is done using the standard ExoMol format
\citep{jt548,jt631}. 
Extracts from the start of the
$^{51}$V$^{16}$O files are given in Table \ref{t:Energy-file} (the states
file) and Table \ref{t:Transit-file} (the transition file). The full line
list are available online\footnote{The full line list can be downloaded from the CDS, via \url{ftp://cdsarc.u-strasbg.fr/pub/cats/J/MNRAS/xxx/yy}, or
\url{http://cdsarc.u-strasbg.fr/viz-bin/qcat?J/MNRAS//xxx/yy} or \url{www.exomol.com}}. The line list
and partition functions together with the auxiliary data including the
potential parameters and dipole moment functions, as well as the
absorption spectrum given in cross-section format \citep{jt542}, can all
be obtained from \url{www.exomol.com}.

\begin{figure*}
\includegraphics[width=0.7\textwidth]{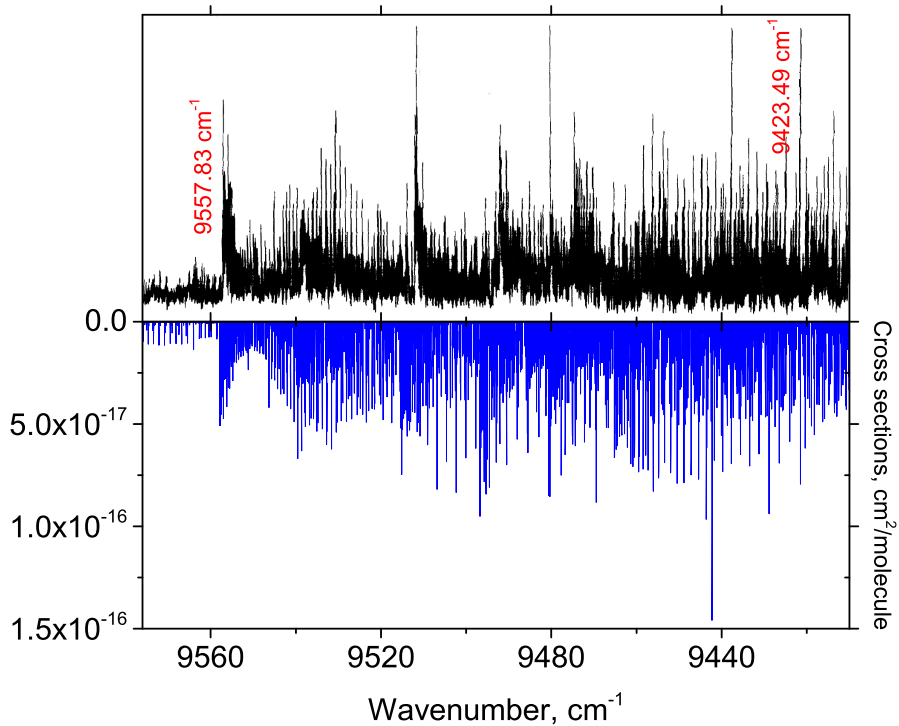}
\caption{\label{fig:AgainstExp1} Comparison of VO X-A band (bottom panel) against laboratory spectra (top panel).}
\end{figure*}

\section{Comparisons}
\label{sec:comparisons}

Figure~\ref{fig:AgainstExp1} compares the A-X 0-0 band in the region 9420 to 9560 \cm{} as observed by \citet{82ChTaMe.VO} against a cross-section produced with our new linelist using ExoCross LTE emission at 1000~K, with HWHM=0.03 \cm. It is clear that the bandhead is very well reproduced (within 0.01 \cm{}). The general structure of the rest of the region is good, though not perfect.

\begin{figure*}
\includegraphics[width=0.7\textwidth]{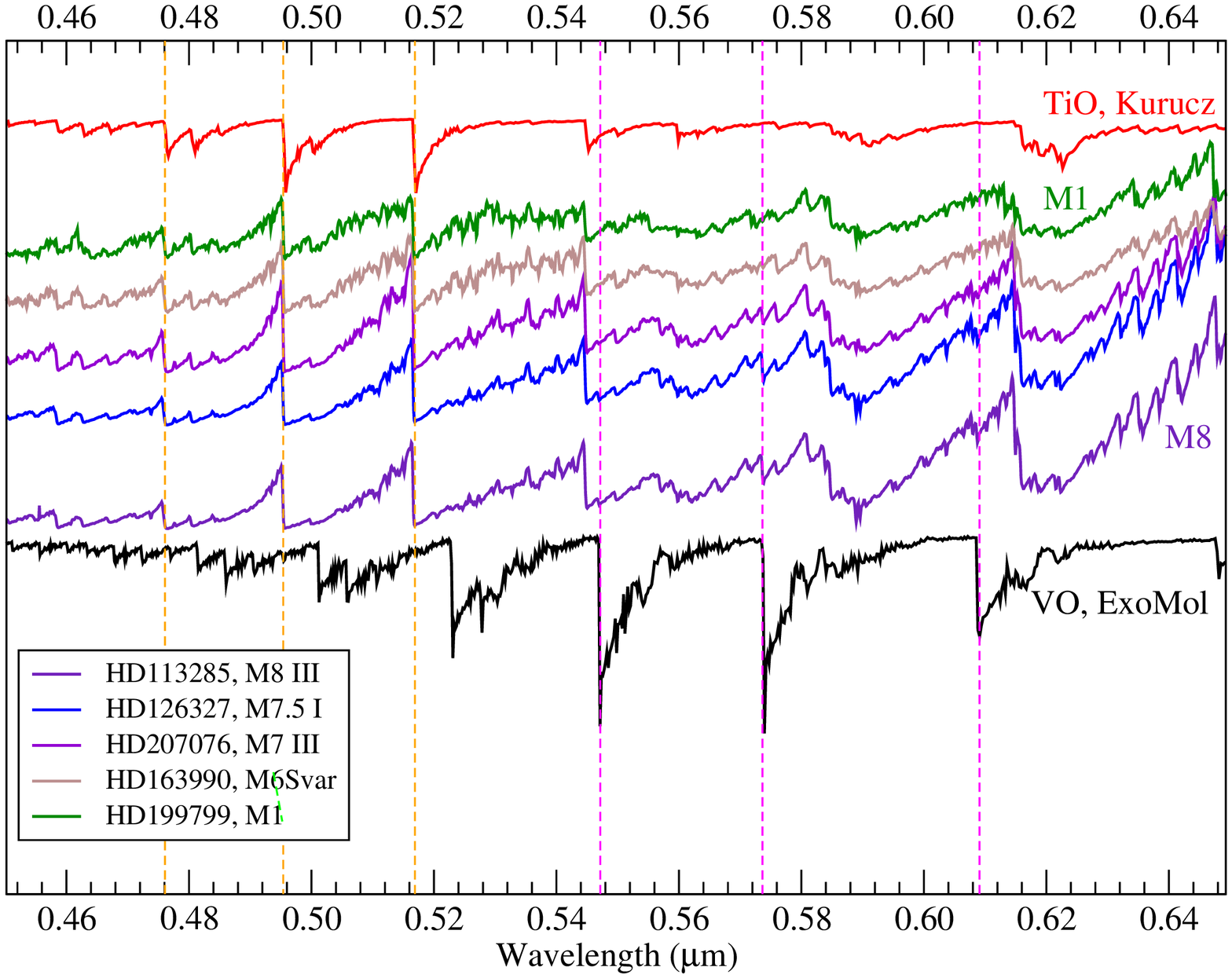}
\includegraphics[width=0.7\textwidth]{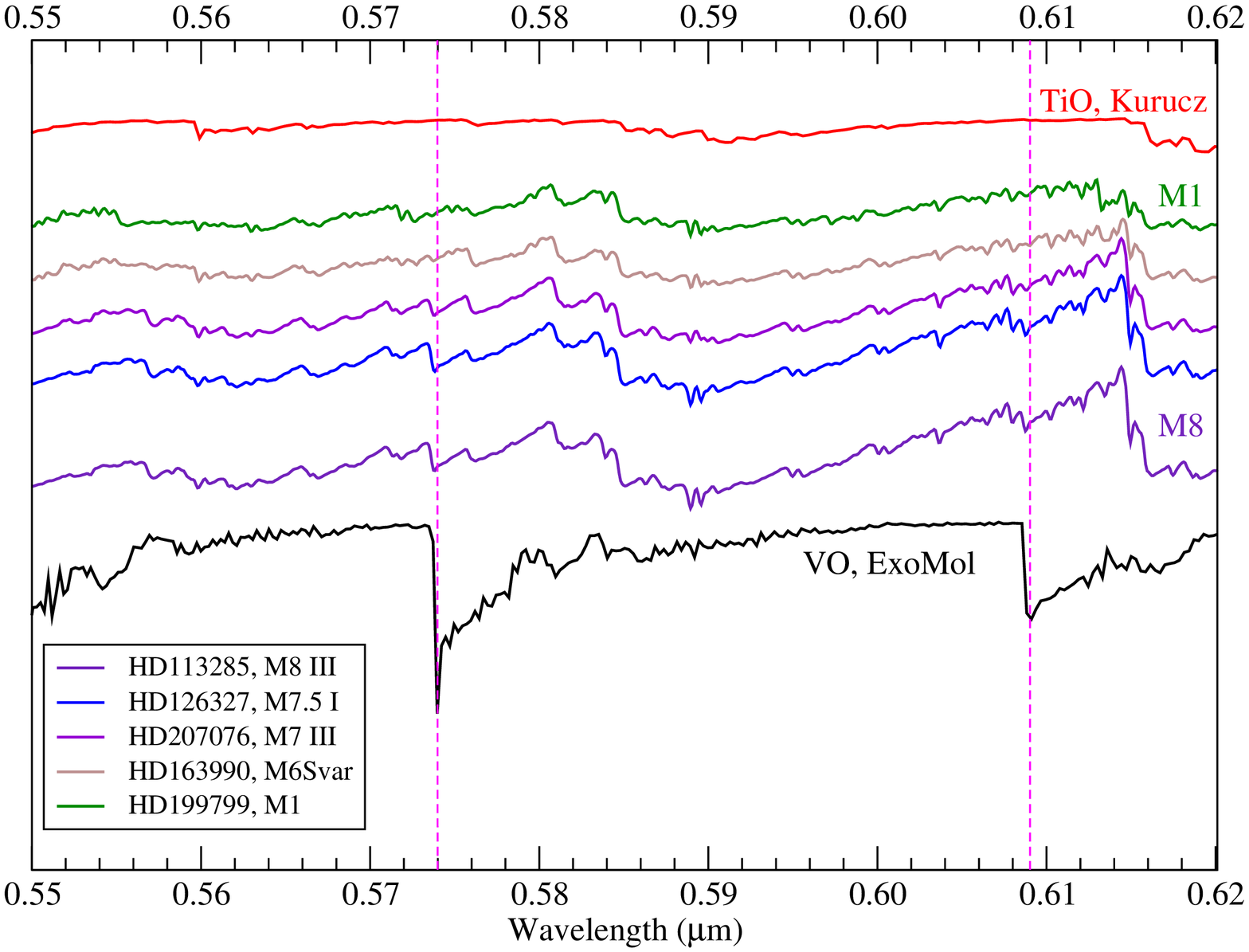}
\caption{\label{fig:Mstars} Comparison of VO spectra against stars of M stellar class (in legend) in the region of the C-X bands. Astronomical data are taken from the MILES stellar library of
\citet{06SaPeJi.coolstars} and \citet{11FaSaVa.coolstars}. Vertical lines indicate bandheads. For both panels, the stars from top to bottom are HD199799 (M1), HD163990 (M6Svar), HD207076 (M7III), HD12327 (M7.5 I) and HD113285, M8III).  }
\end{figure*}


The top plot in Fig. \ref{fig:Mstars} shows the spectra of M-type stars between 0.46 - 0.64 \um{} (15,600 to 22,000 \cm{}) against the new VO ExoMol line list and the pre-existing TiO linelist by Kurucz. The bottom plot is a zoomed-in view of the region from 0.56 to 0.61 \um{} (16400 to 17,900 \cm{}). The absorption by VO in this spectral region is dominated by the C-X transition bandheads. The top plot shows clearly that the main bandheads observed in the M-star spectra arise from TiO, not VO; this is expected as TiO is about an order of magnitude more abundant than VO. However, at approximately 0.548, 0.574 and 0.609 \um{}, the VO bandheads become strong while there are weak or non-existent nearby TiO bandheads. In this regions, spectral absorption features in the M-type spectra do align well with the strong VO bandheads. This is further confirmed by the depth of this absorption feature as the spectral class changes from early M-type to late M-type stars (top to bottom). The absorption feature is significantly more pronounced for the cooler star. This aligns with the fact that late M-type stars have more VO than early M-type stars (in fact, the presence of VO absorption bands is one of the characterising spectral features of late M-type stars).  

\begin{figure}
\includegraphics[width=0.5\textwidth]{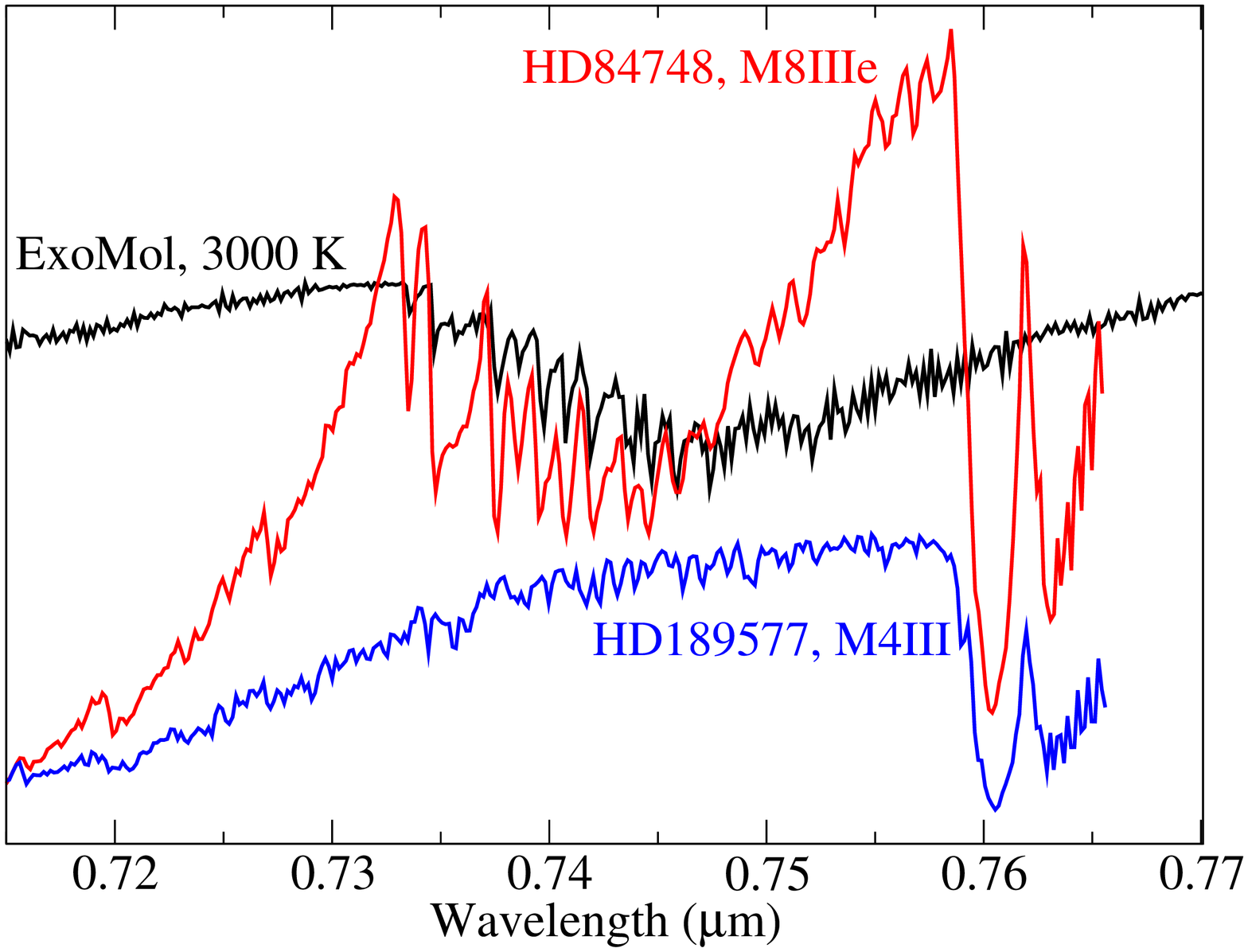}
\includegraphics[width=0.5\textwidth]{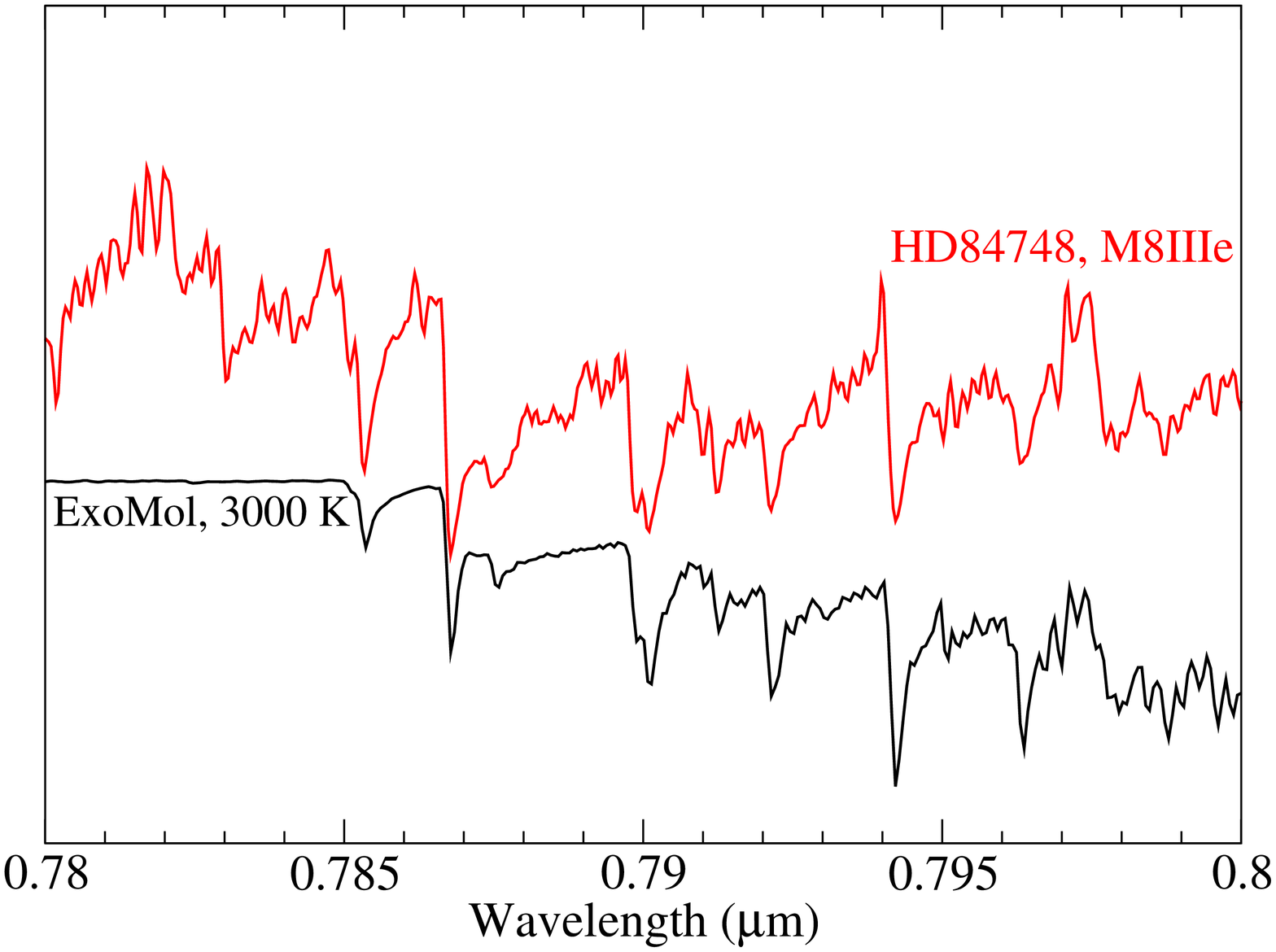}
\includegraphics[width=0.5\textwidth]{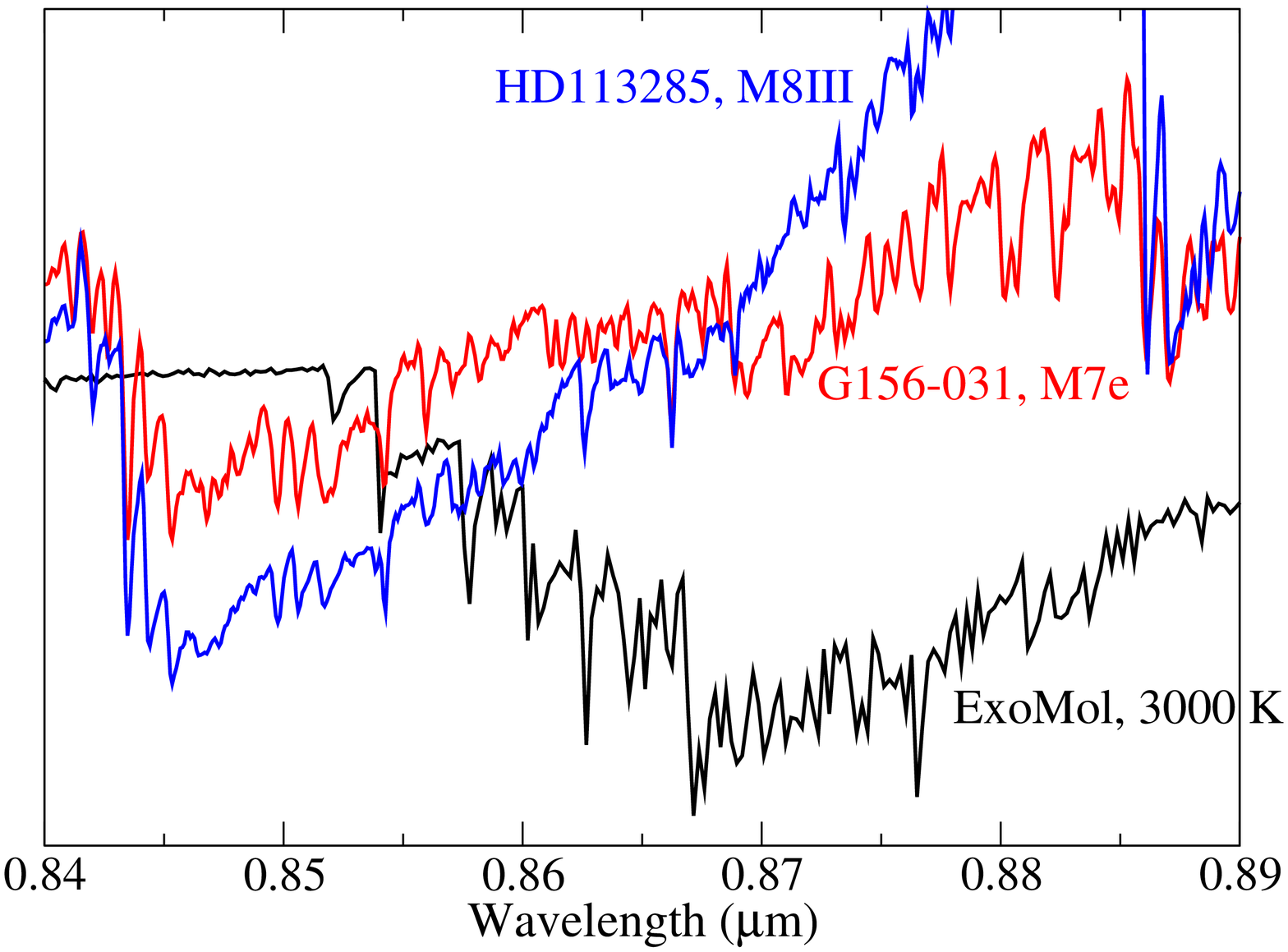}

\caption{\label{fig:Mstars3} Comparison of VO spectra against stars of M stellar class (in legend) in the region B-X 1-0 (top), 0-0 (middle) and 0-1 (bottom) bands. Astronomical data is taken from the MILES stellar library by \citet{06SaPeJi.coolstars},\citet{11FaSaVa.coolstars}  for the top 
two panels  and from CaT stellar library from \citet{01CeCaGo.coolstars} for the bottom panel. Vertical lines indicate VO bandheads. }

\end{figure}

Fig. \ref{fig:Mstars3} compares the cross-section in the B-X region from the new VOMYT linelist against stellar spectra. The top plot is the B-X 1-0 band in the region 0.72 to 0.77 \um{} (12,990 to 13,890 \cm{}).  Note that the M-star spectra are total cross-section, including a black-body contribution that rises steeply in this spectral region, whereas the ExoMol cross-section is a pure absorption spectra on a fixed intensity background. It is expected that the M8 star contains significant VO absorption features which are weak or non-existent in the M4 spectra. This is indeed what is observed. The agreement in the region 0.73 to 0.75 \um{} is particularly strong. This corresponds to the B-X 1-0 band. The blackbody background is probably responsible for at least some of the differences from 0.75 to 0.758 \um{}, while the difference in the 0.758 to 0.764 \um{} is probably caused by absorption by another molecule. 

The middle plot is the 0-0 band, centered around 0.79 \um{} (12660 \cm{}). This figure compares the VO ExoMol cross-section against the spectra of a late M-type star in the region 0.783 to  0.81 \um{} (12,770 to 12,350 \cm{}).  The absorption peaks in the stellar spectra and the VOMYT cross-section are extremely similar, giving high confidence to our VOMYT linelist. 

Finally, the bottom plot of Fig. \ref{fig:Mstars3} is the B-X 0-1 band with origin around 0.862 \um{}. 
This figure compares the VOMYT cross-section at 3000 K against the spectra of a M7 and M8 star in the region 0.844 to 0.890 \um{} (11,230 to 11,850 \cm{}). Some features clearly align between the VOMYT and stellar spectra, as indicated by the vertical lines.

\begin{figure}
\includegraphics[width=0.5\textwidth]{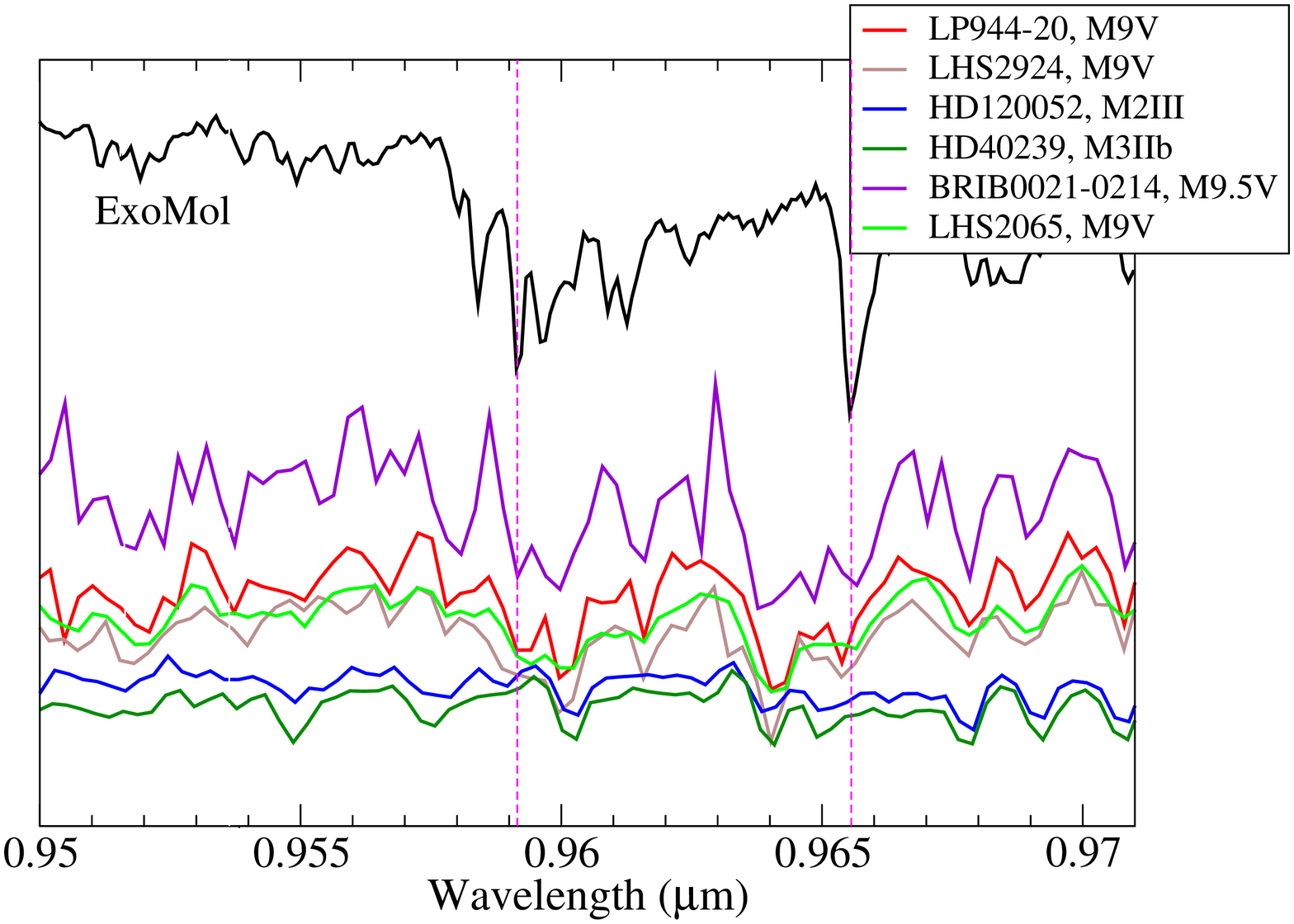}
\includegraphics[width=0.5\textwidth]{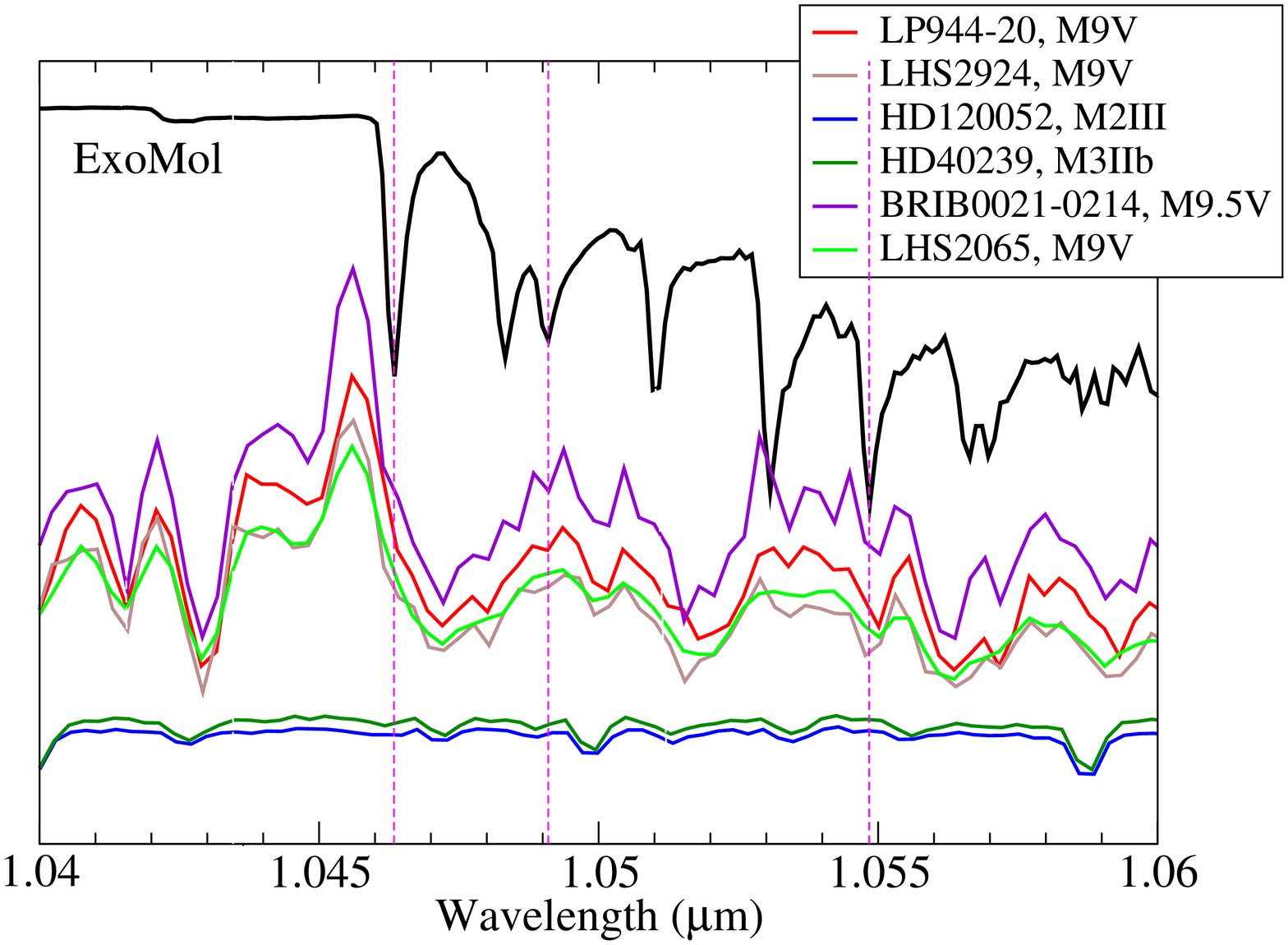}
\includegraphics[width=0.5\textwidth]{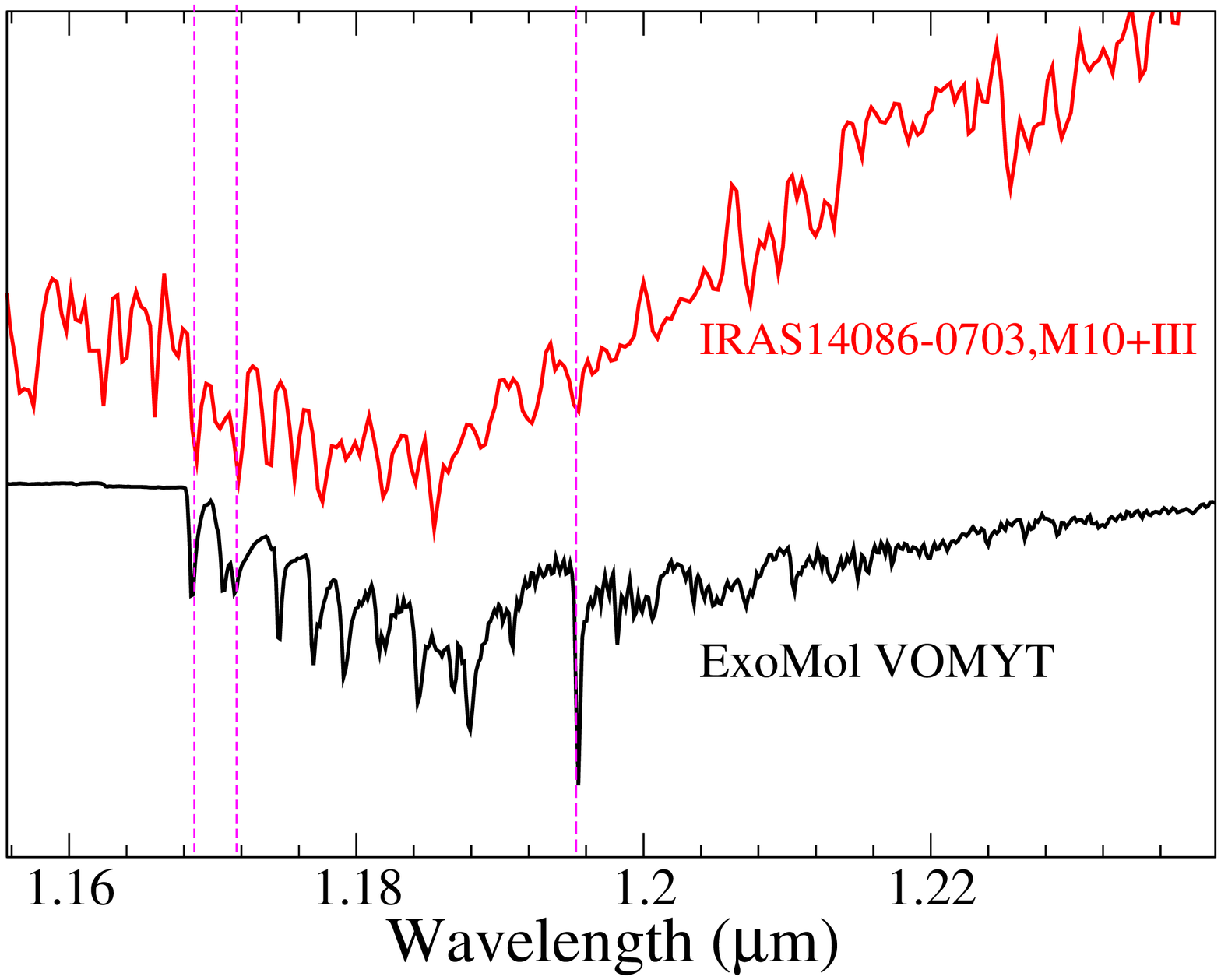}
\caption{\label{fig:Mstars5} Comparison of VO spectra against stars of M stellar class (in legend) in the region of A-X bands.
Astronomical data is taken from \citet{05CuRaVa.TiO,09RaCuVa.coolstars}. Vertical lines indicate VO bandheads. For the top two panels, the stellar curves from top to bottom are BRIB0021-0214 (M9.5V), LP944-20 (M9V), LHS2065 (M9V), LHS2929 (M9V), HD40239 (M3IIb) and HD120052 (M2III).}
\end{figure}

Figure~\ref{fig:Mstars5} compares the VOMYT VO cross-section against stellar spectra in the region of the A-X 0-1 (top), 0-0 (middle) and 1-0 (bottom) bands. Considering first the middle subplot (the 0-0 band). The first major peak in the VO cross-section at around 1.046 \um{} aligns well with a bump in the M9 spectral absorption, though there is obviously another species causing a nearby strong absorption at 1.046 \um{}. The other peaks of the VO spectra all align reasonably well or at least are not inconsistent with the M9 stellar spectra, though there is again many other absorption peaks in this region. Note that the M2 and M3 stellar spectra seem reasonably featureless in this region and their absorption features do not coincide with VO's (as expected). The M2 and M3 absorption features do seem roughly correlated with some other absorption in the M9 stellar spectra. 

The top subplot of Fig. \ref{fig:Mstars5} covers the 0-1 band region and shows similar characteristics to the 0-0 band;
the VOMYT absorption features are present in (or consistent with) the M9 but not the M2/M3 spectra and there are many other absorption features in the M9 and M2/M3 spectra that do not arise from VO absorption, but probably a different molecule. 

The bottom subplot of Fig. \ref{fig:Mstars5} covers the VO 1.2 \um{} band. There is little experimental data on the vibrationally excited \A{} state (only a tentative assignment of the vibrational frequency). Despite this, the agreement between the stellar spectra and the VO linelist is reasonably good. The shape of the band is well reproduced and identification of the spectral features that can be attributed to VO are relatively clear. However,  the positions of the spectral lines is somewhat in error (on order 10 \AA, around 10 \cm{}). This is about the expected error of this band and is a good indicator of the quality of this line list where direct experimental frequency measurements are not available, e.g. many hot and overtone bands. 

Thus, our comparisons in Fig. \ref{fig:Mstars}, \ref{fig:Mstars3} and \ref{fig:Mstars5} show that the VO absorption features from our ExoMol cross-sections are consistent in the major bands to absorption in the late M-type stars, while VO absorption features are absent in the early M-type stars, as expected. 
\begin{figure}
\includegraphics[width=0.5\textwidth]{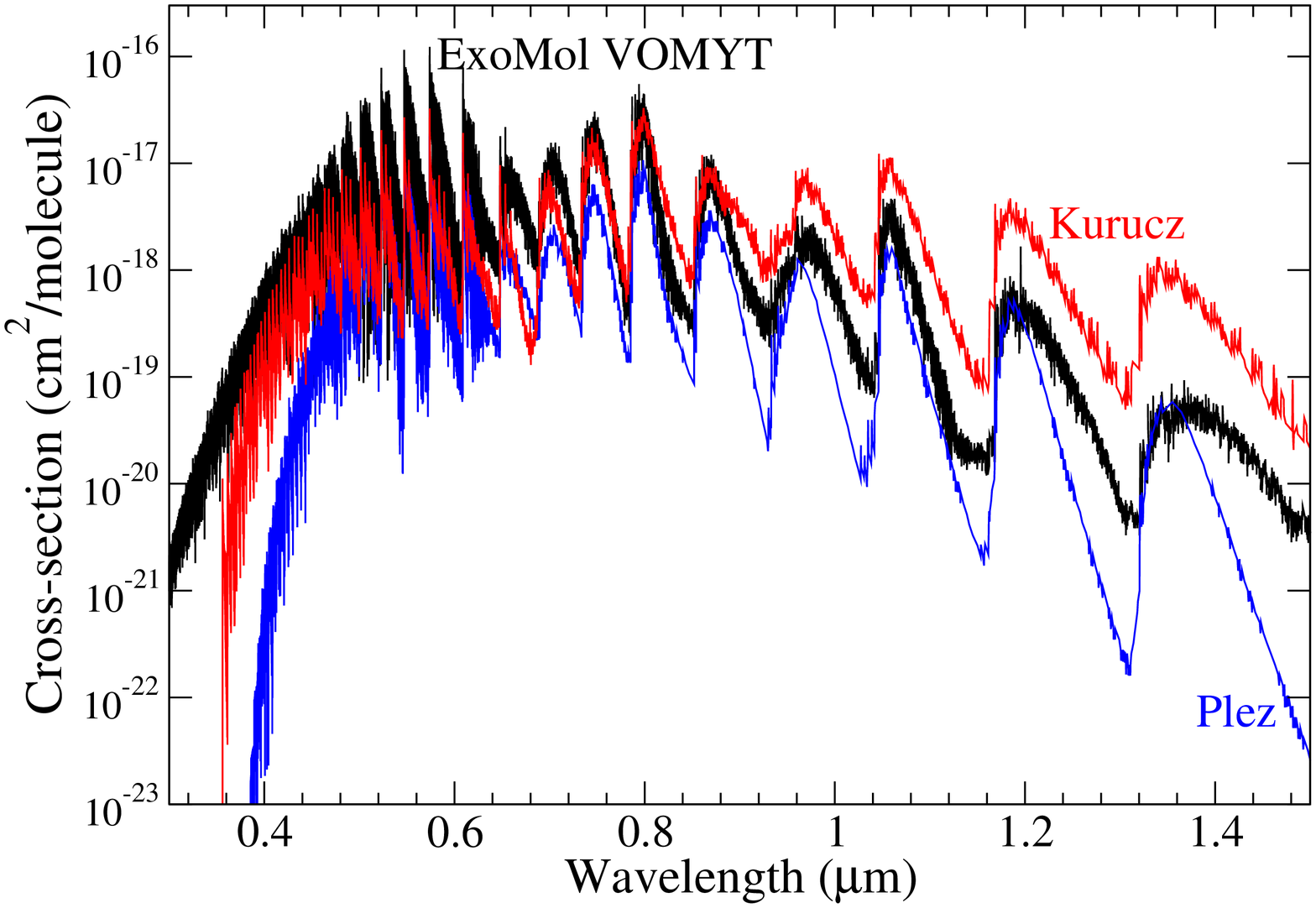}
\includegraphics[width=0.5\textwidth]{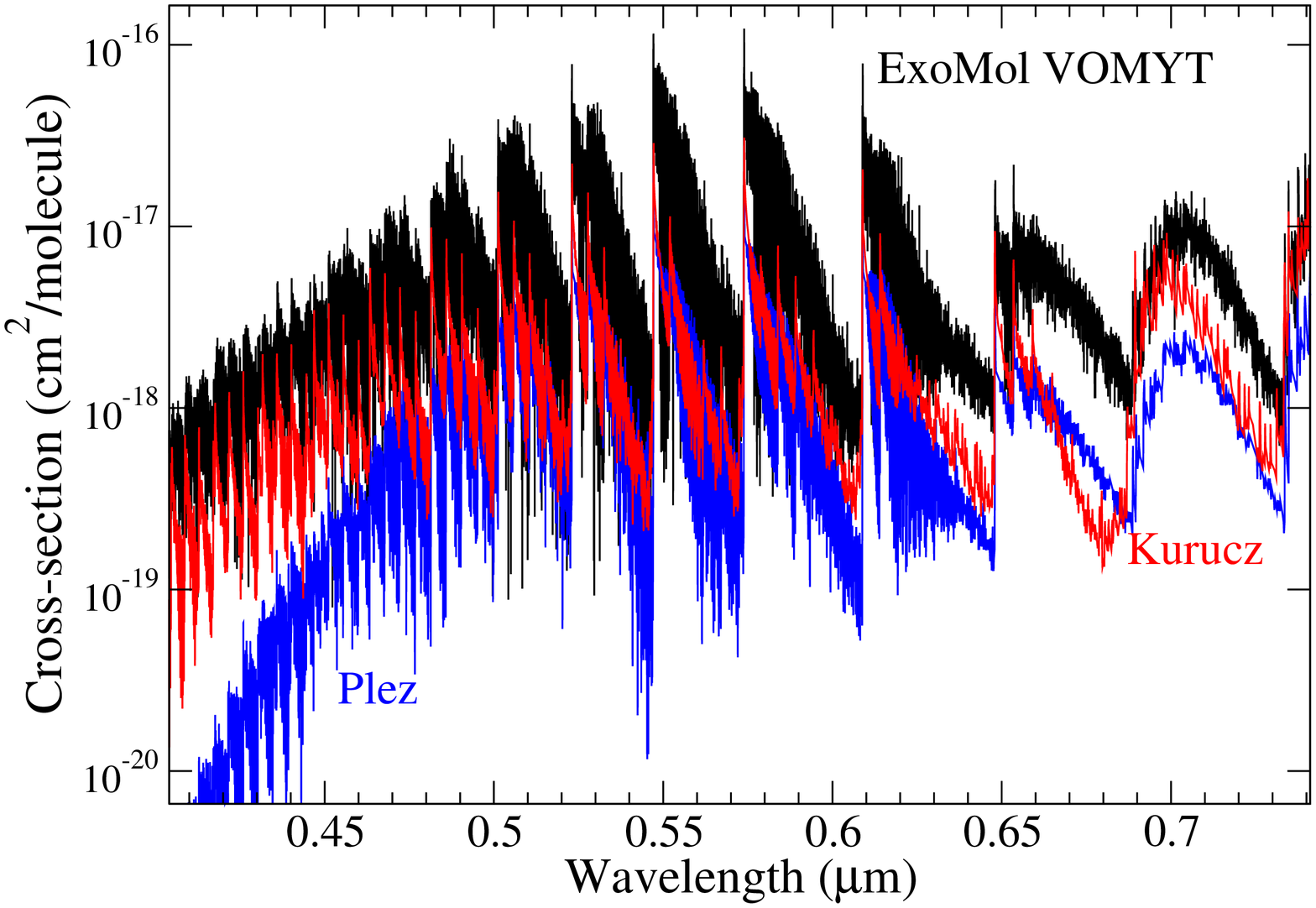}
\includegraphics[width=0.5\textwidth]{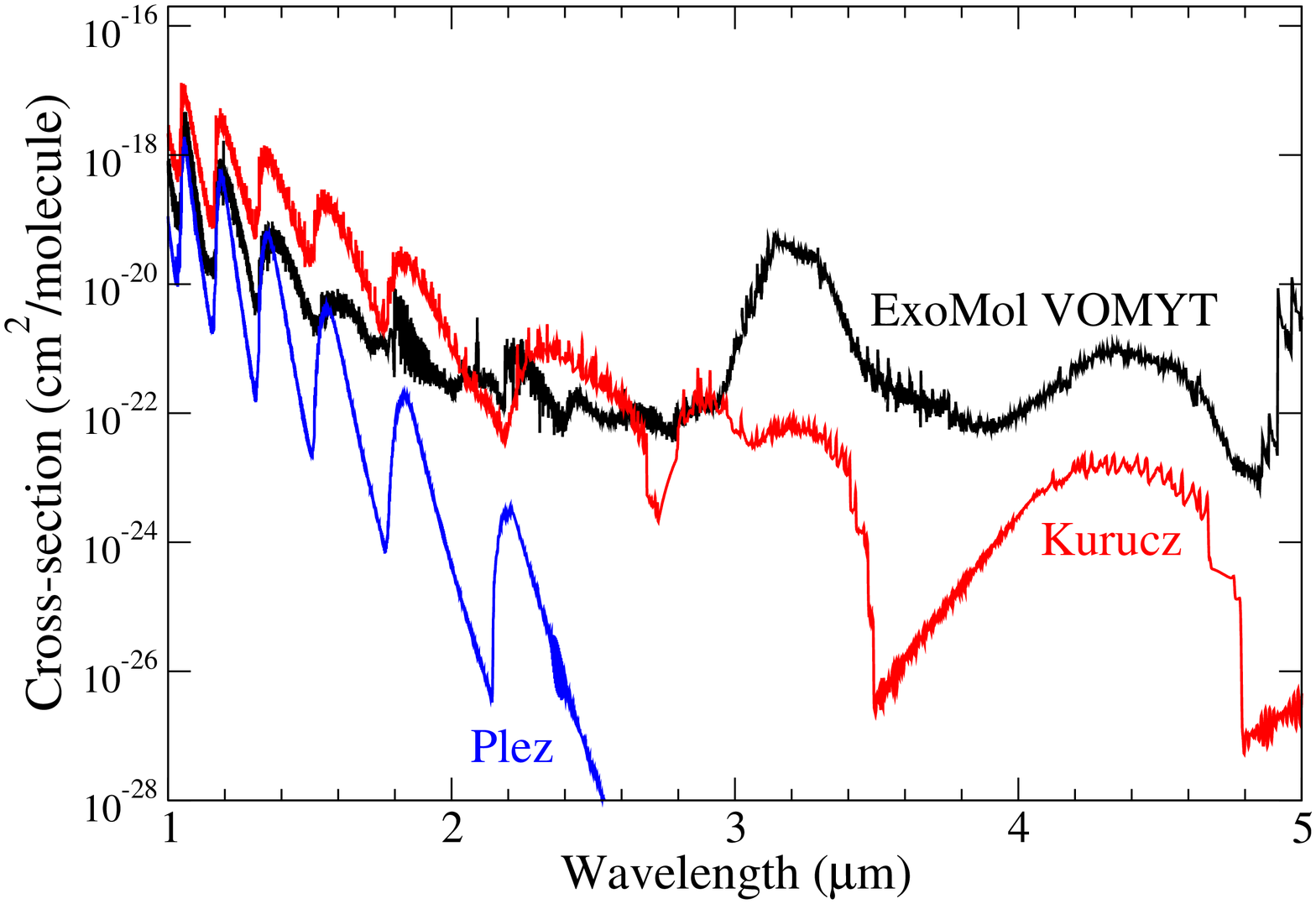}
\caption{\label{fig:Burrows} Comparison between the Plez, Kurucz and new ExoMol VOMYT line lists for VO. All cross-sections are obtained at 2200 K and using a HWHM of 1 \cm{}.}
\end{figure}


\begin{table*}
\def\arraystretch{1.2}
\caption{\label{tab:LineList} Summary of available VO linelists.}
\begin{tabular}{lrrrrrr}
\toprule
&   \citet{99Plxxxx.VO} & \citet{Kurucz} & VOMYT   \\
\midrule
\# Electronic States &  4 & 4 &  13 \\
Max $v$ & 15 & 15 & 59  \\
Max J & 199.5 & 199.5 & 197.5 \\
Max Energy (\cm{}) & 42,387 & &  50,000  \\
Min Freq. (\cm{}) & 3,767 & 0.537& 0.01 \\
Max Freq. (\cm{}) & 25,939 & 28,096 & 35,000 \\
Min WaveLength (\um{}) & 0.386 & 0.355  & 0.2857 \\
Max WaveLength (\um{}) & 2.655 & 18627.4 & 1,000,000 \\
Number of lines &   3,171,552 &  4,509,519 &277,131,624 \\
\bottomrule
\end{tabular}

\end{table*}


 The
 characteristics of the \citet{Kurucz}, \citet{99Plxxxx.VO}  and new ExoMol VOMYT linelists are summarised in Table
 \ref{tab:LineList}. Note that even though the Kurucz line list has very low frequency transitions, all transitions are electronic; the infrared rovibrational transitions within the X state are not included.  It is clear that the new VOMYT linelist is significantly more complete, with more than 60 times more lines covering both shorter and longer wavelengths than either of the previous line lists. There are more electronic states considered, and more vibrational energy levels. The maximum rotational quantum number $J$ is slightly lower due to the fact we used a lower energy cut-off to exclude all spectral lines originating from initial energies greater than 20,000 \cm{}. 
 
Use of our partition function suggests that the
Kurucz and Plez line lists are 90\% complete at 2800 K, but fall to
approximately 50\% complete by 5000 K. This incompleteness means that for high temperatures
significant sources of opacity will be missing.
The Plez line list has used extensively in models of hot Jupiters, brown dwarfs
and cold stars, eg. by \citet{99Plxxxx.VO} as well as within the MARCS
\citep{08GuEdEr.model}, PHOENIX \citep{PHOENIX} and VSTAR \citep{12BaKexx.dwarfs} 
modelling programs.

Fig. \ref{fig:Burrows} compares the Plez, Kurucz and ExoMol VOMYT line lists in the regions 0.3 and 5 \um{}. 
It is clear that the three line lists are broadly similar shorter than about 2 \um, though the ExoMol VOMYT linelist extends to shorter wavelengths (and thus more to the blue). However, there are noticeable differences in intensities. For the A-X bands around 1 \um, the Kurucz linelist is significantly stronger than either the Plez or ExoMol VOMYT line lists. The three line lists have similar strengths around the B-X bands, though the ExoMol line list is generally strongest especially at regions far from the bandhead; this is expected as it is most complete. The ExoMol VOMYT line list has stronger transitions in the C-X bands than either the Kurucz or Plez line lists. 

Looking to the longer wavelength regions, the Plez line list does not represent the region longer than 2 \um{} well. The Kurucz linelist is better with some features, but the ExoMol VOMYT linelist is significantly more complete. This is because the ExoMol linelist for the first time includes transitions from other electronic states, and from excited electronic states. The additional infrared absorption of VO described by our new line list may be important for energy transport and the spectroscopy of stellar and planetary atmospheres which contain VO molecule.
%
%

\begin{figure}
\includegraphics[width=0.5\textwidth]{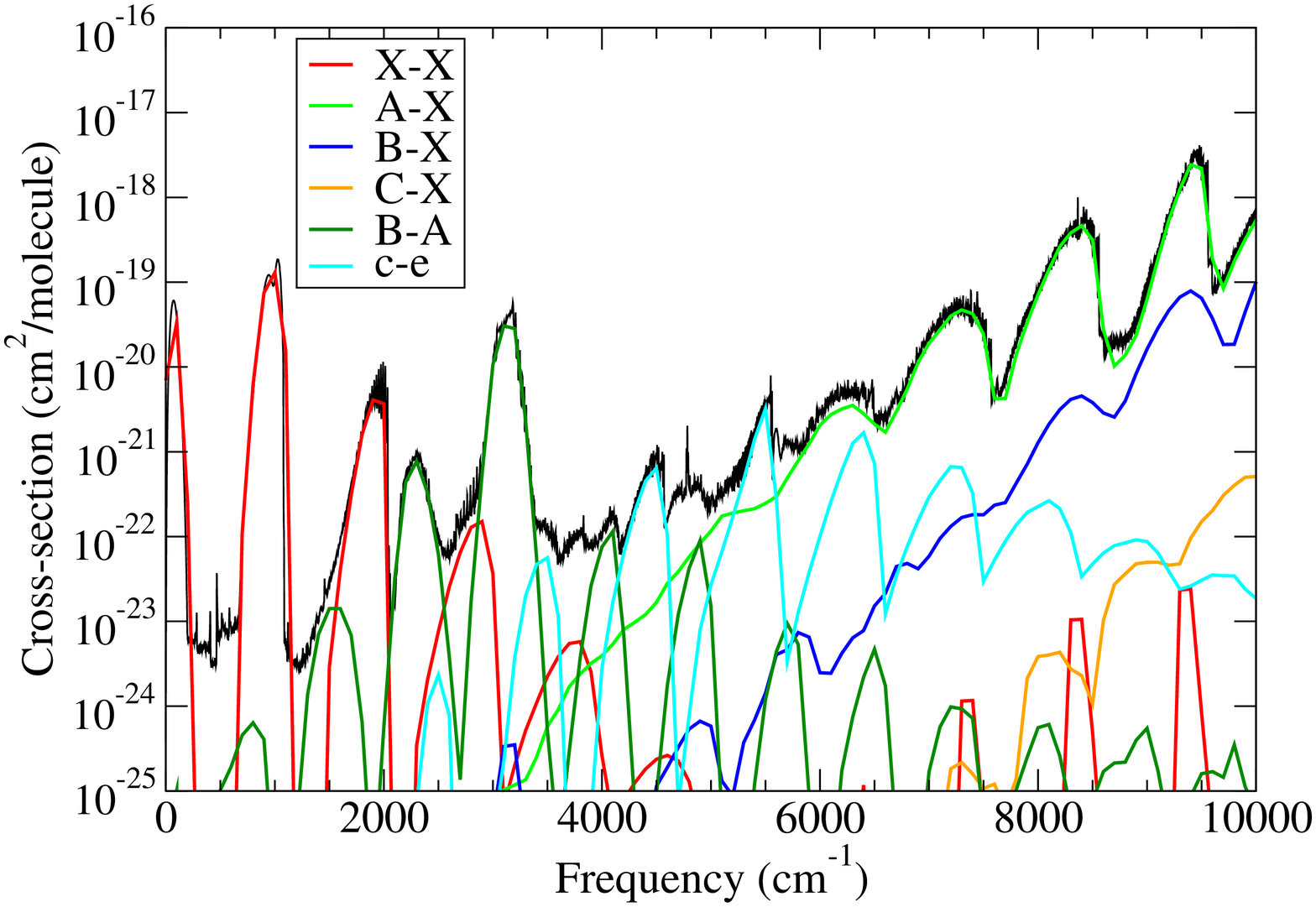}
\includegraphics[width=0.5\textwidth]{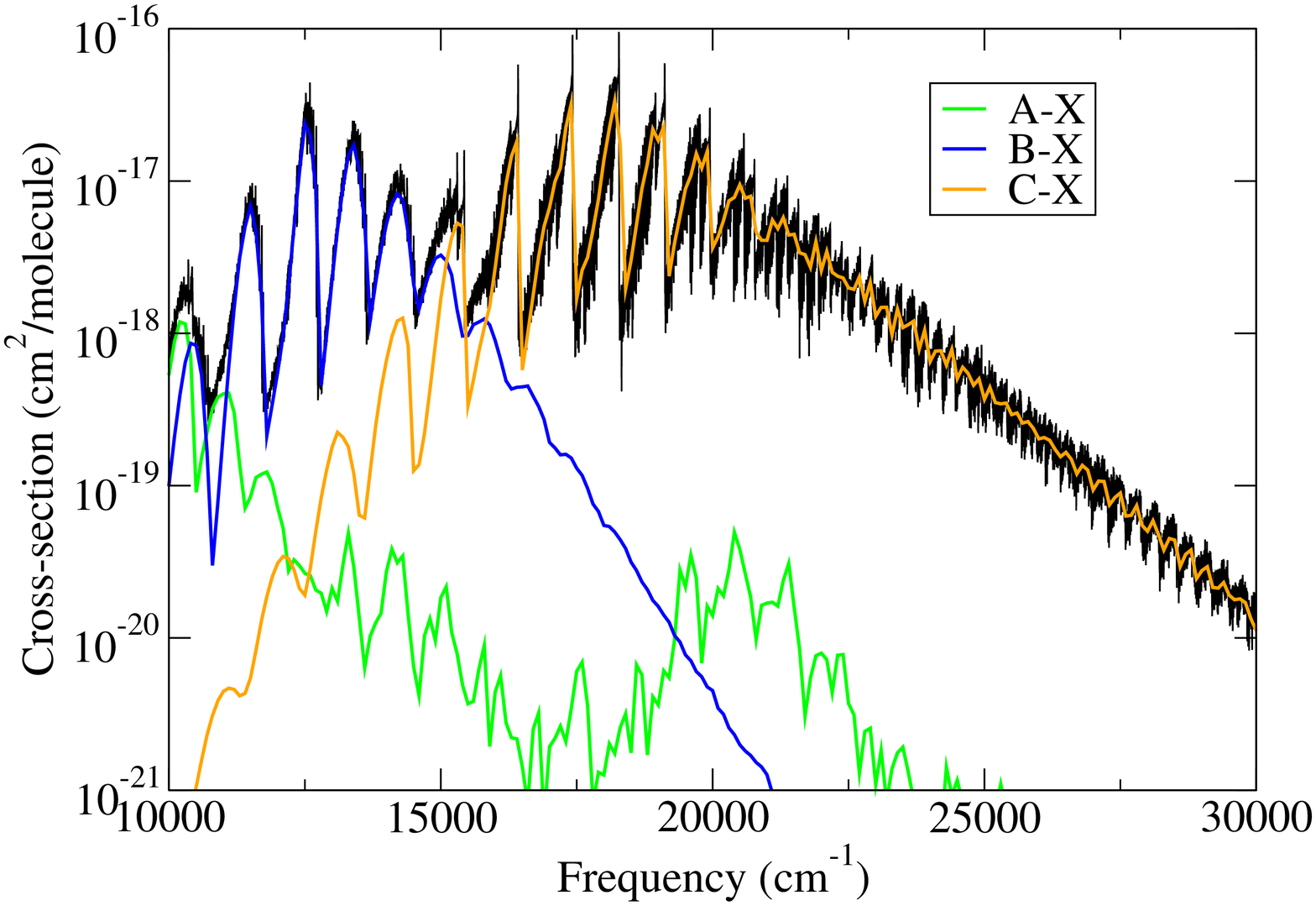}
\caption{\label{fig:cmfull} Decomposition of the total cross-section of VO below 30,000 \cm{} into the main bands. HWHM = 1 \cm{}, T=2200K. In the top panel, the X-X transition is responsible for the first three main peaks. The A-X transition is responsible for all peaks from 6000 - 10,000 \cm{}. The c-e transition produces peaks at around 4500 and 5500 \cm{}. The B-X transition peaks around 2300 and 3100 \cm{}. The C-X and B-X transitions are weak below 10,000 \cm{} and do not contribute significantly to the absorption.  In the bottom panel, the curve peaking at around 10,000 \cm{} is for the A-X band, the curves peaking between 10,000 - 15,000 \cm{} is the B-X band and the curve peaking from 15,000 to 25,000 \cm{} is the C-X band. }
\end{figure}

Figure \ref{fig:cmfull} identifies the main bands in the absorption spectra of VO at 2200 K.  The strongest microwave and infrared bands are about three orders of magnitude weaker than the strongest visible bands.  In visible spectral region, the C-X transition has an inherently stronger intensity than the B-X transition, which is stronger than the A-X transition.
It is clear that the A-X, B-X and C-X transitions contribute almost all of the opacity above 7000 \cm{} (below 1.4 \um).  At low frequencies (longer wavelengths), there are a significant number of contributing transitions, particularly the e-c, B-A and X-X transitions.

\begin{figure}
\includegraphics[width=0.5\textwidth]{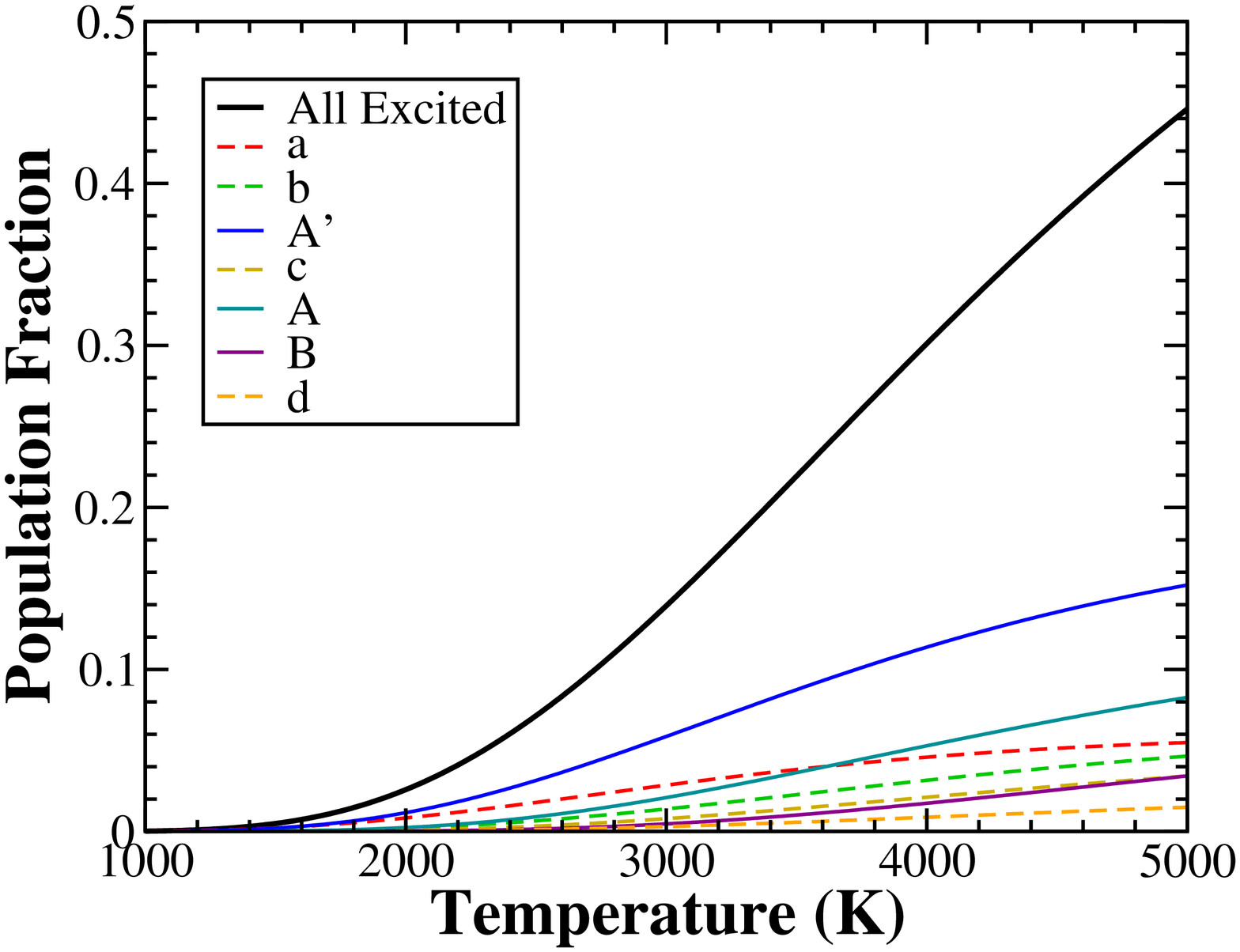}
\caption{\label{fig:PF} Relative population of excited states (where full population is normalised to 1). At 5000 K, the lines from top to bottom refer to All excited states, A', A, a, b, B, c and d states. The solid lines are quartet states, while the dotted lines are doublet states.}
\end{figure}

The importance of these non-ground electronic states can be quantified by considering  the thermal distribution of initial states in an ensemble of VO
molecules. Fig. \ref{fig:PF} shows the relative population of excited
states as a function of temperature. At 2000 K, only 2\% of molecules
are in excited states; by 5000 K, this increases to almost 45\%!
Furthermore, there are many excited states that are populated
significantly; though the A$^{\prime}$ state have the highest population (15\%),
there is significant population (above 1\%) in six other excited
states.

Our calculated transitions associated with the ground state will
generally be more reliable than transitions that occur between
excited states. This is due to existing experimental data. For example, $T_e$ for the \Da{} state is not well known
experimentally and thus the frequencies of all transitions from this
state have inherent large uncertainties. Limitations in the accuracy
of the \abinitio\ calculations also needs to be considered; low lying
states are generally described more accurately than higher electronic
states. 

\begin{figure}
\includegraphics[width=0.5\textwidth]{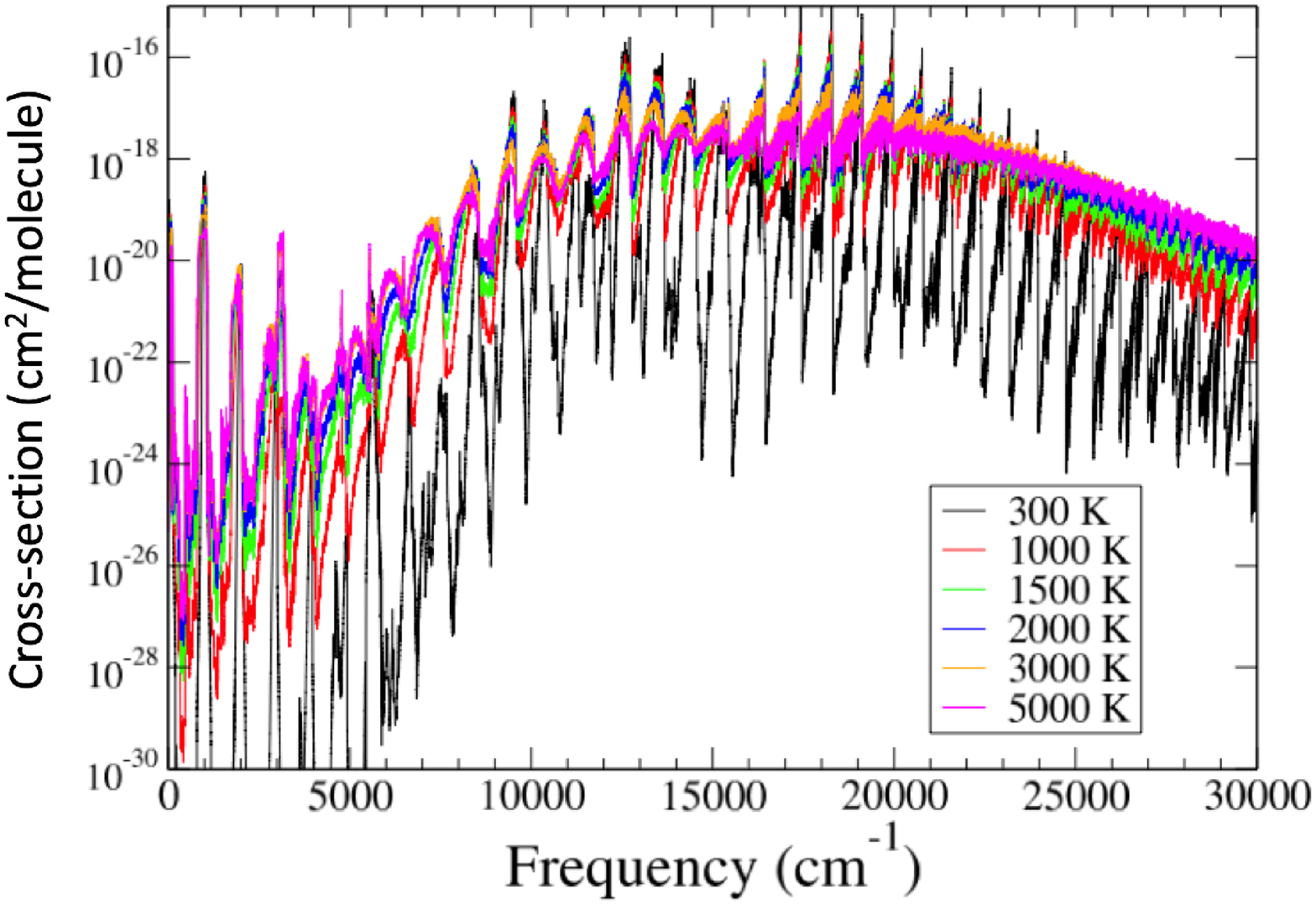}
\caption{Overview of the full spectrum of VO for $T$ = 300, 1000, 2000, 3000 and 5000~K, absorption cross-sections (cm$^2$/molecule) with HWHM = 1 cm$^{-1}$. Looking at the minimum of the spectra, the cross-sections are ordered in increasing temperature.}
\label{fig:TempEffect}
\end{figure}

Fig. \ref{fig:TempEffect} shows the absorption spectra as a function of temperature. At higher temperatures, the structure of the spectra becomes less defined and broader, as expected. 

\section{Conclusion}

Transition metal diatomics are important species in cooler stars and hot
Jupiters. However, the difficulty of the \abinitio\ calculations and the
relative lack of experimental data mean that it is difficult to construct high
quality line lists for these species.

Here we present the first ExoMol line list for a transition metal diatomic
species of astrophysical relevance. Work on CrH, MnH and TiH is in advanced stages and will be published shortly. Using the lessons from the
construction of the VO line list, we are now actively working on an improved
high quality line list for TiO to address much discussed shortcomings in the
existing line list in terms of intensities and at very high resolution. This new
TiO line list will use high quality \abinitio\ results and be fitted to all
available experimental data. Furthermore, a {\sc Marvel}-type analysis \citep{jt412} is currently underway to extract high quality experimental energies
from experimental frequencies.

Our VOMYT rovibronic line list for VO, containing over 277 million
transitions, can be accessed online at www.exomol.com in the extended ExoMol format
described by \citet{jt631}.
It includes the transition energies and Einstein coefficients,
partition functions, lifetimes and temperature-dependent cross-sections. Land\'{e} g factors to describe the splitting of the energy levels due to the Zeeman effect will be added shortly. We have also included the Duo input file with our spectroscopic model for VO.

\section*{Acknowledgements}

This work is supported by ERC Advanced Investigator Project 267219.  The authors acknowledge the use of the UCL Legion High Performance Computing Facility (Legion@UCL), and associated support services, in the completion of this work. 
Thanks to Joanna Barstow,  Adam Burgasser, Jane MacArthur, August Muench and Hannah Wakeford for recommending good sources of astronomical data to compare against our linelist's predictions.


\bibliographystyle{mn2e}
\end{document}